\documentclass[letterpaper, 11pt]{article}
\pdfoutput=1

\usepackage{amsmath,amssymb}
\usepackage{bm,bbm}
\usepackage{graphicx, graphics, color}
\usepackage{cite}
\usepackage{slashed}
\usepackage{caption, subcaption}
\usepackage{hyperref}
\usepackage{multirow}
\usepackage{array}

\usepackage[letterpaper, margin=1in]{geometry}
\numberwithin{equation}{section}

\newcommand{\fref}[1]{Fig.~\ref{f.#1}}

\newcommand{\erefn}[1]{~(\ref{e.#1})}

\newcommand{\sref}[1]{Section~\ref{s.#1}}
\newcommand{\ssref}[1]{Section~\ref{ss.#1}}
\newcommand{\sssref}[1]{Section~\ref{sss.#1}}
\newcommand{\tref}[1]{Table~\ref{t.#1}}

\newcommand{\beq}{\begin{eqnarray}}
\newcommand{\eeq}{\end{eqnarray}}
\newcommand{\beqa}{\begin{equation}\begin{aligned}}
\newcommand{\eeqa}{\end{aligned}\end{equation}}
\newcommand{\beqs}{\begin{eqnarray*}}
\newcommand{\eeqs}{\end{eqnarray*}}

\newcommand{\bra}[1]{\bigl\langle #1 \bigr|}
\newcommand{\ket}[1]{\bigl| #1 \bigr\rangle}

\newcommand{\al}{\alpha}
\newcommand{\be}{\beta}
\newcommand{\ga}{\gamma}
\newcommand{\Ga}{\Gamma}
\newcommand{\de}{\delta}
\newcommand{\ep}{\epsilon}
\newcommand{\la}{\lambda}
\newcommand{\La}{\Lambda}
\newcommand{\sg}{\sigma}
\newcommand{\sgb}{\bar{\sigma}}
\newcommand{\Sg}{\Sigma}

\newcommand{\ca}{{\mathcal A}}
\newcommand{\cb}{{\mathcal B}}
\newcommand{\cc}{{\mathcal C}}

\newcommand{\ci}{{\mathcal I}}
\newcommand{\cj}{{\mathcal J}}
\newcommand{\cl}{{\mathcal L}}
\newcommand{\cm}{{\mathcal M}}
\newcommand{\cn}{{\mathcal N}}
\newcommand{\co}{{\mathcal O}}
\newcommand{\cp}{{\mathcal P}}
\newcommand{\cs}{{\mathcal S}}
\newcommand{\cu}{{\mathcal U}}
\newcommand{\cz}{{\mathcal Z}}

\newcommand{\gev}{\>\mathrm{GeV}}
\newcommand{\tev}{\>\mathrm{TeV}}

\newcommand{\tr}{\mathrm{Tr}}
\newcommand{\simlt}{\lesssim}
\newcommand{\simgt}{\gtrsim}

\newcommand{\del}{\partial}
\newcommand{\dd}{\mathrm{d}}
\newcommand{\DD}{\mathrm{D}}
\newcommand{\I}{\mathrm{i}}
\newcommand{\e}{\mathrm{e}}
\newcommand{\C}{{\mathrm c}}

\newcommand{\dt}{\!\cdot\!}

\newcommand{\SU}{\mathrm{SU}}

\newcommand{\ds}[1]{\displaystyle{#1}}

\newcommand{\PD}{{\phantom\dag}}

\newcommand{\ttc}[1]{{\tt \textsc{#1}}}

\newcommand{\proton}{{\mathrm p}}
\newcommand{\col}{{\mathrm{c}}}
\newcommand{\acol}{{\bar{\mathrm{c}}}}
\newcommand{\s}{{\mathrm{s}}}
\newcommand{\pT}{p_\mathrm{T}}
\newcommand{\ptv}{p_\mathrm{T}^\text{veto}}
\newcommand{\ptvbar}{\bar{p}^\text{veto}_\mathrm{T}}
\newcommand{\Gcusp}{{\Ga^{\text{cusp}}_\mathrm{F}}}
\newcommand{\jv}[1] {{#1}^{\text{veto}}}
\newcommand{\muh}{\mu_\text{h}}
\newcommand{\muf}{\mu_\text{f}}
\newcommand{\UV}{{\text{\tiny UV}}}
\newcommand{\SM}{{\text{\tiny SM}}}
\newcommand{\vetosum}{\sum\raise0.5ex\hbox{$'$}}
\newcommand{\vetosumX}{\sum_X\raise1ex\hbox{$'$}}
\newcommand{\vetosumXcol}{\sum_{X_\col}\raise1ex\hbox{$'$}}
\newcommand{\vetosumXacol}{\sum_{X_\acol}\raise1ex\hbox{$'$}}

\begin{document}

\begin{titlepage}

\setcounter{page}{0}

\vskip 1.25in

\begin{center}

{\Huge\bf An Explanation of the WW Excess at the LHC by Jet-Veto Resummation}

\vskip .5in

{\Large {\bf Prerit Jaiswal}$^1$ and {\bf Takemichi Okui}$^{2}$}

\vskip .5in

$^a$\ {\it Department of Physics, Florida State University,
          Tallahassee, FL 32306, USA}

\vskip 1in

\abstract{
The $W^+ W^-$ production cross section measured at the LHC has been consistently exhibiting a mild excess beyond the SM prediction, in both ATLAS and CMS at both 7-TeV and 8-TeV runs. We provide an explanation of the excess in terms of resummation of large logarithms that arise from a jet-veto condition, i.e., the rejection of high-$\pT$ jets with $\pT^\PD > \ptv$ that is imposed in the experimental analyses to reduce backgrounds. Jet veto introduces a second mass scale $\ptv$ to the problem in addition to the invariant mass of the $W^+ W^-$ pair. This gives rise to large logarithms of the ratio of the two scales that need to be resummed. Such resummation may not be properly accounted for by the Monte Carlo simulations used in the ATLAS and CMS studies. Those logarithms are also accompanied by large $O(\pi^2)$ terms when the standard, positive sign is chosen for the squared renormalization scale, $\mu^2$. We analytically resum the large logarithms including the $\pi^2$ terms in the framework of soft collinear effective theory (SCET), and demonstrate that the SCET calculation not only reduces the scale uncertainties of the SM prediction significantly but also renders the theory prediction well compatible with the experiment. We find that resummation of the large logarithms and that of the $\pi^2$ terms are both comparably important.
}

\end{center}

\vfill
\begin{flushleft}
$^1$~{\tt prerit.jaiswal@hep.fsu.edu}\\
$^2$~{\tt okui@hep.fsu.edu}
\end{flushleft}

\end{titlepage}

\tableofcontents
\vskip .5in

\section{Introduction}
\label{s.intro} \setcounter{equation}{0} \setcounter{footnote}{0}

%
\begin{table}[t]
\begin{center}
\renewcommand{\arraystretch}{1.5}
\begin{tabular}{ | c || c | c | c | }
\hline
  & ATLAS & CMS & Theory ({\tt MCFM}) \\ 
  $\sqrt{s}$  & $\sigma$ [pb] 	& $\sigma$ [pb]  & $\sigma$ [pb]  \\ \hline
 $7 \tev$ & $51.9^{+2.0 +3.9 +2.0}_{-2.0 -3.9 -2.0}$ \cite{WW:ATLAS_7TeV}	
 & $52.4^{+2.0 +4.5 +1.2}_{-2.0 -4.5 -1.2}$ \cite{WW:CMS_7TeV}
 & $47.04^{+2.02+0.90}_{-1.51-0.66}$ \\ \hline
 $8 \tev$ &  $71.4^{+1.2+5.0+2.2}_{-1.2-4.4-2.1}$ \cite{WW:ATLAS_8TeV} 
 & $69.9^{+2.8 +5.6 +3.1}_{-2.8 -5.6 -3.1}$ \cite{WW:CMS_8TeV}
 & $57.25^{+2.35+1.09}_{-1.60-0.80}$ \\ \hline
\end{tabular}
\end{center}
\caption{Comparison of cross-section measurements of ATLAS and CMS experiments and the NLO theory predictions as
obtained from {\tt MCFM} for $ p p \rightarrow W^+ W^-$ at $\sqrt{s} = 7$ and $8 \tev$. The first, second and third 
errors in the experimental results are the statistical, systematic and luminosity errors, respectively, while the first and 
second errors in theory calculations are the scale and PDF uncertainties, respectively.}
\label{t.xsec}
\end{table}

After the discovery of a Higgs boson, two primary objectives of the Large Hadron Collider (LHC) are to test the
electroweak sector more precisely 
and to look for new physics. The former requires
improving measurements on the Higgs couplings with other Standard Model (SM) 
particles as well as anomalous couplings of the electroweak gauge bosons. 
The latter must carefully exclude the possibilities of SM processes mimicking the signals in question. 
In both cases, it is in particular of utmost importance that the $\proton \proton \rightarrow W^+ W^-$ SM background is theoretically well understood. 
For example,  $\proton \proton \rightarrow W W \rightarrow \ell \nu \ell \nu$ is the dominant background in the measurement of the 
Higgs decay channel,  $h \rightarrow W W^* \rightarrow \ell \nu \ell \nu$ \cite{HWW_LHC_ATLAS, HWW_LHC_CMS}. 
Although the backgrounds are normalized to data in the control region, their extrapolation to the signal region requires theoretical inputs of differential cross-sections.
The SM $\proton \proton \rightarrow W^+ W^-$ production can also be a significant background for certain new physics processes, where the problem of separating signals from backgrounds is exacerbated by the loss of information due to 
invisible neutrinos. Without a proper theoretical understanding of the total and differential cross-sections, the $WW$ background
samples could be contaminated beyond expectation.  

Recently, ATLAS and CMS experiments have presented their results for the $W^+ W^-$ total inclusive cross-sections using the
leptonic decay channels of the $W$ bosons, which are summarized in \tref{xsec}. In the same table, we have also shown the next-to-leading-order (NLO) theoretical
predictions for the total inclusive cross-sections obtained from a Monte-Carlo (MC) program {\tt MCFM} \cite{MCFM1, MCFM2}. 
The numerical results from {\tt MCFM} include contributions from the $gg$ channel,%
\footnote{
not including the higgs contributions, which would be at most $\sim 3$~pb for the 8-TeV run even without considering lower lepton efficiencies due to softer leptons from an off-shell $W^\pm$.
See discussions in \ssref{results}. 
}
which is 
formally higher order, $\co(\al_\s^2)$, as compared to the $q \bar{q}$ channel, which is $\co(\al_\s^0)$.
The theoretical results use {\tt MSTW2008nlo} PDFs~\cite{MSTW}%
\footnote{
The PDF uncertainties are considerably higher ($ \sim 3$ -- $3.5$\%) when \ttc{CT10} PDFs \cite{CT10} are used instead of our default choice of 
{\tt MSTW2008nlo} PDFs ($\sim 1.5$ -- $2$\%).
}
with both renormalization and factorization scales set to $W$ boson mass ($\mu_\text{r} = \mu_\text{f} = m_W$) and the scale 
uncertainties were obtained by varying the scales in the range $m_W/2 < \mu_\text{r} = \mu_\text{f} < 2 m_W$.

It is interesting to note that, while compatible within $2 \sigma$,
the experimental results are nonetheless consistently higher than the NLO predictions for both ATLAS and CMS and for both 7- and 8-TeV runs, by as much as $10$--$20\%$. 
The two experiments seem more consistent with each other than with the NLO prediction.
This mild 
discrepancy  between the measured and predicted $W^+ W^-$ cross-sections have led to speculations that new physics 
with $2 \ell + \slashed{E}_\mathrm{T}$ signatures could be hiding in the $W^+ W^-$ measurement~\cite{WW:NewPhysics1, 
WW:NewPhysics2, WW:NewPhysics3, WW:NewPhysics4, WW:NewPhysics5, WW:NewPhysics6}. 
It is therefore imperative to assess higher-order corrections to the SM predictions.

Let us briefly review the status of higher order corrections to the process $\proton \proton \rightarrow W^+ W^-$. The QCD NLO
corrections to $q \bar{q} \rightarrow W^+ W^-$ have been known for a long time \cite{WW:nlo1, WW:nlo2}, including the full
leptonic decays of the $W$ bosons \cite{WW:nloLepton1, WW:nloLepton2}. 
The $K$-factor for the total inclusive cross-section is approximately $1.6$ and stays roughly the same for $\sqrt{s}$ between 7 and 14 TeV\@. 
The $gg$-initiated contribution $g g \rightarrow W^+ W^-$ (without the higgs) to the total inclusive cross-section is $\sim 3$ -- $4 \%$ in the same energy range, 
and higher-order QCD corrections for the $gg$ channel  
are discussed in~\cite{ggWW1, ggWW2, ggWW3}.  
The NLO electroweak corrections are investigated
in~\cite{WW:ew1} and found to be $\simlt 2 \%$, 
while it is shown in~\cite{WW:ew2} that the inclusion of initial-state photons to the electroweak 
corrections cancels the virtual contributions so that the net electroweak corrections are in fact negligible. 
The NNLL threshold-resummed cross section and approximate NNLO cross section  
have been recently calculated in~\cite{WW:threshold}, 
and found to increase the total cross-section by at most  $\sim 3 \%$ 
for $\sqrt{s}$ between 7 and 14 TeV\@. Less-inclusive NLO (and partial NNLO) cross-sections using realistic phase-space cuts can be found in~\cite{MCFM1, MCFM2}, while fixed-order calculations matched/merged to MC + parton shower are 
extensively studied in~\cite{WW:MC1, WW:MC2} as well as in a more recent study~\cite{WW:MC3} that includes the $gg$ and $q(\bar{q})g$ initial states. 
The $\pT$ distribution of $W$ pair is analytically derived in~\cite{WW:pTresum1}
using the $b$-space resummation technique%
\footnote{
$b$ is a variable conjugate to $\pT$ through Fourier 
transformation.
}
and the results are found to be in good agreement with MC predictions. 
Transverse-momentum resummation has also been performed in the SCET formalism~\cite{WW:pTresum2}, and the results are found to be in reasonable 
agreement with the previous approach. 
Nevertheless, the slight discrepancy between the theoretical and experimental summarized in \tref{xsec} calls for more theoretical investigations on the effects of cuts employed by the experimental analyses.

In particular, while both ATLAS and CMS experiments present the inclusive cross-sections for $W^+ W^-$ production, 
they actually reject jets with $\pT$ greater than a prescribed value $\ptv$ (``jet veto''). 
The primary purpose of jet veto in their event selection is to reduce QCD backgrounds from single top
and top pair production. 
The inclusive cross-sections are extrapolated by folding in the jet-veto efficiencies, 
$\epsilon(\ptv) \equiv \sigma(\ptv)/\sigma_\text{inclusive}$, 
where $\sg(\ptv)$ is the cross section without jets with $\pT > \ptv$.
However, their estimates of $\epsilon(\ptv)$ are obtained using MC simulations for the signal and 
some of the background processes. In \fref{ATLAS_WW}, taken from the ATLAS $WW$ cross-section measurement
results for $\sqrt{s}=8\tev$ run \cite{WW:ATLAS_8TeV}, the MC predictions are compared with the data for different 
jet-multiplicities in the $e \mu \nu \nu$ channel. As noted in \cite{WW:ATLAS_8TeV} itself, 
there is a clear discrepancy in the \emph{zero-}jet bin, 
while the data agrees well with MC predictions at higher jet-multiplicities. 
This naturally casts doubt upon the validity of MC (+ parton shower) predictions in the zero-jet bin, that is, in the presence of a jet veto. 

\begin{figure}
        \centering
                \begin{subfigure}[b]{0.42\textwidth}
                \includegraphics[trim = 0.7cm 0cm 0.9cm 0cm,clip=true,width=\textwidth]{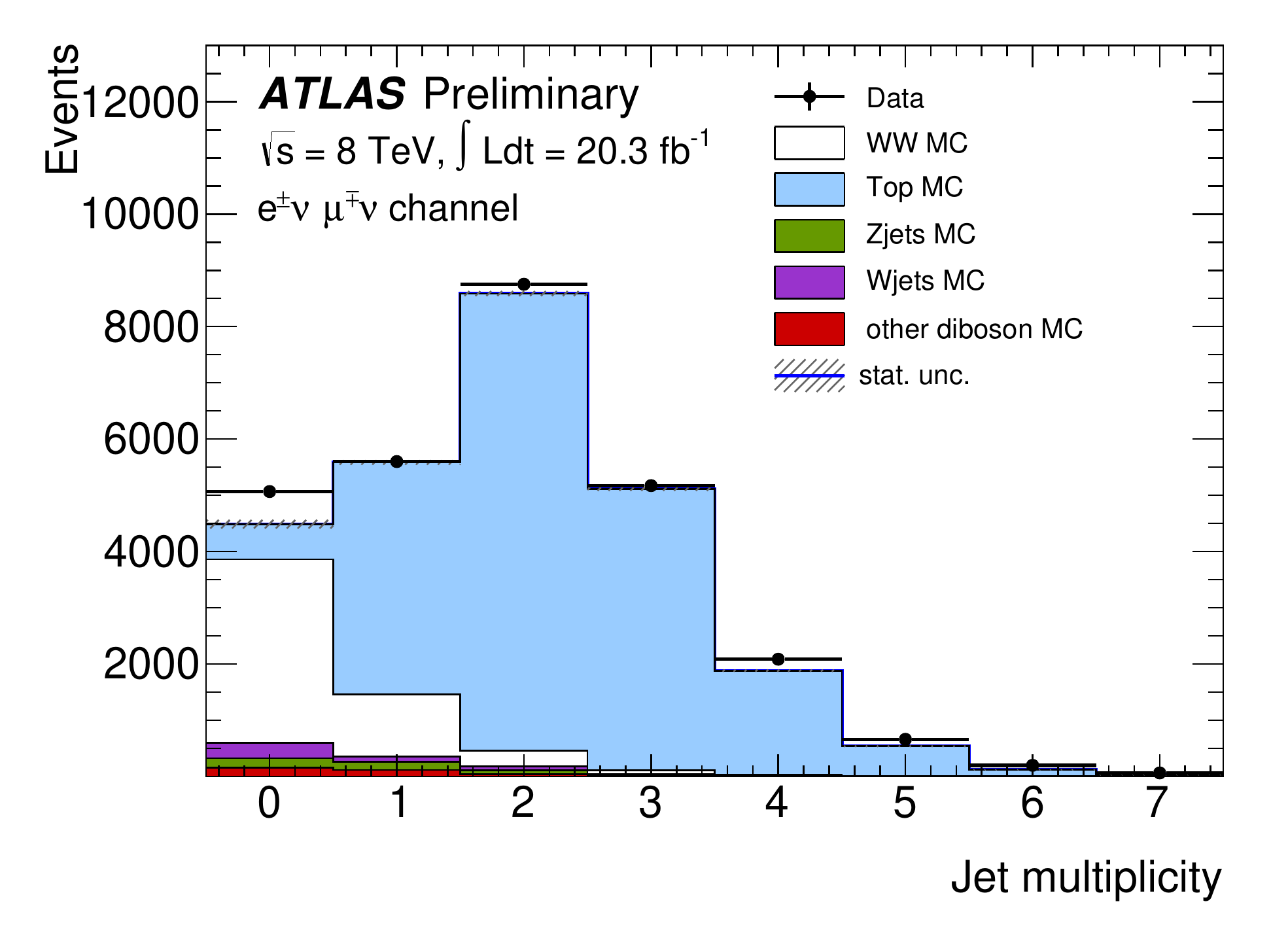}
                \caption{}
                \label{f.ATLAS_WW}
        \end{subfigure}%
        \begin{subfigure}[b]{0.29\textwidth}
                \includegraphics[width=\textwidth]{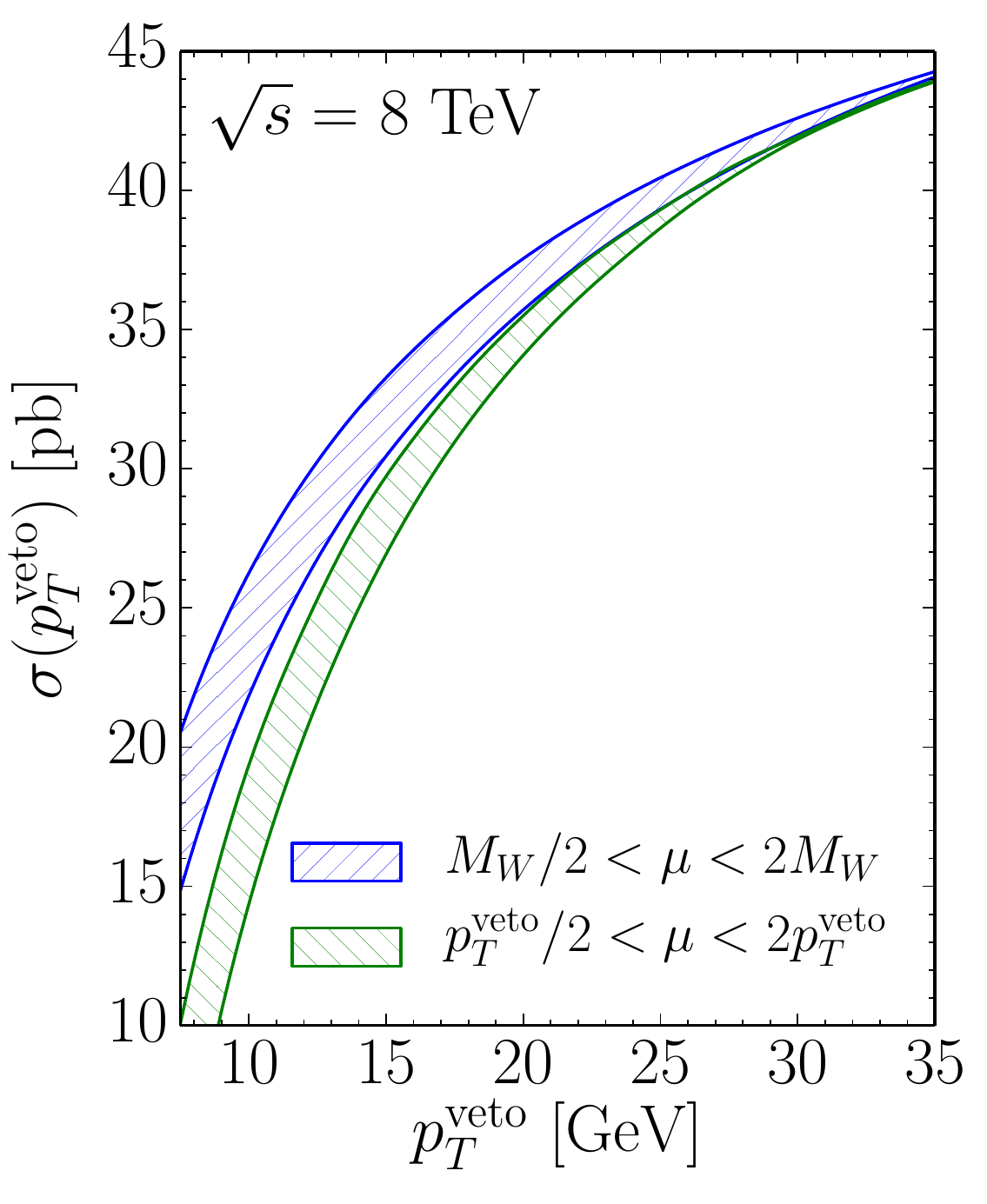}
                \caption{}
                \label{f.NLO_scale}
        \end{subfigure}%
        \begin{subfigure}[b]{0.29\textwidth}
                \includegraphics[width=\textwidth]{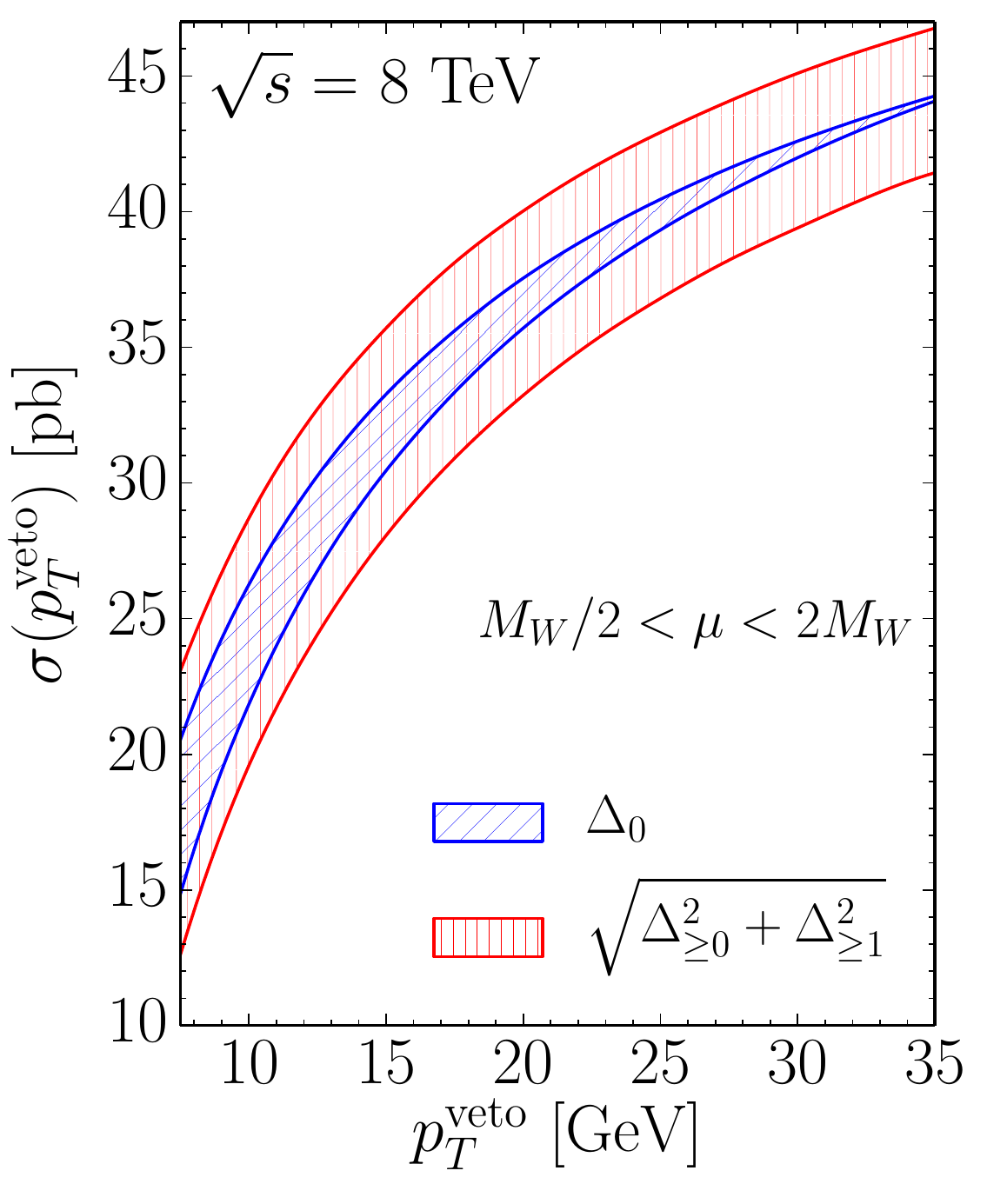}
                \caption{}
                \label{f.NLO}
        \end{subfigure}%
        \caption{(a) Comparison of data and MC as a function of jet multiplicity by the ATLAS experiment at $\sqrt{s}=8\tev$ run 
        \cite{WW:ATLAS_8TeV}, for events passing the selection criteria (except jet-veto) as required by the $W^+ W^-$ 
        cross-section measurement. \\
        (b) NLO cross-sections for $q \bar{q} \rightarrow W^+ W^-$ at $\sqrt{s}=8 \tev$ LHC run, as a function of 
        $\ptv$ obtained using {\tt MCFM}\@. The blue (or green) hatched region corresponds to scale variation by a factor of 
        $1/2$ and $2$ around the central value of $\mu_\text{r} = \mu_\text{f} = m_W$ (or $\ptv$). \\
        (c) Same as (b) but the scale variation in the red hatched region is calculated using the procedure described in \cite{0Jet}.
}
\end{figure}

While the $K$-factor for inclusive $W$-pair production at NLO is $\sim 1.6$, 
fixed-order calculations with a jet veto \cite{MCFM2, WW:nloVeto} show significant reductions of the $K$-factor by as much as $40 \%$.  
This calculation suggests large cancellations between the higher-order virtual corrections and 
the real corrections from jets that are emitted collinear to the beam axis as required by the jet-veto constraint. 
Such cancellations, if not properly taken into account, could lead to a deceivingly underestimated scale uncertainty of the theoretical prediction.
Jet veto also introduces a new mass scale $\ptv$ to the problem, in addition to the invariant mass $M$ of the $W^+ W^-$ pair.
The presence of more than one scale is always problematic for a 
fixed-order calculation, since it is not obvious where to set the scale $\mu$ that appears in renormalization. In the process at hand, the two possible 
scale choices are $M$ and $\ptv$. In \fref{NLO_scale}, cross-sections from {\tt MCFM} (ignoring $gg$ initiated contribution)
are presented by varying the scale $\mu = \mu_\text{r} = \muf$ by factors of $1/2$ and $2$ around $\mu = M_W$ (as a proxy for $M$) and $\mu = \ptv$.  
Clearly, for 
smaller values of $\ptv$, the perturbation theory fails miserably, and the two choices of scales not only have large 
uncertainties but also yield results incompatible with each other. 
One might be tempted to say the error bands are much smaller and the results for the two scale choices seem to converge 
in the range $\ptv \sim 20$ -- $30 \gev$ that is actually used by the aforementioned ATLAS and CMS studies.
However, as we have already warned above, this seemingly small uncertainty is just an artifact of cancellations between the virtual corrections and real emissions. 

Large cancellations of this kind are well known in the literature,
and we briefly summarize the arguments presented in \cite{0Jet}. Defining $\sigma_{\geq N}$ to be the cross-section with the number of 
jets $\geq N$, one may parametrize the total inclusive cross-section $\sigma_{\geq 0}$ and the 1-jet inclusive cross-sections $\sigma_{\geq 1} (\ptv)$ with at least one jet with $\pT > \ptv$ as
\beqa
\sigma_{\geq 0} 
&= 
\sigma_\mathrm{B} \, 
\biggl( 1 + \sum_{n = 1}^\infty c_n \al_\s^n \biggr) 
\,,\\
\sigma_{\geq 1} (\ptv) 
&= 
\sigma_\mathrm{B} 
\sum_{n = 1}^\infty 
\sum\limits_{m=0}^{2n} 
d_{n,m} \, \al_\s^n  L^m 
\,,
\eeqa
where $\sigma_\mathrm{B}$ is the tree-level cross-section and 
$L \equiv \ds{\log\bigl[ M^2 / (\ptv)^2 \bigr]} \gg 1$. 
The 0-jet inclusive cross-section $\sigma_{\geq 0}$ does not have any 
large logarithms, as there is only one mass scale $M$ in the problem so we can simply set $\mu \sim M$. 
Since the inclusive NLO $K$-factor to $WW$ production is $\sim 1.6$, the coefficient $c_1$ is large. 
On the other hand,
the 1-jet inclusive cross-section $\sigma_{\geq 1}$ 
is given at NLO by 
$\sigma_{\geq 1} 
\simeq 
\sigma_\mathrm{B} \, \al_\s \, ( d_{1,2} L^2 + d_{1,1} L + d_{1,0} )$, 
which can again be large due to the presence of large logarithms $L^n$. 
However, the jet-veto cross-section is given by the difference 
$\sigma_{\geq 0} - \sigma_{\geq 1}$, 
where the large logarithm term $d_{1,2} L^2$ is subtracted from the large virtual correction term $c_1$. 
The significant reduction of the $K$-factor for the jet-veto cross section mentioned above implies that there is a substantial cancellation in this subtraction.
Such cancellation then suggests that the scale uncertainties in the jet-veto cross-sections as shown in \fref{NLO_scale} are gross underestimations.
To more properly assess the scale uncertainties, 
Ref.~\cite{0Jet} suggests the use of 
$\sqrt{\Delta_{\geq 0}^2 + \Delta_{\geq 1}^2}$ for the 
estimate of scale uncertainty in the 0-jet bin, where $\Delta_{\geq N}$ is the scale uncertainty in 
$\sigma_{\geq N}$. The basic idea behind this formula is to assume that the scale uncertainties in $\sigma_{\geq 0}$ and $\sigma_{\geq 1}$ are uncorrelated 
as they begin at different orders in $\al_\s$. 
In \fref{NLO}, we compare the naive method of estimating scale uncertainty 
with the refined method just described, 
and we clearly see that the refined scale-uncertainty bands not only properly includes both error bands of \fref{NLO_scale} at low $\ptv$ but also does not exhibit a fake convergence at higher $\ptv$. 
We will make extensive use of this refined measure of scale uncertainties later to properly estimate scale uncertainties in the MC jet-veto efficiencies. 

However, in order to actually reduce the scale uncertainty itself, 
it is clear that we must go beyond fixed-order calculations and resum the large logarithms $L^n$ 
or, equivalently, resolve the ambiguity in the choice of $\mu$ by arranging the calculation in such a way that only one mass scale appears at each and every step of the calculation. 
The impact of large logarithms is larger than one might expect from the fact that $M$ is a few hundred GeV and $\ptv$ is a few tens of GeV, 
because the logarithm actually appears in the form $\ds{\log\bigl[ (-M^2-\I 0^+) / (\ptv)^2 \bigr]} = \ds{\log\bigl[ M^2 / (\ptv)^2 \bigr]} - \I \pi$.
In particular, the $d_{1,2}$ term in $\sigma_{\geq 1}$ contains not only $\ds{\log^2\bigl[ M^2 / (\ptv)^2 \bigr]} \sim \co(10)$ but also $\pi^2 \sim 10$.
In this paper, 
we will perform resummations of the large logarithms including the $\pi^2$ terms,
and obtain refined predictions of the jet-veto $\proton \proton \to WW$ cross-sections for the $\sqrt{s}=7$ and $8$~TeV LHC runs with significantly reduced scale uncertainties. 
We will then compare our predictions with MC results as well as the ATLAS and CMS measurements.

While $\pi^2$ resummation has a long history~\cite{pisquare_1, pisquare_2, pisquare_3},
resummation of jet-veto logarithms is a fairly new subject. 
Since parton distribution functions (PDFs) are fundamentally defined for fully inclusive processes in which all hadronic final states are summed over,  
the description of jet-vetoed processes requires new objects analogous to PDFs corresponding to summing over only the hadronic final states that satisfy the jet-veto condition. 
Such objects, called \emph{beam functions}, were first introduced and developed in~\cite{beam-func-1, beam-func-2} (and will be reviewed in \ssref{factorization_and_partons}).
The framework for jet-veto resummation using effective field theory was first laid out in~\cite{Higgs:veto1}, 
and the jet-veto higgs production cross section in the $gg$ channel at the next-to-next-to-leading-logarithmic (NNLL) accuracy was calculated in~\cite{Higgs:veto2, Becher:2012qa, logR:1, Higgs:veto}, and the effective field theory calculations~\cite{Becher:2012qa, logR:1} are in agreement with the results from direct perturbative QCD calculations~\cite{Higgs:veto2, Higgs:veto}.
Partial N$^3$LL jet-veto cross sections for the higgs production were computed in~\cite{Becher:2013xia, most-complete},
and the latest result~\cite{most-complete} in particular exhibits a clear improvement of perturbative convergence owing to jet-veto resummation.  
At NNLL and beyond, jet-veto cross sections depend on the jet-clustering algorithm, which was first studied in~\cite{Higgs:veto2, logR:1} and more extensively analyzed in~\cite{logR:2}.
The effects of logarithms of a finite quark mass (e.g., $m_{b,t}$) and $\ptv$ were studied in~\cite{mass-effects}.
Jet-veto resummation has been also successfully applied to the higgs production in association with a vector boson~\cite{VH-1, VH-2}, to the processes with tagged jets~\cite{tagged-jets-1, tagged-jets-2, tagged-jets-3}, and to the higgs production at future colliders~\cite{future-jet-veto}.

This paper is organized as follows. 
In \sref{set-up-SCET}, we carefully set up an effective field theory that is suitable for performing the resummation of the large logarithms including the $\pi^2$ terms and is free of scale-choice ambiguities.
Analytical calculations and results in the effective field theory will then be presented in \sref{analytical}.
In \sref{numerical}, we numerically evaluate our analytical results and compare them with MC simulations and experimental results. We will see that the resummation not only improves the scale uncertainties significantly but also renders the theory prediction consistent with the ATLAS and CMS measurements at $\lesssim 1\sigma$.

\section{Setting up a SCET}
\label{s.set-up-SCET}

The process of our interest is 
$\ds{\proton + \proton \to W^+ + W^- + \vetosum X}$, 
where the $'$ on $\sum'$ indicates that we are only summing over jets satisfying the jet-veto condition, $\pT(X) < \ptv$.
After summing over $X$ 
and integrating over the rapidity and angular orientation of the $W^+ W^-$ system, 
we are left with two scales in the problem: 
the invariant mass $M$ of the $W$ pair, 
and the jet-veto scale, $\ptv$. 
Experimentally, we are interested in the situation in which $\ptv \ll M$. 
Since this hierarchy of scales introduces 
a new dimensionless parameter $M / \ptv \gg 1$ 
to perturbative calculations of the cross section for this process,
integrations over loop momenta will yield 
large logarithms of the form $\ds{\bigl( \log\bigl[ (-M^2 - \I 0^+) / (\ptv)^2 \bigr] \bigr)^n} = \ds{\bigl( \log\bigl[ M^2 / (\ptv)^2 \bigr] - \I\pi \bigr)^n}$. 
Thus, a fixed-order calculation of this process with a jet veto 
should be assigned a larger uncertainty 
than that of the same process without a jet veto. 
To resum those large logarithms as well as the $\pi^2$ terms to improve the accuracy of the theoretical prediction, 
we employ the formalism of \emph{soft collinear effective theory} 
(SCET), which was originally formulated in~\cite{Bauer_SCET_1, Bauer_SCET_2, Bauer_SCET_3, Bauer_SCET_4} in what is now referred to as the ``label SCET'' formalism, 
which was then quickly reformulated as the ``multipole expansion'' formalism in~\cite{Beneke_SCET_1, Beneke_SCET_2}.
More recently, further alternative formulations of SCET have been developed in~\cite{Luke_SCET} and~\cite{Matt_SCET_1, Matt_SCET_2}.
In this section, we present a brief review of SCET, adopting the multipole-expansion formulation,  
to highlight main conceptual ingredients relevant for our calculations 
as well as to establish our notation.
The details of our analytical and numerical results will be presented 
in \sref{analytical} and \sref{numerical}.

\subsection{The Degrees of Freedom, Power Counting, and Symmetries}
\label{ss.dof_pc_symm}

\subsubsection{A Practical View on (Effective) Field Theories}
\label{sss.EFT}
In diagrammatic calculations in quantum field theories, 
vertices are simple and propagators are complicated. 
Vertices are polynomials of momenta and hence analytic in momenta, 
while propagators have poles and give rise to complicated singularities in scattering amplitudes upon loop integration, 
including singularities that \emph{cannot actually occur 
for the values of external momenta under consideration.}
This considerably complicates calculations and obscures the physics in question.

The fundamental principle of \emph{effective field theory} (EFT) is 
to capture as much physics as possible at the lagrangian level 
by writing down an effective lagrangian whose vertices are designed 
to reproduce the non-singular part of amplitudes as much as possible, 
\emph{including} those singularities that do not actually occur and should actually be regarded as analytic, 
while striving to keep \emph{only} the propagators that are necessary to reproduce the singularities that can actually occur.
Clearly, the construction of such effective theory is possible 
only if we kinematically restrict the set of processes we consider 
(e.g., only initial states with energy less than $1\tev$, 
and/or only final states with $\pT < \ptv$). 
The key step in the construction of an EFT is 
to anticipate in the full theory the class of diagrams relevant to the restricted class of processes in question (without actually doing any loop integrals), 
and identify momentum modes that \emph{cannot} go on-shell for any values of external momenta under consideration. 
Those modes will never lead to singularities that can actually occur, 
so we \emph{integrate them out}, i.e., write effective vertices to reproduce their effects as analytic functions of momenta.
We are then only left with modes that \emph{may} go on-shell for some values of external momenta under consideration.

The above procedure is reminiscent of the construction of so-called reduced diagrams for analyzing the IR divergence structure of a given amplitude.
(For a nice review, see e.g.~Ref.\cite{Sterman:2004pd}.)
The essential difference is that 
the construction of reduced diagrams strives to identify the propagators that \emph{will} produce singularities, 
which is a laborious task that must be carried out diagram-by-diagram.
In EFTs, in contrast, 
we strive to identify the modes that \emph{will not} go on-shell 
and then write an effective \emph{lagrangian} without those ``guaranteed-off-shell'' modes.
There may still be some singularities that cannot actually occur in EFT amplitudes, 
but significantly fewer of them than in the corresponding full-theory amplitudes, 
and we just try our best by converting as many guaranteed-off-shell modes as possible into effective vertices. 
The ease and benefit of working at the lagrangian level greatly outweighs 
not completely identifying the necessary and sufficient conditions for singularities.    

A second principle of EFT follows naturally from 
this ``vertices-heavy'' nature of effective lagrangians. 
By construction, 
EFT lagrangians have many more vertices than the corresponding full lagrangians, usually an infinite number of them. 
This necessitates a well-defined organization principle that permits us to truncate the infinite series of vertices to achieve a desired accuracy.
Thus, in addition to the small coupling constants it inherits from the full theory,
a useful EFT must be equipped with a set of new small parameters with \emph{well-defined power-counting rules} that govern how each propagator and each vertex should scale with those small parameters. 
This lets us not only truncate the lagrangian but also discard quantitatively irrelevant diagrams at the outset. 
In addition to the absence of guaranteed-off-shell modes discussed earlier, 
the presence of more small expansion parameters with well-defined power-counting rules is another feature of EFTs that enables us to see the physics more clearly and easily. 

There is a third aspect of EFT that greatly facilitates calculations in EFTs. 
To separate modes into guaranteed-off-shell modes to integrate out and ``can-be-on-shell'' modes to keep, 
we must introduce \emph{cutoffs}, i.e., artificial parameters that define the boundaries between the guaranteed-off-shell modes and the can-be-on-shell modes.
In addition, 
the principle of well-defined power counting often requires a further division of can-be-on-shell modes into multiple subgroups 
such that each subgroup can be assigned definite power-counting rules.
This division also requires cutoff parameters to define the boundaries between the subgroups.
Since both types of cutoffs are just artificial separators introduced by us solely for our own convenience, 
all physical observables such as cross sections must be independent of the cutoffs. 
Cutoff independence can be expressed as a set of differential equations (i.e., renormalization group (RG) equations) for the parameters of the theory.
The RG equations permit us to work with only one scale at each and every step of calculations 
so that no large logarithms like $\ds{\log(M / \ptv)}$ ever appear in the calculations, effectively resumming the large logarithms.

Finally, while the cutoffs are necessary for the separation of modes, 
it is technically cumbersome at the loop level to literally implement the cutoffs by bounding the limits of loop integrations, 
because the integration limits other than $0$ and $\pm\infty$ 
tend to make the integrals considerably harder to evaluate or even just to estimate.
The standard trick is to deliberately make ``mistakes'' by keeping integration limits \emph{un}bounded and letting integrals formally diverge,
and (re-)regulate the divergences without bounding the integration limits, e.g., 
by dimensional regularization (DR),
and correct for the ``mistakes'' at the very end by renormalization.
Therefore, 
corresponding to every artificial boundary of modes we introduce to an EFT, 
there is a divergence and a regulator.
(For example, our SCET will have two types of divergences and two regulators, as we will discuss in \sssref{rap_div}.)
The regulators reintroduce auxiliary parameters (like $\mu$ in DR), 
and the RG equations can be (re)derived by demanding that physical observables should be independent of those auxiliary parameters.
Like cutoffs, 
the divergences are artificial features of the theory introduced by us for our convenience, and they must cancel out in physical observables. 
As cutoff independence leads to convenient RG equations, 
the cancellation of divergences can also be exploited to facilitate EFT calculations.

\subsubsection{Collinear and Anticollinear Modes}
\label{sss.col_acol_modes}
To identify the degrees of freedom to include in our SCET lagrangian, 
we must first ask what modes can go on-shell in the process $\ds{q_1 + q_2 \to W^+ + W^- + \vetosum X}$, 
where $q_1$ and $q_2$ are a quark and an antiquark or vice versa.%
\footnote{\label{ftnote:onlyqqbar}
We ignore the $gg$ and $qg$ channels in our SCET calculations 
as they are small and would thus receive little benefit from the resummation of large logarithms. 
For $gg$, this is because they are absent at the tree level in the SM\@.
The $qg \to WWq$ channel exists at tree level in the SM,
but their jet-veto cross sections are highly suppressed as it is very difficult for the final-state $q$ to satisfy a jet veto. 
Therefore, the channels other than $q\bar{q}$ are only relevant for power corrections, which we will discuss later in \ssref{power_corrections}.
}
To characterize the kinematics of the initial state,
we introduce two lightlike 4-vectors $n_+^\mu$ and $n_-^\mu$ such that 
the 4-momenta of $q_1$ and energetic gluons emitted by $q_1$ are all nearly parallel to $n_+$ (which is ensured by the jet veto condition). 
Likewise, $n_-$ is associated with $q_2$ and its energetic radiations.  
Being lightlike, they satisfy $\ds{n_\pm \dt n_\pm} = 0$, 
and we choose their relative normalization as
\beqa
n_\pm \dt n_\mp = 2
\,.\label{e.n_norm}
\eeqa
(There is no Lorentz invariant way to fix their individual normalizations.)
The kinematics of our process is most conveniently described by lightcone coordinates spanned by $n_\pm$ in which
\beqa
\del_\pm \equiv n_\pm \dt \del 
\,,\quad
x^\pm \equiv \frac12 \, n_\mp \dt x
\,,\label{e.lightcone_def}
\eeqa
so that $\del_\pm x^\pm = 1$ and $\del_\pm x^\mp = 0$.
The lower- and upper-indexed lightcone components of a general 4-vector $a$ must be defined in the same way as those of $\del$ and $x$, respectively,
so we have 
\beqa
a_\pm \equiv n_\pm \dt a \equiv 2 a^\mp 
\,.\label{e.+-components}
\eeqa
We then define a 4-vector $a_\perp^\mu$ to be the projection of $a^\mu$ onto the plane orthogonal to both $n_+$ and $n_-$, i.e., 
\beqa
a_\perp \dt n_+ = a_\perp \dt n_- = 0  
\,.
\eeqa
This together with\erefn{n_norm} and\erefn{+-components} then implies that any 4-vector $a^\mu$ can be decomposed as
\beqa
a^\mu 
= a_\parallel^\mu + a_\perp^\mu
\label{e.4-vec-decomposition}
\eeqa
with
\beqa 
a_\parallel^\mu 
\equiv 
a^+ n_+^\mu + a^- n_-^\mu 
\,.
\eeqa
From\erefn{+-components}, we see that the metric and inverse metric%
\footnote{
We adopt the $+$$-$$-$$-$ sign convention for the spacetime metric.
}
in the $x^+$-$x^-$ subspace are given by
\beqa
(g_{\mu\nu}) 
= 
\left(\begin{array}{cc}
0 & 2 \\
2 & 0
\end{array}
\right),
\quad
(g^{\mu\nu}) 
= 
\left(\begin{array}{cc}
0   & 1/2 \\
1/2 &   0
\end{array}
\right)
,\label{e.metric}
\eeqa
so, for arbitrary 4-vectors $a^\mu$ and $b^\mu$, we have
\beqa
a \dt b
\equiv 
a_+ b^+ + a_- b^- + a_\perp \dt b_\perp
&= 
2 (a^- b^+ + a^+ b^-) + a_\perp \dt b_\perp  \\
&= 
\frac12 (a_+ b_- + a_- b_+) + a_\perp \dt b_\perp
\,.
\eeqa

Now, by definition and without loss of generality, 
we let the 4-momentum of the initial quark $q_1$ be dominantly in the $p_+$ component.%
\footnote{
Unless otherwise noted, we always index 4-momenta by a lower index,
as they are associated with a spacetime derivative $\del$, whose index is naturally lowered.
}
So, we parametrically have $p_+ \sim \co(M)$, where $M$ is the invariant mass of the $W^+ W^-$ pair.
The $p_\perp$ component, on the other hand, is parametrically never larger than $\co(\ptv)$, 
because the jet veto condition prevents the $\perp$ component of momentum of a gluon radiated off of $q_1$ from being larger than $\ptv$. 
We express this parametrics as $|p_\perp| \sim \co(\la M)$, where $\la \equiv \ptv / M$.
The parametric size of the $p_-$ component then follows from requiring that the quark \emph{can} be on-shell, that is, $p^2 = \ds{p_+ p_- + p_\perp \dt p_\perp}$ \emph{can} be zero. 
(If it cannot, this quark mode should not be in the effective theory.)
This determines that $p_- \sim \co(\la^2 M)$. 
Therefore, the components of $p$ of the initial quark must have the following parametric scaling behavior in terms of $M$ and $\la$:
\beqa
(p_+, p_-, p_\perp) \sim (1, \la^2, \la) M  \,.
\eeqa
We refer to this scaling behavior as the \emph{collinear scaling.}
Similarly, the $p$ of the initial quark $q_2$ should scale as 
\beqa
(p_+, p_-, p_\perp) \sim (\la^2, 1, \la) M  \,,
\eeqa
which we refer to as the \emph{anticollinear scaling.}
Note that generic collinear and anticollinear modes have virtuality of order $\co(\la M) \sim \ptv$, that is,
\beqa
p^2 \sim \co(\la^2 M^2) 
\,,
\eeqa
which is the square of the size of the $\perp$ component. 
It will be useful to remember the virtualities of collinear and anticollinear modes are given by their $\pT$.

Next, let us look at a gluon radiated off of a collinear quark with 4-momentum $p$, 
where the collinear quark splits into a quark with 4-momemtum $q$ and a gluon with $k$.
We would like to find the condition on $k$ such that $q$ can remain (nearly) on-shell, 
because otherwise the $q$ mode should not be in the effective theory.
To find that condition, 
let $k$ scale as $(k_+, k_-, k_\perp) \sim (\la^a, \la^b, \la^c) M$.
In order for the $k$ mode to be in the theory,
we must ensure that $k$ can be on-shell, i.e., $k^2 = \ds{k_+ k_-\! - |\vec{k}_\perp|^2} = 0$.
This can happen to nonzero $k$ only if $a + b = 2c$.
If $c < 1$, then $k_\perp$ is parametrically larger than $\ptv$ 
so the gluon would be rejected by jet veto.
Hence we do not have to consider gluons with $c < 1$. 
If $c > 1$, then $k_\perp$ is parametrically smaller than $\ptv$
so the gluon would pass the jet veto condition. 
This implies that contributions from real gluons with $c > 1$ would completely cancel out with those from virtual gluons with $c > 1$.
Therefore, we do not have to consider the $c > 1$ case either.%
\footnote{
In particular, we do not have to consider the so-called ultra-soft modes $\sim (\la^2, \la^2, \la^2) M$ for our purpose.
}
Only the $c=1$ case needs careful analysis, as some of those gluons may pass jet veto and some others may not. 

Having fixed $c$ by looking at $k$, 
we can determine $a$ and $b$ 
from the condition that $q$ can be on-shell, 
$q^2 = \ds{(p-k)^2} = \ds{(p_+\! - k_+) (p_-\! - k_-) - |\vec{p}_\perp\! - \vec{k}_\perp|^2} = 0$. 
Since $c=1$, 
the $\ds{|\vec{p}_\perp\! - \vec{k}_\perp|^2}$ term scales as $\la^2$,
so $q$ can be on-shell 
only if the $\ds{(p_+\! - k_+) (p_-\! - k_-)}$ term also scales as $\la^2$.
In this term, $p_+ p_-$ scales as $\la^2$, 
so both $k_+ p_-$ and $p_+ k_-$ must scale as $\la^2$ or higher. 
Thus, we must have $a + 2 \geq 2$ and $b \geq 2$ in order that $q^2$ can be on-shell.
Recalling the relation $a + b = 2c =2$, we determine that $a=0$ and $b=2$.
Therefore, in order for a nearly on-shell collinear quark to remain nearly on-shell after emitting a nearly on-shell gluon,
both the gluon and post-radiation quark must have a collinear momentum $\sim (1, \la^2, \la) M$.
A similar statement clearly holds for the anticollinear sector with relabelling $+ \leftrightarrow -$.

\subsubsection{Rapidity Divergences and Collinear Anomalies}
\label{sss.rap_div}
As discussed in \sssref{EFT}, the separation of modes in an EFT requires the introduction of cutoffs to define the boundaries between different groups of modes.
In our SCET, there are two cutoffs. 
One is the scale $\La$ that separates the guaranteed-off-shell modes and the can-be-on-shell modes. 
That is, 
if a mode has 4-momentum $p$ with $|p^2 - m^2| > \La^2$, then integrate it out, 
or else include it in the effective theory. 
In particular, 
our SCET must be first defined (or matched to the SM) with $\La \sim M$, 
as we begin by removing guaranteed-off-shell modes with virtuality of $\co(M)$, such as the $s$-channel photon propagator leading up to the $W^+ W^-$ production vertex. 
Then, $\La$ must be run down to $\La \sim \ptv$ via RG equations, 
because the actual scale of virtuality of can-be-on-shell modes in our process is $\co(\ptv)$ due to the jet veto condition.
This RG running is what resums the large logarithms $\ds{\log\left[ M^2 / (\ptv)^2 \right]}$ as we will see explicitly later.

The second cutoff in our effective theory is for separating collinear modes and anticollinear modes.
The only difference between collinear and anticollinear modes is their rapidities.
Collinear modes have a large negative rapidity 
$\eta = (1/2) \log(p^+ / p^-) = (1/2) \log(p_- / p_+) \sim \log\la$, 
while anticollinear modes have a large positive rapidity 
$\eta \sim \log(1/\la)$. 
So, we actually need two rapidity cutoffs $\eta_\col$ and $\eta_\acol$ with $\eta_\col \ll -1$ and $\eta_\acol \gg 1$, 
such that negative-rapidity modes with $\eta < \eta_\col$ are classified as collinear 
while positive-rapidity modes with $\eta > \eta_\acol$ as anticollinear. 
While the symmetry of our problem obviously suggests the choice $\eta_\col = -\eta_\acol$,  
it can be useful to remember that they are independent in principle,
as we will see in \sssref{col_anom}. 

As discussed in \sssref{EFT}, 
the standard trick in EFT is to ignore cutoffs and let loop integrals formally diverge, 
and (re-)regulate them by convenient regulators like dimensional regularization (DR)\@. 
We will employ DR to regulate the ``UV'' divergences due to ignoring $\La$, 
and the so-called \emph{analytic regularization} \cite{analytic_reg} to regulate \emph{rapidity divergences}~\cite{Chiu:2012ir} 
due to ignoring $\eta_\col$ and $\eta_\acol$.
Analytic regularization will be defined in terms of a regulator parameter $\al$ and an auxiliary scale $\nu$, 
where the $\al \to 0$ limit will correspond to removing the regulator.
This is conceptually completely analogous to DR, 
which is defined in terms of a regulator parameter $\ep$ and an auxiliary scale $\mu$ in the $\ep \to 0$ limit.
The \emph{apparent} dependence of an amplitude on the artificial scale $\nu$ is called a \emph{collinear anomaly}~\cite{Becher:2010tm}, 
and the requirement that physical observables should be free of collinear anomalies leads to RG equations with respect to $\nu$,%
\footnote{\label{ftnote:soft-modes}
By letting all $\nu$ dependence be carried by collinear and anticollinear fields 
without including the so-called \emph{soft} gluons (the gluons with momenta scaling as $\sim (\la, \la, \la) M$, i.e., those with a \emph{small} rapidity), 
we have implicitly chosen a renormalization scheme for rapidity divergences 
in which the only role of the soft modes is to provide renormalization constants to absorb the $1 / \al$ poles of rapidity divergences. 
The same scheme was adopted in, e.g., a similar calculation for the higgs production with a jet veto~\cite{Becher:2012qa}.
We therefore will not discuss soft modes in this paper.
}
in addition to standard RG equations with respect to $\mu$ associated with DR\@.
We will see how this works explicitly in \sssref{col_anom}.

Finally, since all rapidity integrals go from $-\infty$ to $\infty$ 
whether we are dealing with collinear or anticollinear modes, 
how do we actually distinguish the two modes inside loop integrals? 
Equivalently, how do we avoid double-counting the modes at the loop level (the so-called \emph{zero-bin subtraction} problem~\cite{Manohar:2006nz})?
Those two modes can be distinguished 
because they are assigned different scaling laws.
Even if we have two integrals, 
one for collinear and the other for anticollinear, 
with apparently the same integrands and the same integration limits, 
the two integrands should be expanded differently in powers of $\la$.
(Note that everything \emph{must} be expanded in EFTs for consistent and manifest power counting.)  
Thus, order-by-order in $\la$, 
their integrands differ, 
lead to different divergences, 
and yield different results.
The principle of well-defined power counting is precisely what resolves the ambiguity/double-counting problem. (This point was particularly well elucidated in Ref.~\cite{Becher:2013xia}.)

\subsubsection{Nonlocality on the Lightcone}
\label{sss.nonlocality}
Let $\phi_\col(x)$ be a field that interpolates a collinear particle,
that is, 
let $\phi_\col(x)$ consist only of Fourier modes scaling as $\sim (1, \la^2, \la) M$.
The components of a spacetime derivative acting on $\phi_\col$ then scale as
\beqa
\del_+ \phi_\col \sim M \phi_\col
\,,\quad
\del_- \phi_\col \sim \la^2 M \phi_\col
\,,\quad
\del_\perp \phi_\col \sim \la M \phi_\col
\,.
\eeqa
Since the effective theory is an expansion in terms of two small dimensionless parameters $\la$ and $\alpha_\mathrm{s}$, with only one dimensionful scale $M$,%
\footnote{
Strictly speaking, we also have $m_W$ and $m_Z$. 
For parametrics/scaling discussions, we treat them as $\sim \co(M)$.
}
the scaling behavior $\del_+ \phi_\col \sim M \phi_\col$ implies that 
a Taylor expansion of $\phi_\col$ in powers of $\del_+ / M$ cannot be truncated at any finite order. 
Therefore, there are no small parameters in the collinear sector that imply locality in the $x^+$ coordinate~\cite{Bauer_SCET_1, Bauer_SCET_2, Bauer_SCET_3}.
On the other hand, $\del_\perp / M$ and $\del_- / M$ acting on collinear fields are suppressed by $\la$ and $\la^2$, respectively, 
so the lagrangian can be truncated at some finite orders in $\del_{\perp} / M$ and $\del_- / M$, 
giving rise to locality in the $x_\perp$ and $x^-$ coordinates.
(In contrast, in familiar Lorentz-invariant Wilsonian EFTs, 
the fact that we have $\del / \La \ll 1$ in all directions at low energy implies an isotropically local lagrangian.)
Similarly, in the anticollinear sector, 
the effective lagrangian is nonlocal in the $x^-$ coordinate 
while local in $x^+$ and $x_\perp$. 
Intuitively, these nonlocalities make a perfect sense. 
Since the $p_+$ component of a collinear momentum is $\co(M)$, 
we can form a wave packet of length $\sim M^{-1}$ in the $x^+$ direction, 
so we can actually resolve the intrinsic nonlocality of the effective theory arising from integrating out off-shell propagators at distances of $\co(M^{-1})$.
A nonlocal EFT can be just as useful as local EFTs 
as long as it possesses well-defined power-counting rules and symmetries to ensure that there are only a finite number of operators we can write down at any given order in the power-counting parameters.
This is indeed the case for our SCET lagrangian, as we will see later.

\subsubsection{Collinear and Anticollinear Gauge Invariances}
\label{sss.gauge_groups}
The requirement that a collinear field should only contain collinear Fourier modes imposes a significant restriction on the space of $\SU(3)_\mathrm{C}$ gauge transformations~\cite{Bauer_SCET_3, Bauer_SCET_4}, 
because gauge transformations should map a collinear modes to a collinear mode in order for gauge invariance to be compatible with power counting.
We thus define \emph{collinear gauge transformations} to be the $\SU(3)_\mathrm{C}$ gauge transformations that map collinear modes to collinear modes:
\beqa
\phi_\col(x) 
\> \stackrel{U_\col}{\longmapsto} \>
\phi_\col'(x) = U_{\!\col}(x) \, \phi_\col(x)
\,.
\eeqa
This implies that $U_\col(x)$ itself should only consist of collinear modes.
Hence, $U_\col(x)$ must be associated with the \emph{collinear gluon field} $G_{\!\col\mu}(x)$, which itself should be a collinear field and transform under collinear gauge transformations as
\beqa
G_{\col\mu} 
\> \stackrel{U_\col}{\longmapsto} \>
G_{\col\mu}'
=
U_{\!\col} G_{\col\mu} U_{\!\col}^\dag + \frac{\I}{g_\col} (\del_\mu U_{\!\col})  U_{\!\col}^\dag 
\,.\label{e.col_trans}
\eeqa
We must check that 
such restricted gauge transformations do not defeat the very purpose of gauge invariance 
as a redundancy of the theory to remove the gauge boson's unphysical polarization whose polarization 4-vector is parallel to the gauge boson's 4-momentum.
Since $U_\col(x)$ contains precisely the same set of Fourier modes as $G_{\col\mu}(x)$,
the removal of the unphysical polarization from $G_{\col\mu}(x)$ works if and only if we require that the \emph{polarization} components of $G_{\col\mu}$ should scale in the same way as its \emph{momentum} components~\cite{Bauer_SCET_2, Bauer_SCET_3, Bauer_SCET_4}, i.e., 
\beqa
(G_{\col+}, G_{\col-}, G_{\col\perp}) \sim (1, \la^2, \la) M  
\,.\label{e.col_gluon_scaling}
\eeqa

Having introduced the associated gauge field and gauge transformation laws, 
collinear gauge invariance can be accounted for in the usual manner.
We define the collinear covariant derivative as
\beqa
\DD_{\col\mu} \equiv \del_\mu + \I g_\col G_{\col\mu}  
\label{e.col_cov_dev}
\eeqa
so that $\DD_{\col\mu} \phi_\col(x) \longmapsto \DD'_{\col\mu} \phi_\col'(x) = U_\col(x) \, \DD_{\col\mu} \phi_\col(x)$.
We also define the collinear field strength tensor as
\beqa
G_{\col\mu\nu} \equiv \frac{1}{\I g_\col} [\DD_{\col\mu}, \DD_{\col\nu}]  
\,.\label{e.col_field_strength}
\eeqa
Thanks to the nonlocality of SCET discussed in \sssref{nonlocality}, 
there is a third gauge covariant object that can be used to construct the effective lagrangian.
We define a Wilson line $W_\col(x,y)$ as
\beqa
W_{\!\col}(x,y)
\equiv
\cp_{\!z} 
\exp\!\left[
-\I g_\col \int_y^x \!\!
\dd z^\mu \, G_{\col\mu}(z)
\right],
\eeqa
where $\cp_z$ denotes a path-ordered product along a path $z^\mu$ in which the factors associated with points $x$ and $y$ appear at the left-most and right-most positions, respectively. 
Being a Wilson line, it transforms covariantly as
\beqa
W_{\!\col}(x,y) 
\>\longmapsto\> 
W'_{\!\col}(x,y) = U_{\!\col}(x)\, W_{\!\col}(x,y)\, U_{\!\col}^\dag(y) 
\eeqa
under collinear gauge transformations\erefn{col_trans}.
As discussed in \sssref{nonlocality}, the collinear sector is permitted to have nonlocality in the $x^+$ coordinate but not in other directions.
So, a collinear Wilson line must be a straight line in the $x^+$ direction.
Such collinear Wilson lines only pick up the $+$ component of the collinear gauge field $G_{\col\mu}$, 
because $\ds{\dd z \dt G_{\col}} = \dd z^+ \, G_{\col+}$ along a straight path in the $x^+$ direction. 
The $G_{\col+}$ component is also the largest of $G_{\col\mu}$ 
as one can see from\erefn{col_gluon_scaling}.
A convenient choice of the initial point $y$ is somewhere at $y^0 \to -\infty$, 
because there are no collinear gluons in the initial state of our problem
so the boundary condition of our path integral is that $G_{\col\mu}$ should vanish in the infinite past.
Therefore, we define a \emph{collinear Wilson line}~\cite{Bauer_SCET_2, Bauer_SCET_3, Bauer_SCET_4} as
\beqa
W_{\!\col}(x)
\equiv
\cp_{\!s} 
\exp\!\left[
-\I g_\col \int_{-\infty}^0 \!\!
\dd s \, G_{\col+}(z(s))
\right]
\label{e.col-Wilson}
\eeqa
with $\cp_s$ denoting path ordering in the increasing order in $s$ from right to left, 
where the path $z^\mu(s)$ starts out from a point in the past infinity at $s=-\infty$, 
moves straight up in the $x^+$ direction as $s$ increases, 
and arrives at point $x$ when $s=0$:
\beqa
z^+(s) = x^+ \!+ s
\,,\quad
z^-(s) = x^-  
\,,\quad
\vec{z}_\perp(s) = \vec{x}_\perp 
\,.\label{e.z}
\eeqa
Because of the boundary condition at the past infinity, 
this Wilson line transforms as
\beqa
W_{\!\col}(x) 
\> \stackrel{U_\col}{\longmapsto} \>
W'_{\!\col}(x) = U_{\!\col}(x)\, W_{\!\col}(x) 
\,.\label{e.Wtrans}
\eeqa
In particular, this implies that the combination $W_\col^\dag(x) \, \phi_\col(x)$ is gauge invariant.
The collinear Wilson line $W_\col(x)$ also allows us to construct a gauge covariant 4-vector operator $\ca_{\col\mu}(x)$ as~\cite{Hill:2002vw}
\beqa
\ca_{\col\mu}(x) 
\equiv 
\frac{\I}{2g_\col} 
\Bigl(
\bigl[ \del_\mu W_\col(x) \bigr] W_\col^\dag(x)
-
W_\col(x) \bigl[ \del_\mu W_\col^\dag(x) \bigr]
\Bigr)
\,,\label{e.covariant-gluon}
\eeqa
which transforms as an object in the adjoint representation:
\beqa
\ca_{\col\mu}(x) 
\> \stackrel{U_\col}{\longmapsto} \>
U_\col(x) \, \ca_{\col\mu}(x) \, U_\col^\dag(x)
\,.
\eeqa

The fact that $U_\col(x)$ consists of collinear modes implies that 
it would \emph{not} map anticollinear modes to anticollinear modes.
Therefore, we must define anticollinear 
fields to be invariant under collinear transformations:
\beqa
\phi_\acol(x) 
\> \stackrel{U_\col}{\longmapsto} \>
\phi_\acol(x)
\,.
\eeqa
Of course, this does not imply that anticollinear fields are completely gauge invariant. 
Clearly, in the anticollinear sector, 
we must introduce the anticollinear gluon field $G_{\acol\mu}(x)$ with associated anticollinear gauge transformations:
\beqa
G_{\acol\mu} 
\> \stackrel{U_\acol}{\longmapsto} \>
G_{\acol\mu}'
=
U_{\acol} G_{\acol\mu} U_{\acol}^\dag + \frac{\I}{g_\acol} (\del_\mu U_{\acol}) \, U_{\acol}^\dag 
\,,\label{e.acol_trans}
\eeqa
where $U_\acol(x)$ only contains anticollinear Fourier modes $\sim (\la^2, 1, \la) M$. 
The anticollinear gauge coupling $g_\acol$ is an independent parameter from the collinear gauge coupling $g_\col$, 
because $G_{\col\mu}$ and $G_{\acol\mu}$ are separate fields with separate gauge transformations in the effective theory. 
The scaling law for the polarization components of $G_{\acol\mu}$ should clearly be given by 
\beqa
(G_{\acol+}, G_{\acol-}, G_{\acol\perp}) \sim (\la^2, 1, \la) M  
\,.\label{e.acol_gluon_scaling}
\eeqa
An anticollinear Wilson line $W_\acol(x)$ is defined as 
\beqa
W_{\acol}(x)
\equiv
\cp_{\!s} 
\exp\!\left[
-\I g_\acol \int_{-\infty}^0 \!\!
\dd s \, G_{\acol-}(\bar{z}(s))
\right]
\label{e.acol-Wilson}
\eeqa
with
\beqa
\bar{z}^+(s) = x^+ 
\,,\quad
\bar{z}^-(s) = x^- \!+ s 
\,,\quad
\vec{\bar{z}}_\perp(s) = \vec{x}_\perp 
\,.\label{e.zbar}
\eeqa
%

\subsection{The Effective Lagrangian}

\subsubsection{The Collinear and Anticollinear Quark Fields}
\label{sss.col_acol_fields}
In \sssref{col_acol_modes}, 
we have analyzed how different components of collinear and anticollinear \emph{4-momenta} scale with $\la$.
Here, we ask how collinear and anticollinear \emph{fields} should scale. 
Let $\psi_\col(x)$ be a left-handed Weyl-spinor field%
\footnote{\label{ftnote:spinor}
Since we are ignoring quark masses and dealing only with gauge interactions,  chirality is completely conserved.
(For ignoring the top quark mass, see a discussion at the end of \sssref{matching}.
For ignoring the higgs contribution, see footnote~\ref{ftnote:onlyqqbar}.)
The chirality conservation suggests that 
we should use Weyl spinors rather than Dirac spinors. 
We therefore describe all spin-$1/2$ fermions in terms of left-handed Weyl spinors until the very end of \ssref{factorization_and_partons}.
We adopt a widely used notation of indexing the left-handed and right-handed spinors as ${}_\al$ and ${}^{\dot{\al}}$, respectively, 
with their contractions going as ${}^\al {}_\al$ 
and ${}_{\dot{\al}} {}^{\dot{\al}}$, 
where ${}^\al \equiv \ep^{\al\be} \, {}_\be$ 
and ${}_{\dot{\al}} \equiv \bar{\ep}_{\dot{\al}\dot{\be}} \, {}^{\dot{\be}}$
with $\ep \equiv \I \sg^2$ and $\bar{\ep} \equiv -\I \sg^2$.
The matrices $\sg^\mu$ and $\sgb^\mu$ are defined 
as $\sg^0 \equiv \sgb^0 \equiv\mathbbm{1}$ 
and $\sg^{1,2,3} \equiv -\sgb^{1,2,3} \equiv \sg_{x,y,z}$.
}
interpolating a collinear quark, 
that is, 
let $\psi_\col(x)$ consist only of Fourier modes that scale as $\sim (1, \la^2, \la) M$. 
Hence, we have
\beqa
\del_+ \psi_\col \sim M \psi_\col
\,,\quad
\del_- \psi_\col \sim \la^2 M \psi_\col
\,,\quad
\del_\perp \psi_\col \sim \la M \psi_\col
\,.\label{e.psi_col_scaling}
\eeqa
This does \emph{not} imply that the kinetic term 
$\ds{\psi_\col^\dag \, \sgb \dt \del \, \psi_\col^\PD} 
= 
\ds{\psi_\col^\dag (\sgb^+ \del_+} 
+ \sgb^- \del_- 
+ \ds{\vec{\sg}_\perp \dt \vec{\del}_\perp) \psi_\col}$ 
can be approximated by $\ds{\psi_\col^\dag \, \sg^+ \del_+ \psi_\col}$ at the leading order in $\la$,
because the underlying Lorentz invariance requires that 
all the three terms in $\ds{\psi_\col^\dag \, \sgb \dt \del \, \psi_\col^\PD}$ should be present.%
\footnote{
The underlying Lorentz invariance manifests itself in SCET as \emph{reparameterization invariance} (RPI)~\cite{Manohar:2002fd}, 
i.e., the invariance of action under different choices of lightcone axes $n_\pm$ satisfying $\ds{n_\pm \dt n_\pm} = 0$ and $\ds{n_\pm \dt n_\mp} = 2$. 
RPI holds only if all the three terms in $\ds{\psi_\col^\dag \, \sgb \dt \del \, \psi_\col^\PD}$ have the same coefficient.
}
Rather, the scaling\erefn{psi_col_scaling} only implies that 
upon projecting the spinor space onto the subspaces annihilated by $\sgb^+$ or $\sgb^-$, 
the different projected components of $\psi_\col$ should scale differently 
such that all the terms in $\ds{\psi_\col^\dag \, \sgb \dt \del \, \psi_\col}$ scale in $\la$ homogeneously. 

To construct projection operators $\cp$ and $\bar{\cp}$ onto the subspaces annihilated by $\sgb^+$ and $\sgb^-$, respectively, 
notice that the Dirac algebra in the lightcone metric\erefn{metric} takes the form
\beqa
\sg^+ \sgb^- + \sg^- \sgb^+ = 2 g^{+-} \mathbbm{1} = \mathbbm{1}
\,,\quad 
\sgb^+ \sg^+ = g^{++} \mathbbm{1} = 0
\,,\quad 
\sgb^- \sg^- = g^{--} \mathbbm{1} = 0
\,. 
\eeqa
The projection operators are hence given by
\beqa
\cp \equiv \sg^+ \sgb^- 
\,,\quad
\bar{\cp} \equiv \sg^- \sgb^+
\,.
\eeqa
Just like chirality is an eigenvalue of $\ga_5$, 
we can define the \emph{collinearness operator} $\cc$ as
\beqa
\cc \equiv \cp - \bar{\cp}
\eeqa
so that the subspaces projected by $\cp$ and $\bar{\cp}$ have $\cc$-eigenvalues $+1$ and $-1$, respectively. 
With $\cp$ and $\bar{\cp}$, 
we can now obtain the desired decomposition of $\psi_\col$ as
\beqa
\psi_\col = \xi_\col + \Xi_\col
\qquad\text{with}\quad
\xi_\col \equiv \cp \, \psi_\col  
\,,\quad
\Xi_\col \equiv \bar{\cp} \, \psi_\col
\label{e.psi_col} 
\eeqa
so that
\beqa
&\sgb^+ \xi_\col = 0
\,,
&&\cp \, \xi_\col = \xi_\col
\,,
&&\cc \, \xi_\col = +\xi_\col
\,,\\
&\sgb^- \Xi_\col = 0
\,,
&&\bar{\cp} \, \Xi_\col = \Xi_\col
\,,
&&\cc \, \Xi_\col = -\Xi_\col
\,.\label{e.xi_col_const}
\eeqa
To see that only $\xi_\col$ is a dynamical degree of freedom,
consider a situation in which
the initial quark is on-shell and exactly collinear, 
i.e., its 4-momentum only has the $p_+$ component.
In this case, 
the Dirac equation gives 
$0
= 
\ds{\sgb \dt \del \psi_\col} 
\propto 
\sgb^+ \psi_\col 
= 
\sgb^+ \Xi_\col$. 
So, the $\Xi_\col$ component is forced to identically vanish by the equation of motion and thus never goes on-shell.
On the other hand, 
$\xi_\col$ can be nontrivial even on-shell and thus describes a dynamical, physical degree of freedom.%
\footnote{
Being always off-shell, 
$\Xi_\col$ should ultimately be integrated out from the effective theory. 
Nevertheless, it can be convenient to keep it in the theory as an auxiliary field for the purpose of analyzing the constraints of RPI or the underlying Lorentz invariance.
This is exactly reminiscent of heavy quark effective theory (HQET) 
in which keeping the ``small'' component of a heavy quark field (the one annihilated by $(1 + \slashed{v}) / 2$, typically denoted by $H_v(x)$) makes the bookkeeping of RPI far more transparent~\cite{Sundrum:1997ut}.
Yet other examples of non-dynamical fields that are convenient for symmetry purposes are the $D$ and $F$ fields in supersymmetric field theories.
The benefit of keeping $\Xi$ in SCET is particularly emphasized in Ref.~\cite{Luke_SCET}.
}

We are now ready to derive the scaling laws for $\xi_\col$ and $\Xi_\col$.
In terms of these fields, the kinetic term of $\psi_\col$ becomes
\beqa
\psi_\col^\dag \, \sgb \dt \del\,  \psi_\col^\PD 
=
 \xi_\col^\dag \, \sgb^- \del_- \xi_\col  
+\Xi_\col^\dag \, \sgb^+ \del_+ \Xi_\col
+\xi_\col^\dag \, \vec{\sg}_\perp \dt \vec{\del}_\perp \Xi_\col 
+\Xi_\col^\dag \, \vec{\sg}_\perp \dt \vec{\del}_\perp \xi_\col 
\,.\label{e.psi_col_pure_kin}
\eeqa
Since the scaling\erefn{psi_col_scaling} implies that 
$(\dd x^+, \dd x^-, \dd x_\perp) \sim (1, \la^{-2}, \la^{-1}) M^{-1}$,
the $\dd^4 x$ integration measure in the action 
$\int\! \dd^4 x\, \ds{\psi_\col^\dag \, \sgb \dt \del \, \psi_\col}$ 
scales as $\la^{-4} M^{-4}$.
Since kinetic terms are treated as the leading terms in our perturbation theory, 
the scaling dimensions of fields should be governed by the kinetic terms.
In particular,  
demanding that the action 
$\int\! \dd^4 x\, \ds{\psi_\col^\dag \, \sgb \dt \del \, \psi_\col}$ 
should not scale with $\la$ nor $M$,
we see that the kinetic term as a whole must scale as $\la^4 M^4$. 
Then, the underlying Lorentz invariance requires that each of the four terms in\erefn{psi_col_pure_kin} should scale as $\la^4 M^4$.
We therefore conclude that~\cite{Bauer_SCET_1, Bauer_SCET_2}
\beq
\xi_\col \sim \la M^{3/2} 
\,,\quad 
\Xi_\col \sim \la^2 M^{3/2} 
\,.\label{e.col_fermion_scaling}
\eeq
Thus, while $\xi_\col$ and $\Xi_\col$ both have the canonical mass dimension of fermion fields, 
their $\la$ dimensions differ and $\Xi_\col$ is subdominant in $\la$ expansion.
Nevertheless, we have already noted above that 
the underlying Lorentz invariance (or RPI of SCET) forbids us from simply dismissing $\Xi_\col$ as subdominant.
Fortunately, 
precisely because $\Xi_\col$ is always an off-shell, non-dynamical degree of freedom,
we can resolve this dilemma by treating $\Xi_\col$ as an auxiliary field and integrating it out by using its equation of motion as needed. 
This is again very reminiscent of the ``small'' component of a heavy field in HQET\@.

The initial anticollinear quark clearly works in a similar way.
Let $\psi_\acol(x)$ be a left-handed Weyl spinor field that interpolates an anticollinear quark.
Then, the expressions\erefn{psi_col} and\erefn{xi_col_const} should be translated for $\psi_\acol$ as
\beqa
\psi_\acol = \xi_\acol + \Xi_\acol
\qquad\text{with}\quad
\xi_\acol \equiv \bar{\cp} \, \psi_\acol  
\,,\quad
\Xi_\acol \equiv \cp \, \psi_\acol
\label{e.psi_acol} 
\eeqa
and
\beqa
&\sgb^- \xi_\acol = 0
\,,
&&\bar{\cp} \, \xi_\acol = \xi_\acol
\,,
&&\cc \, \xi_\acol = -\xi_\acol
\,,\\
&\sgb^+ \Xi_\acol = 0
\,,
&&\cp \, \Xi_\acol = \Xi_\acol
\,,
&&\cc \, \Xi_\acol = +\Xi_\acol
\,.\label{e.xi_acol_const}
\eeqa

Finally, let us discuss the interactions of the collinear and anticollinear quarks.
Besides through the collinear covariant derivative\erefn{col_cov_dev} and collinear field strength\erefn{col_field_strength}, 
collinear gluons can interact with collinear quark modes via the collinear Wilson line $W_\col(x)$ defined in\erefn{col-Wilson}. 
A similar statement clearly applies to the anticollinear quark.
For later use, we define the following gauge-invariant versions of $\xi_\col$ and $\xi_\acol$~\cite{Bauer_SCET_4} using the Wilson lines\erefn{col-Wilson} and\erefn{acol-Wilson}:
\beqa
\chi_\col(x) \equiv W_\col^\dag(x) \, \xi_\col(x)
\,,\quad
\chi_\acol(x) \equiv W_\acol^\dag(x) \, \xi_\acol(x) 
\,.\label{e.chi}
\eeqa
Since Wilson lines carry no spinor indices, the constraints\erefn{xi_col_const} and\erefn{xi_acol_const} for $\xi_\col$ and $\xi_\acol$ apply to $\chi_\col$ and $\chi_\acol$ in the same manner.
In particular, we have
\beqa
&\sgb^+ \chi_\col = 0
\,,
&&\cp \chi_\col = \chi_\col
\,,
&&\cc \chi_\col = +\chi_\col
\,,\\
&\sgb^- \chi_\acol = 0
\,,
&&\bar{\cp} \chi_\acol = \chi_\acol
\,,
&&\cc \chi_\acol = -\chi_\acol
\,.\label{e.chi_col-chi_acol-constraints}
\eeqa

Finally, let us briefly comment on the so-called Glauber or Coulomb modes~\cite{Collins:1981ta, Bodwin:1981fv}, 
which were shown to be necessary in SCET for consistency~\cite{Bauer:2010cc}.
However, since the Glauber/Coulomb modes are always off-shell, they should be integrated out from the effective theory in accord with our principle that guaranteed-off-shell modes should be integrated out, 
which is expected to give rise to some (nonstandard) nonlocal interactions between SCET modes~\cite{Bauer:2010cc}.
Calculating the effects of those operators is beyond the scope of this paper, 
and we simply assume that the effects of Glauber/Coulomb modes cancel out in the final results as they do in the inclusive Drell-Yan process~(see \cite{glauber_cancel_1} for the pioneering work and also \cite{glauber_cancel_2, glauber_cancel_3, glauber_cancel_4}).

\subsubsection{The Effective Lagrangian}
\label{sss.lagrangian}
The splitting of modes into collinear and anticollinear modes means that the lagrangian should also be split.    
So, the entire effective lagrangian is given at the leading order in $\la$ (i.e., $\co(\la^0)$) by
\beqa
\cl_\text{eff}
=
\cl_\col + \cl_\acol + \cl_\text{hard}
\,,\label{e.Leff}
\eeqa
where $\cl_\col$ is a SCET lagrangian for the collinear sector:
\beqa
\cl_\col
=
-\frac{1}{2 g_\col^2} \tr\left[ G_{\col\mu\nu} G_\col^{\mu\nu} \right]
+\I \bigl(
 \xi_\col^\dag \, \sgb^- \DD_{\col-} \xi_\col  
+\Xi_\col^\dag \, \sgb^+ \DD_{\col+} \Xi_\col
+\xi_\col^\dag \, \vec{\sg}_\perp \dt \vec{\DD}_{\col\perp} \Xi_\col 
+\Xi_\col^\dag \, \vec{\sg}_\perp \dt \vec{\DD}_{\col\perp} \xi_\col
\bigr)
\,.\label{e.col_lagrangian} 
\eeqa
There are no other terms we can write at the leading order in $\la$
except that in principle the terms in $\cl_{\col, \acol}$ could display the nonlocalities discussed in \sssref{nonlocality}.
For example, at $\co(\la^0)$, 
the most general operator bilinear in $\xi_\col^\dag$ and $\xi_\col$ is a nonlocal operator 
$\int \dd t\, f(t) \, \ds{\xi_\col^\dag(x^+\! + t, x^-\!, x_\perp)} \, \sgb^- \DD_{\col-} \xi_\col(x)$ 
with a Wilson coefficient $f(t)$.
However, a matching calculation onto QCD tells us that $f(t) = \de(t)$, thus giving us\erefn{col_lagrangian}. 
In fact, $\cl_\col$ is exact to all orders in $\la$, because the collinear sector taken \emph{in isolation} must be identical to a QCD by Lorentz invariance.
In other words, $\cl_\col$ can be just viewed as a QCD lagrangian written in the lightcone coordinates to make power counting manifest so that it can be readily used in EFT calculations.

The second term $\cl_\acol$ in\erefn{Leff} is the obvious anticollinear counterpart of $\cl_\col$, which again is identical to QCD in isolation and exact to all orders in $\la$.
The third term $\cl_\text{hard}$ is the only place where the collinear and anticollinear fields come together (to produce $W^+ W^-$) 
and consequently where $\la$ expansion is nontrivial. 
We will now describe $\cl_\text{hard}$ at the leading order in $\la$. 

\subsubsection{The Hard Interaction}
\label{sss.hard-int}
In our calculation, 
we treat electroweak gauge interactions only at the tree level. 
In particular, 
the $W^\pm$ bosons appear only as external final states in both the full and effective theories. 
The $Z$ and $\ga$ appear in the full theory only as an $s$-channel propagator leading up to the $W^+ W^-$ production vertex.
Since those $Z$ and $\ga$ propagators are always highly off-shell, 
they cannot appear in the effective theory and their effects must be incorporated in an effective vertex. 
The $W^+ W^-$ pair can also be produced via a $t$-channel quark exchange in the full theory. 
This $t$-channel propagator is also always far off-shell 
so must be integrated out into the effective vertex.
Therefore, in our SCET, the $W^+ W^-$ production must be described by an operator of collinear and anticollinear fields coupled to \emph{external} $W^+$ and $W^-$ states:%
\beqa
\cl_\text{hard} 
= 
\frac{1}{M} \, 
\ep^*_\mu(p_3, s_3) \, \ep^*_\nu(p_4, s_4) \, \e^{\I (p_3 + p_4) \cdot x} \, \cj^{\mu\nu}(x)  
\,,\label{e.Lint}
\eeqa
where $\cj^{\mu\nu}(x)$ is a SCET operator that destroys the initial collinear and anticollinear quarks, 
while $(p_3, s_3)$ and $(p_4, s_4)$ are the 4-momenta and polarizations of the $W^+$ and $W^-$, respectively.
We have already substituted the final-state wavefunctions 
$\ds{\ep^*_\mu(p_3, s_3) \, \e^{\I p_3 \cdot x}}$ and 
$\ds{\ep^*_\mu(p_4, s_4) \, \e^{\I p_4 \cdot x}}$, 
because the $W^\pm$ bosons appear only as external states,
simply acting as a source for the SCET operator $\cj^{\mu\nu}(x)$. 

As discussed in \sssref{nonlocality},  
collinear and anticollinear fields constituting the SCET operator $\cj^{\mu\nu}(x)$ are allowed to be nonlocal in the $x^+$ and $x^-$ coordinates, respectively, while they must be local in the remaining coordinates. 
Hence, at the leading order in $\la$ (that is, $\co(\la^0)$), the operator $\cj^{\mu\nu}(x)$ can be written in terms of $\chi_\col$ and $\chi_\acol$ defined in\erefn{chi} as 
\beqa
\cj^{\mu\nu}(x) 
= 
\int\! \dd t_1 \, \dd t_2 \,
\chi_\acol^{i\al} (x^- \!+ t_2, \vec{x}_\perp) \,
\bigl[ \Ga^{\mu\nu}(t_1, t_2, p_{3+4\parallel}, p_{3-4}^\PD, \mu) \bigr]_\al^{~\be} \, 
{\chi_\col}_{i\be} (x^+ \!+ t_1, \vec{x}_\perp) \,,
\label{e.current} 
\eeqa 
where $p_{3\pm4} \equiv p_3 \pm p_4$, 
while $\al$ and $\be$ are spinor indices (see footnote~\ref{ftnote:spinor} for our convention), 
and $i$ an $\SU(3)_\mathrm{C}$ index. 
As we are working at the leading order in $\la$,
no $\Xi_\col$ or $\Xi_\acol$ should appear in\erefn{current}, 
as they scale with a higher power of $\la$ than $\xi_\col$ and $\xi_\acol$,
as shown in\erefn{col_fermion_scaling}.
Similarly, 
we have not considered any $\del_-$ or $\del_\perp$ acting on $\chi_\col$, 
nor any $\del_+$ or $\del_\perp$ on $\chi_\acol$, 
as they are subdominant in $\la$. 
The absence of derivatives then excludes the possibility of 
collinear and anticollinear gluon fields entering $\cj^{\mu\nu}$ 
through covariant derivatives.
We cannot insert a gluon field strength ($G_{\col \mu \nu}$ or $G_{\acol \mu \nu}$) or a gauge covariant 4-vector field\erefn{covariant-gluon} between $\chi_\acol$ and $\chi_\col$, as that would no longer correspond to a hard $q\bar{q} \to WW$ process.  
Therefore, gluon couplings can only be through 
the Wilson lines\erefn{col-Wilson} and\erefn{acol-Wilson},
and the separate collinear and anticollinear gauge invariances tell us 
that they can only appear through the gauge-invariant combinations $\chi_\col$ and $\chi_\acol$ defined in\erefn{chi}.
While gauge invariant, $\chi_{\col}$ and $\chi_\acol$ both transform under the \emph{common, global} $\SU(3)_\mathrm{C}$ that they inherit from the global part of the original $\SU(3)_\mathrm{C}$. 
Since the $W^+ W^-$ state is a color singlet,  
the operator $\cj^{\mu\nu}$ must be a singlet under the global $\SU(3)_\mathrm{C}$, 
which is why the index $i$ is contracted in\erefn{current}.

To ensure a well-defined power counting in every single step of the calculation, 
we have only kept the leading, $\co(\la^0)$ arguments of the fields $\chi_\col$ and $\chi_\acol$ and of the function $\Ga^{\mu\nu}$ in\erefn{current}.
Observe that while the momenta $p_3$ and $p_4$ individually scale as $\sim (1,1,1) M$, 
their sum $p_3 + p_4$ scales as $\sim (1,1,\la) M \sim (1,1,\ptv)$ because of the jet veto. 
Hence, the $(p_{3+4})_\perp^\PD$ arguments of $\Ga^{\mu\nu}$ are dismissed as subleading, $\co(\la)$ effects, while $(p_{3+4})_\parallel^\PD$ and all components of $p_{3-4}$ should be fully kept.
The scaling of $p_{3+4}$ also tells us that,
upon integrating the interaction term\erefn{Lint} over the whole spacetime, 
the exponential $\e^{\I (p_3 + p_4) \cdot x}$ in $\cj$ oscillates rapidly in both $x^+$ and $x^-$ directions with short wavelengths of $\co(M^{-1})$, 
while slowly with long wavelengths of $\co(\la^{-1} M^{-1})$ in the $x_\perp$ directions. 
On the other hand, 
consisting only of Fourier modes scaling as $\sim (1,\la^2, \la) M$,
the collinear field $\chi_\col(x)$ varies slowly in the $x^-$ direction with a long wavelength of $\co(\la^{-2} M^{-1}) \gg M^{-1}$, 
while it varies as fast as $\e^{\I (p_3 + p_4) \cdot x}$ in the $x^+$ and $x_\perp$ directions. 
Thus, the variation of $\chi_\col(x)$ in the $x^-$ direction is a subleading, $\co(\la^2)$ effect and must be discarded at the leading order in $\la$.
Similarly, the variation of $\chi_\acol(x)$ in the $x^+$ direction must be neglected.

The function $[\Ga^{\mu\nu}(t_1,t_2, p_{3+4\parallel}, p_{3-4}^\PD, \mu)]_{\al}^{~\be}$ is a Wilson coefficient that encodes the effects of guaranteed-off-shell modes with virtuality $\gtrsim \mu$ that have been integrated out.
So, $\Ga^{\mu\nu}$ should be first determined at $\mu \sim M$ by matching SCET amplitudes to the full-theory counterparts, 
where the latter involves the $s$- and $t$-channel propagators with virtuality of $\co(M)$ as discussed above.
This matching calculation will be presented in \sssref{matching}.
Once $\Ga^{\mu\nu}$ is matched, 
we must integrate out guaranteed-off-shell modes with virtuality between $\co(M)$ and $\co(\ptv)$ before we calculate the cross section using the interaction\erefn{Lint},
because the actual scale of virtuality of our process, 
that is, the scale of virtuality of \emph{can-be-on-shell} modes, 
is $\co(\la M) \sim \ptv$ due to the jet veto.
This is the step that resums the large logarithms $\sim \log (M / \ptv)$, 
which will be discussed in \sssref{hard-RG}.

$\Ga^{\mu\nu}$ as a matrix in the spinor space can actually have only one nonzero component.
This can be made obvious by spanning the spinor space in terms of the eigenstates of the collinearness operator $\cc$.
Since the constraints\erefn{chi_col-chi_acol-constraints} tell us that 
$\chi_\col$ and $\chi_\acol$ are eigenstates of $\cc$ with eigenvalues $+1$ and $-1$, respectively,
only one entry of $\Ga^{\mu\nu}$ that corresponds to those eigenvalues can be nonzero.
To locate this non-vanishing component in a basis independent way,
we define two left-handed Weyl spinors $u_\col$ and $u_\acol$ solving the constraints
\beqa 
&\cc u_\col = +u_\col
\,,\quad
u_\col u_\col^\dag = \sg^+
\,,\\
&\cc u_\acol = -u_\acol
\,,\quad
u_\acol u_\acol^\dag = \sg^-
\,.\label{e.ucol_uacol}
\eeqa
These conditions completely determine $u_\col$ and $u_\acol$ up to overall phases.
Then, since $u_\col$ and $u_\acol$ satisfy the same constraints as $\chi_\col$ and $\chi_\acol$ of the form\erefn{chi_col-chi_acol-constraints},
the product ${u_\col}_\al u_\acol^\be$ is nonzero 
precisely for the $\al$ and $\be$ 
for which $[\Ga^{\mu\nu}]_\al^{~\be}$ can be nonzero.
We therefore write $\Ga^{\mu\nu}$ as
\beqa
\bigl[ \Ga^{\mu\nu}(t_1, t_2, p_{3+4\parallel}, p_{3-4}^\PD, \mu) \bigr]_\al^{~\be}
= C^{\mu\nu}(t_1, t_2, p_{3+4\parallel}, p_{3-4}^\PD, \mu) \, \Ga_\al^{~\be} 
\,,\label{e.Gamma_munu}
\eeqa
where  
\beqa
\Ga_\al^{~\be} 
\equiv 
{u_\col}_\al u_\acol^\be
\,.\label{e.Gamma}
\eeqa
Therefore, 
we just need to match one number, $C^{\mu\nu}(t_1, t_2, p_{3+4\parallel}, p_{3-4}^\PD, \mu)$. 
Needless to say, this non-vanishing component of $\Ga$ is picking up the on-shell, physical polarizations of the initial collinear and anticollinear quarks.
The remaining polarizations are always off-shell and hence do not appear in the SCET\@.

\subsection{Factorization and the Emergence of the Parton Picture}
\label{ss.factorization_and_partons}

\subsubsection{Factorization of Matrix Elements}
To calculate the cross section for the process  $\proton(P_1) + \proton(P_2) \to W^+ (p_3, s_3) + W^- (p_4, s_4) + X$, 
we need to evaluate the matrix element
\beqa
J^{\mu\nu}_X (x, P_1,P_2, p_3,p_4)
\equiv
\bra{X} \cj^{\mu\nu}(x) \ket{\proton(P_1) \, \proton(P_2)}
\,,\label{e.JX}
\eeqa
where dependences on the proton spins are implicit.
It is also understood 
that the fields inside $\cj^{\mu\nu}(x)$ are time-ordered 
and that $J_X^{\mu\nu}$ is only the connected part of the matrix element.
Substituting\erefn{current} for $\cj^{\mu\nu}(x)$ together with\erefn{Gamma_munu}, we get
\beqa
&
J^{\mu\nu}_X (x, P_1,P_2, p_3,p_4)
\\
&=
\int\! \dd t_1 \, \dd t_2 \,
C^{\mu\nu}(t_1, t_2, p_{3+4\parallel}, p_{3-4}^\PD, \muf) \,
\bra{X} 
\chi_\acol^{i\al} (x^- \!+ t_2, \vec{x}_\perp) \;  
\Ga_\al^{~\be} \, 
{\chi_\col}_{i\be} (x^+ \!+ t_1, \vec{x}_\perp) 
\ket{\proton(P_1) \, \proton(P_2)}
\,,
\eeqa
where the Wilson coefficient $C^{\mu\nu}$ is now evaluated at the scale $\mu = \muf \sim \ptv$, 
because the actual scale of virtuality in the process in question is $\co(\la M) \sim \ptv$, as we already noted above.

Now, since $\chi_\col$ can only create collinear states and $\chi_\acol$ only anticollinear states, 
and also since the remnants of the colliding protons are collinear or anticollinear, 
the hadronic state $\ket{X}$ must be composed of only collinear and anticollinear states, i.e.,  
\beqa
\ket{X} = \ket{X_\acol X_\col}
\,,
\eeqa
where $\ket{X_\col}$ consists only of collinear particles, and $\ket{X_\acol}$ only of anticollinear particles.
For the initial state, 
we let $P_1$ be collinear and $P_2$ anticollinear by definition and without loss of generality, 
so $\chi_\col$ and $\chi_\acol$ must act on $\ket{\proton(P_1)}$ and $\ket{\proton(P_2)}$, respectively.
Moreover, we cannot form a gluon loop connecting $\chi_\col$ and $\chi_\acol$, 
because $\chi_\col$ can only emit collinear gluons and $\chi_\acol$ only anticollinear gluons, as they are charged under separate gauge groups as discussed in \sssref{gauge_groups}.
Therefore, we have
\beqa
&
\bra{X} 
\chi_\acol^{i\al} (x^- \!+ t_2, \vec{x}_\perp) \;  
\Ga_\al^{~\be} \,
{\chi_\col}_{i\be} (x^+ \!+ t_1, \vec{x}_\perp) 
\ket{\proton(P_1) \, \proton(P_2)}
\\
&=
\bra{X_\acol} \chi_\acol^{i\al} (x^- \!+ t_2, \vec{x}_\perp) \ket{\proton(P_2)} \; 
\Ga_\al^{~\be} \, 
\bra{X_\col} {\chi_{\col}}_{i\be} (x^+ \!+ t_1, \vec{x}_\perp) \ket{\proton(P_1)} 
\,.
\eeqa
Using the momentum operator to relocate the fields $\chi_\acol$ and $\chi_\col$ to the same point $x$, this becomes
\beqa
=
\e^{-\I p_{2-}^\PD t_2^\PD} \,
\e^{-\I p_{1+}^\PD t_1^\PD} \,
\bra{X_\acol} \chi_\acol^{i\al} (x^-, \vec{x}_\perp) \ket{\proton(P_2)} \; 
\Ga_\al^{~\be} \,
\bra{X_\col} {\chi_\col}_{i\be} (x^+, \vec{x}_\perp) \ket{\proton(P_1)} 
\,,\label{e.relocated}
\eeqa
where 
\beqa
p_{1+}^\PD \equiv (P_1^\PD - P_{X_\col}^\PD)_+^\PD 
\,,\quad
p_{2-}^\PD \equiv (P_2^\PD - P_{X_\acol}^\PD)_-^\PD
\label{e.p1p2}  
\eeqa
with $P_{X_\col}$ and $P_{X_\acol}$ being the 4-momenta of the states $\ket{X_\col}$ and $\ket{X_\acol}$, respectively.
By unpacking $\Ga_\al^{~\be}$ using\erefn{Gamma}, 
the matrix element\erefn{relocated} becomes
\beqa
=
\e^{-\I p_{2-}^\PD t_2^\PD} \,
\e^{-\I p_{1+}^\PD t_1^\PD} \,
\bra{X_\acol} \chi_\acol^{i}(x^-\!, \vec{x}_\perp) \, u_\col \ket{\proton(P_2)} \; 
\bra{X_\col} u_\acol \, {\chi_\col}_{i}(x^+\!, \vec{x}_\perp) \ket{\proton(P_1)}
\,, 
\eeqa
where the spinor indices are now implicit and just contracted within each bra-ket.
Therefore, we obtain
\beqa
&
J^{\mu\nu}_X (x, P_1,P_2, p_3,p_4)
\\
&=
\bra{X_\acol} \chi_\acol^{i}(x^-\!, \vec{x}_\perp) \, u_\col \ket{\proton(P_2)} \; 
\bra{X_\col} u_\acol \, {\chi_\col}_{i}(x^+\!, \vec{x}_\perp) \ket{\proton(P_1)} \;
\tilde{C}^{\mu\nu}(p_{1+}, p_{2-}, p_{3+4\parallel}, p_{3-4}^\PD, \muf)
\,,\label{e.factorized}
\eeqa
where
\beqa
\tilde{C}^{\mu\nu}(p_{1+}, p_{2-}, p_{3+4\parallel}, p_{3-4}^\PD, \muf)
\equiv
\int\! \dd t_1 \, \dd t_2 \,
\e^{-\I p_{1+}^\PD t_1^\PD} \,
\e^{-\I p_{2-}^\PD t_2^\PD} \,
C^{\mu\nu}(t_1, t_2, p_{3+4\parallel}, p_{3-4}^\PD, \muf)
\,.\label{e.Ctilde}
\eeqa
The matrix element\erefn{factorized} is now manifestly \emph{factorized,} 
i.e., we can \emph{separately} compute
the hard matrix element (i.e., the $\tilde{C}$ function), 
the purely collinear matrix element (the one with $\ket{X_\col}$), 
and the purely anticollinear matrix element (the one with $\ket{X_\acol}$).
The only subtlety here is that the individual matrix elements have rapidity divergences and display collinear anomalies, all of which must cancel out. 
We will analyze this subtlety in \sssref{col_anom}

\subsubsection{The Factorized Cross Section}
Using the effective interaction\erefn{Lint},
the spin-averaged cross-section for the process  $\proton(P_1) + \proton(P_2) \to \ds{W^+ (p_3, s_3) + W^- (p_4, s_4) + \vetosum X}$ 
is given by
\beqa
\sg 
=\>&  
\frac{1}{2s} \frac{1}{M^2}
\int\!
\frac{\dd^3 \vec{p}_3}{(2 \pi)^3 \, 2 E_3}
\frac{\dd^3 \vec{p}_4}{(2 \pi)^3 \, 2 E_4} \,
\Sg(P_1,P_2, p_3,p_4)
\,,
\label{e.xs1}
\eeqa
where the proton mass has been neglected and hence $s = 2 \ds{P_1 \dt P_2}$,
while the integrand $\Sg(P_1,P_2, p_3,p_4)$ is given by
\beqa
\Sg(P_1,P_2, p_3,p_4)
\equiv 
\frac14 \sum_{\text{p spins}} 
\vetosumX 
\sum_{s_3, s_4} 
\int & \dd^4 x \, \e^{-\I (p_3 + p_4) \cdot x} \, 
\ep_{\rho}(p_3, s_3) \, \ep_{\sg}(p_4, s_4) \,
\ep_{\mu}^*(p_3, s_3) \, \ep_{\nu}^*(p_4, s_4) 
\\
&\times
J^{*\rho\sg}_X (x, P_1,P_2, p_3,p_4) \,
J^{\mu\nu}_X (0, P_1,P_2, p_3,p_4)
\,,
\label{e.Sigma}
\eeqa
where the $'$ on $\sum'$ indicates two things:
(i) because of the jet veto, the sum over $X$ goes only over the states satisfying the jet-veto condition, and
(ii) $P_X$ must be consistent with 4-momentum conservation, $P_1 + P_2 = p_3 + p_4 + P_X$. 
The average over the proton spins is being implied by $\frac14 \sum_{\text{p spins}}$, although proton spin dependences are not explicit. 

Since it is our interest to express the cross section in terms of the $WW$ invariant mass $M$, 
we define a new 4-vector $q \equiv p_3 + p_4$ 
and eliminate either $p_3$ or $p_4$ in favor of $q$. 
We choose to eliminate $p_4$.  
We thus have 
\beqa
\int\!\frac{\dd^3 \vec{p}_4}{2E_4} 
&= 
\int\! \dd^4 p_4 \, 
\de\bigl( p_4^2 - m_W^2 \bigr) \, 
\theta( p_4^0 - m_W) 
\\
&= 
\int\! \dd^4 q \, 
\de\bigl( (q - p_3)^2 - m_W^2 \bigr) \, 
\theta\bigl( q^0 - E_3 - m_W \bigr)
\,.\label{e.p4integral}
\eeqa
We then change integration variables from $q^0$ and $q^3$ to $M$ and $\eta$ defined as  
\beqa
q^0 \equiv \sqrt{M^2 + |\vec{q}_\perp|^2}\, \cosh\eta  
\,,\quad
q^3 \equiv \sqrt{M^2 + |\vec{q}_\perp|^2}\, \sinh\eta  
\,.\label{e.M_and_eta}
\eeqa
These definitions imply $q^2 = (q^0)^2 - (q^3)^2 - |\vec{q}_\perp|^2 = M^2$ 
and $\eta = \frac12 \log\frac{q^0 + q^3}{q^0 - q^3}$, 
so $M$ and $\eta$ are the invariant mass and rapidity of the $WW$ system, respectively.
Upon the change of variables\erefn{M_and_eta}, 
we have $\dd q^0 \, \dd q^3 = M \dd M \, \dd \eta$,
and the requirement $q^0 \geq E_3 + m_W$ translates to $M \geq 2m_W$, 
so\erefn{p4integral} becomes
\beqa
= 
M\! \int\! \dd^2 \vec{q}_\perp \, \dd M \, \dd\eta \; 
\de(M^2 - 2 q \dt p_3) \, \theta(M - 2 m_W)
\,.
\eeqa
Therefore, the cross section\erefn{xs1} can be rewritten as
\beqa
\frac{\dd \sg}{\dd M} 
=\>&  
\frac{1}{4\pi s M} 
\int\!
\frac{\dd^3 \vec{p}_3}{(2\pi)^3 \, 2 E_3} \,
\frac{\dd^2 \vec{q}_\perp}{(2\pi)^2} \, 
\dd\eta \;
\de(M^2 - 2 q \dt p_3) \,
\Sg(P_1,P_2, p_3,p_4)\bigr|_{p_4 = q - p_3}
\label{e.xs2}
\eeqa
where it is understood that the components $q^{0,3}$ are dependent variables
and related to the independent variables $M$, $y$, and $\vec{q}_\perp$ through\erefn{M_and_eta}. 
The constraint $M \geq 2m_W$ is also understood.

Returning to the calculation of $\Sg(P_1, P_2, p_3, p_4)$, 
we perform the summation over the $W^\pm$ polarizations in\erefn{Sigma}
and substitute\erefn{factorized} there, 
and we get
\beqa
\Sg(P_1,P_2, p_3,p_4)
&=
\frac14 \sum_{\text{p spins}} 
\vetosumX
\int\! \dd^4 x \, \e^{-\I (p_3 + p_4) \cdot x} \, 
C(p_{1+}, p_{2-}, p_{3+4\parallel}, p_{3-4}^\PD, \muf) 
\\
&\phantom{=}\times
\left[
\bra{X_\acol} \chi_\acol^{j} (x^-\!, \vec{x}_\perp) \, u_\col \ket{\proton(P_2)} \; 
\bra{X_\col} u_\acol \, {\chi_\col}_{j}(x^+\!, \vec{x}_\perp) \ket{\proton(P_1)}
\right]^*
\\
&\phantom{=}\times
\bra{X_\acol} \chi_\acol^{i}(0,0) \, u_\col \ket{\proton(P_2)} \; 
\bra{X_\col} u_\acol \, {\chi_\col}_{i}(0,0) \ket{\proton(P_1)}
\label{e.polarization_summed}
\eeqa
where
\beqa
C(p_{1+}, p_{2-}, p_{3+4\parallel}, p_{3-4}^\PD, \muf)
\equiv\>&
\bigl[ \tilde{C}^{\rho\sg}(p_{1+}, p_{2-}, p_{3+4\parallel}, p_{3-4}^\PD, \muf) \bigr]^{\!*} \,
\tilde{C}^{\mu\nu}(p_{1+}, p_{2-}, p_{3+4\parallel}, p_{3-4}^\PD, \muf) 
\\
&\times
\left(\! -g_{\rho\mu} + \frac{p_{3\rho} p_{3\mu}}{m_W^2} \right) \!
\left(\! -g_{\sg\nu} + \frac{p_{4\sg} p_{4\nu}}{m_W^2} \right)
.\label{e.C}
\eeqa
Regrouping the objects into the collinear and anticollinear groups, 
we get
\beqa
\Sg(P_1,P_2, p_3,p_4)
&= 
\int\! \dd^4 x \, \e^{-\I (p_3 + p_4) \cdot x} \, 
C(p_{1+}, p_{2-}, p_{3+4\parallel}, p_{3-4}^\PD, \muf) 
\\
&\phantom{=}\times
\frac12 \sum_{\text{p spins}}
\vetosumXcol
\bra{\proton(P_1)} {\chi_\col^\dag}^{j\!} (x^+\!, \vec{x}_\perp) \, u_\acol^\dag \ket{X_\col} \;
\bra{X_\col} u_\acol \, {\chi_\col}_{i} (0,0) \ket{\proton(P_1)} 
\\
&\phantom{=}\times
\frac12 \sum_{\text{p spins}}
\vetosumXacol
\bra{\proton(P_2)} {\chi_\acol^\dag}_{j} (x^-\!, \vec{x}_\perp) \, u_\col^\dag \ket{X_\acol} \; 
\bra{X_\acol}  u_\col \, \chi_\acol^{i} (0,0) \ket{\proton(P_2)}
\,.
\eeqa
Notice that the line with $\sum'_{X_\col}$ contains only collinear fields and collinear states 
without any dependence on anticollinear fields or states. 
Therefore, due to the $\SU(3)_\mathrm{C}$ invariance, 
it must be proportional to $\de^{j}_i$, 
so the whole line can be replaced by $\de^{j}_i / 3$ 
times the original expression with $j = i = k$ with summation over $k$.  
This $\de^{j}_i$ then contracts the ${}_j$ and ${}^i$ indices in the line with $\sum'_{X_\acol}$.
Finally, from the defining relations\erefn{ucol_uacol} of $u_\col$ and $u_\acol$, 
we have 
${u_\col^\dag}{}^{\dot{\al}} \, u_\col^\be = (\sgb^+)^{\dot{\al} \be}$ 
and
${u_\acol^\dag}{}^{\dot{\al}} \, u_\acol^\be = (\sgb^-)^{\dot{\al} \be}$.
Putting all together, we get 
\beqa
\Sg(P_1,P_2, p_3,p_4)
&=
\int\! \dd^4 x \, \e^{-\I (p_3 + p_4) \cdot x} \; 
C(p_{1+}, p_{2-}, p_{3+4\parallel}, p_{3-4}^\PD, \muf) 
\cdot \frac{1}{N_\C} \cdot
\\
&\phantom{=}\times
\frac12 \sum_{\text{p spins}}
\vetosumXcol
\bra{\proton(P_1)} {\chi_\col^\dag}^{k}\! (x^+\!, \vec{x}_\perp) \ket{X_\col} \,
\sgb^- 
\bra{X_\col} {\chi_\col}_{k} (0,0) \ket{\proton(P_1)} 
\\
&\phantom{=}\times
\frac12 \sum_{\text{p spins}}
\vetosumXacol
\bra{\proton(P_2)} {\chi_\acol^\dag}_{\ell} (x^-\!, \vec{x}_\perp) \ket{X_\acol} \,
\sgb^+  
\bra{X_\acol} \chi_\acol^{\ell} (0,0) \ket{\proton(P_2)}
\label{e.color-summed}
\eeqa
with $N_\C = 3$.
Upon substituting this $\Sg(P_1,P_2, p_3 ,p_4)$ back into the cross section\erefn{xs2}, 
the $\dd^2\vec{q}_\perp$ integral is trivial:
$\int \dd^2 \vec{q}_\perp \, \e^{-\I (p_3 + p_4)_{\!\perp}^\PD \cdot\, x_{\!\perp}^\PD} 
= \int \dd^2 \vec{q}_\perp \, \e^{-\I q_{\!\perp}^\PD \cdot\, x_{\!\perp}^\PD}
= (2\pi)^2 \, \de^2(\vec{x}_\perp)$. 
This then gets rid of the $\dd^2 \vec{x}_\perp$ integral
and sets $\vec{x}_\perp$ to zero,  
and we are left with only the $\dd x^+$ and $\dd x^-$ integrals.
Therefore, the cross section\erefn{xs2} becomes
\beqa
\frac{\dd \sg}{\dd M} 
=\>&
2 \cdot 
\frac{1}{4\pi M s} 
\int\!
\frac{\dd^3 \vec{p}_3}{(2\pi)^3 \, 2 E_3} \,
\dd\eta \,
\de(M^2 - 2 q_\parallel \dt p_{3\parallel}) \,
C(p_{1+},p_{2-}, p_{3+4\parallel}, p_{3-4}^\PD, \muf) \Bigr|_{p_4 = q - p_3} 
\cdot \frac{1}{N_\C} \cdot
\\
&\times
\frac12 \sum_{\text{p spins}}
\int\! \dd x^+ \, \e^{-\I q_+^\PD x^+} 
\vetosumXcol
\bra{\proton(P_1)} {\chi_\col^\dag}^{i}\! (x^+\!, 0) \ket{X_\col} \,
\sgb^- 
\bra{X_\col} {\chi_\col}_{i} (0,0) \ket{\proton(P_1)} 
\\
&\times
\frac12 \sum_{\text{p spins}}
\int\! \dd x^- \, \e^{-\I q_-^\PD x^-}
\vetosumXacol
\bra{\proton(P_2)} {\chi_\acol^\dag}_{j} (x^-\!, 0) \ket{X_\acol} \,
\sgb^+ 
\bra{X_\acol} \chi_\acol^{j} (0,0) \ket{\proton(P_2)}
\,,\label{e.factorized_xsec}
\eeqa
where the factor of $2$ in front is due to the fact that $\det(g_{\mu\nu}) = 2$ in the lightcone coordinates\erefn{metric}.
We have replaced $\ds{\de(M^2 - 2 q \dt p_3)}$ by its leading-order expression $\ds{\de(M^2 - 2 q_\parallel \dt p_{3\parallel})}$ to have a consistent $\co(\la^0)$ expression.
We have thus obtained a factorized form of the differential cross section at the leading order in $\la$ (and to all orders in $\al_\s$) for our process.

\subsubsection{The Parton Picture and the Beam Functions}
\label{sss.parton_picture}
Among the arguments of $C(p_{1+}, p_{2-}, p_{3+4\parallel}, p_{3-4}^\PD, \mu)$ in the factorized cross section\erefn{factorized_xsec},
the meanings of $p_3$ and $p_4$ are clear, 
while the definitions\erefn{p1p2} of $p_{1+}$ and $p_{2-}$ are rather unintuitive.
To understand the physical interpretation of $p_{1+}$ and $p_{2-}$,
notice that $P_{2+} = 0$ and $P_{X_\acol+} \sim \co(\la^2 M)$, 
so we can rewrite $p_{1+}$ as
\beqa
p_{1+}
\equiv
(P_1 - P_{X_\col})_+
= 
(P_1 - P_{X_\col} + P_2  - P_{X_\acol})_+^\PD 
\label{e.p1+}
\eeqa
at the leading order in $\la$.
The right-hand side is actually just equal to $(p_3 + p_4)_+$ by 4-momentum conservation. 
With a similar exercise for $p_{2+}$, we thus arrive at the relations
\beqa
p_{1+}^\PD = (p_3 + p_4)_+^\PD = q_+^\PD 
\,,\quad
p_{2-}^\PD = (p_3 + p_4)_-^\PD = q_-^\PD
\label{e.p=q}
\eeqa
at the leading order in $\la$.
Therefore, $p_{1+}$ and $p_{2-}$ are the momenta of the collinear and anticollinear quarks \emph{right before} they annihilate into $W^+ W^-$, i.e., the momenta \emph{after} they have emitted all collinear and anticollinear gluons.  
The parton picture has thus emerged naturally from the SCET formalism, 
where the function $C(p_{1+}, p_{2-}, \ldots)$ is describing the hard interaction of the partons with momenta $p_{1+}$ and $p_{2-}$.
Then, the expressions under the $\dd x^+$ and $\dd x^-$ integrals in\erefn{factorized_xsec} must be interpreted as the distributions of the collinear and anticollinear quark partons inside the corresponding protons right before the collision, with the jet veto condition imposed.
As first introduced in~\cite{beam-func-1, beam-func-2}, 
we are thus led to define the \emph{beam function} of a quark parton $\psi$ inside the proton (``$\psi / \proton$''):
\beqa
\cb_{\psi / \proton}^{(h)}(\xi, \ptv, \mu, \nu)
\equiv
\frac12 \sum_{\text{p spins}}
\frac{1}{2\pi}
\int\! \dd t \, \e^{-\I t \xi (n \cdot P)} 
\vetosumX
\bra{\proton(P)} \chi_\psi^{\dag i\!} (t n) \ket{X} \,
\frac{\slashed{n}}{2} \cp_h \,
\bra{X} \chi_{\psi i}^{\PD\!} (0) \ket{\proton(P)} 
\,,\label{e.beam-func}
\eeqa
where $n$ is an arbitrary lightlike 4-vector, and
\beqa
\chi_\psi(x) \equiv \ds{W_n^\dag(x) \, \psi(x)}
\,,\quad
W_n(x) \equiv W_\col(x) \Bigr|_{n_+ = n}
\,,
\eeqa
and the ultra-relativistic limit, $\ds{n \dt P / m_\proton} \to \infty$, is understood.
We have switched to the 4-component Dirac spinor notation for $\psi$  
to make connections with the literature.
In particular, $h = \pm 1$ denotes the chirality of $\psi$,
and $\cp_h \equiv (\mathbbm{1} + h \ga_5) / 2$. 
The $2 \times 2$ matrix $\sgb^-$ appearing in\erefn{factorized_xsec}, 
which is equal to $\ds{n_+ \dt \sgb} / 2$ due to\erefn{lightcone_def},
has been translated to $(\slashed{n} / 2) \cp_{-1}$ here. 
Since the collinear and anticollinear sectors are factorized from each other, 
there is no longer any need for the labels ${}_\col$ and ${}_\acol$,%
\footnote{
Except for the collinear anomalies, which take different forms in the collinear and anticollinear sectors. 
We will see this in detail in \sssref{col_anom}.
}
which is the reason we have opted for using a generic $n$ instead of $n_+$ or $n_-$.
In fact, $\cb$ is independent of $n$ and $P$ in the ultra-relativistic limit, 
justifying the absence of $n$ and $P$ in the arguments of $\cb$. 
Finally, as we will see in \ssref{beam-func}, 
beam functions suffer from rapidity divergences and hence depend on the scale $\nu$ from analytic regularization of rapidity divergences, 
in addition to the scale $\mu$ from DR\@. 

As noted above, the beam function $\cb$ is a parton distribution function (PDF) in the presence of a jet veto.
Indeed, the factor of $1/2\pi$ in the definition of $\cb$ is introduced so that $\cb$ would be exactly equal to the PDF in the $\ptv \to \infty$ (i.e., no jet veto) limit.
In particular, if $\ket{\proton(P)}$ were just $\ket{\psi(P)}$ and there were no jet veto nor any interactions, then $\cb$ would be exactly equal to $\de(1-\xi)/2$, i.e., 
the PDF for finding a $\psi$ with momentum $\xi P$ and chirality $h$ 
inside a $\psi$ with momentum $P$ and its spins averaged over. 

In terms of the beam functions,  
and summing over all fermion species and chiralities, 
the cross section\erefn{factorized_xsec} can be written as
\beqa
\frac{\dd \sg}{\dd M} 
=\>&
\frac{2 (2\pi)^2}{4\pi M s} 
\int\!
\frac{\dd^3 \vec{p}_3}{(2\pi)^3 \, 2 E_3} \,
\dd\eta \,
\de(M^2 - 2 q_\parallel \dt p_{3\parallel}) \,
\sum_f
\sum_{h = \pm 1} 
C_f^{(h)}(\xi_1 P_1, \xi_2 P_2, p_{3+4\parallel}, p_{3-4}^\PD, \muf) \Bigr|_{p_4 = q - p_3} 
\\
&\times
\frac{1}{N_\C}
\left[
\cb_{f / \proton}^{(h)}(\xi_1, \ptv, \muf, \nu) \;
\cb_{\bar{f} / \proton}^{(h)}(\xi_2, \ptv, \muf, \nu) 
+ (f \leftrightarrow \bar{f})
\right] 
,\label{e.the_xsec}
\eeqa
where $\bar{f}$ is the antiparticle of $f$, and $f = u, d, s, c, b$,%
\footnote{
The top quark contributions will be included through the $gg$ channel, which we will add separately at the end. See also footnote~\ref{ftnote:onlyqqbar}.
}
and $C_f^{(h)}$ is the $C$ function for flavor $f$ with chirality $h$.
The beam functions are evaluated at the scale $\mu = \muf \sim \ptv$ as the scales of virtuality of all states involved in $\cb$ are $\co(\ptv)$ due to the jet veto. 
Because we are ignoring quark masses and Yukawa interactions with the higgs boson (see footnotes~\ref{ftnote:onlyqqbar} and~\ref{ftnote:spinor} on how good those approximations are), 
the different chiralities of an $f$ never mix, 
justifying treating the $h = \pm 1$ contributions separately. 
The neglect of quark masses also justifies our not taking into account any quark mixings.
Finally, the amplitudes with different choices of $f$ and/or $h$ do not interfere with each other, 
because they inevitably have different $\ket{X}$. 

The parton momentum fractions $\xi_{1,2}$ in\erefn{the_xsec} can be determined by using\erefn{p=q} and then\erefn{M_and_eta} with $q_\pm = q^0 \pm q^3$ (that is, by choosing $(n_\pm^\mu) = (1, 0, 0, \mp 1)$):
\beqa
\xi_1 
\equiv 
\frac{p_{1+}}{P_{1+}}
=
\frac{q_+}{P_{1+}}
= 
\frac{M\e^{\eta}}{P_{1+}}
\,,\quad
\xi_2 
\equiv 
\frac{p_{2-}}{P_{2-}}
=
\frac{q_-}{P_{2-}}
= 
\frac{M\e^{-\eta}}{P_{2-}}
\eeqa
at the leading order in $\la$. 
At the LHC, we are in the center-of-momentum frame of the colliding protons, so $P_{1+} = P_{2-} = \sqrt{s}$. Thus,
\beqa
\xi_1 = \sqrt{\tau} \, \e^\eta
\,,\quad 
\xi_2 = \sqrt{\tau} \, \e^{-\eta}
\,\quad\text{with}\quad
\tau \equiv \frac{M^2}{s}
\,.
\eeqa

As we will elaborate in \ssref{beam-func}, 
the beam functions can be related to PDFs.
However, currently available PDF sets such as those used by ATLAS and CMS experiments do not differentiate $h = \pm 1$.
Although quantifying the errors associated to this approximation is beyond the scope of this paper,
we nonetheless expect that 
the relation $\cb^{(-1)}_{f / \proton} = \cb^{(+1)}_{f / \proton}$ should hold well 
to the extent that weak interactions can be neglected inside the proton 
and in parton evolution.
Therefore, instead of the helicity-dependent beam function\erefn{beam-func}, 
we use
\beqa
\cb_{\psi / \proton}(\xi, \ptv, \mu, \nu)
&\equiv
\sum_{h = \pm 1} \cb_{\psi / \proton}^{(h)}(\xi, \ptv, \mu, \nu)
\\
&=
\frac12 \sum_{\text{p spins}}
\frac{1}{2\pi}
\int\! \dd t \, \e^{-\I t \xi (n \cdot P)} 
\vetosumX
\bra{\proton(P)} \chi_\psi^{\dag i\!} (t n) \ket{X} \,
\frac{\slashed{n}}{2} \,
\bra{X} \chi_{\psi i}^{\PD\!} (0) \ket{\proton(P)} 
\,.\label{e.summed-beam-func}
\eeqa
Then, assuming 
$\cb^{(-1)}_{f / \proton} = \cb^{(+1)}_{f / \proton}$ 
(which thus equals $\cb_{f / \proton} / 2$),
the cross section\erefn{the_xsec} becomes
\beqa
\frac{\dd \sg}{\dd M} 
=\>&
\frac{2 (2\pi)^2}{4\pi M s} 
\int\!
\frac{\dd^3 \vec{p}_3}{(2\pi)^3 \, 2 E_3} \,
\dd\eta \,
\de(M^2 - 2 q_\parallel \dt p_{3\parallel}) \,
\sum_f
\sum_{h = \pm 1} 
C_f^{(h)}(\xi_1 P_1, \xi_2 P_2, p_{3+4\parallel}, p_{3-4}^\PD, \muf) \Bigr|_{p_4 = q - p_3}
\\
&\times
\frac{1}{N_\C} \,
\frac{1}{2 \cdot 2} \!
\left[
\cb_{f / \proton}(\xi_1, \ptv, \muf, \nu) \;
\cb_{\bar{f} / \proton}(\xi_2, \ptv, \muf, \nu) 
+ (f \leftrightarrow \bar{f})
\right] 
,\label{e.averaged_xsec}
\eeqa
where $f=u, d, s, c, b$ as before.

\subsection{The Dependence on the Jet-Clustering Algorithm} 
\label{ss.jet-algorithm}
Since our SCET calculation depends crucially on separating modes into collinear and anticollinear modes, 
it is necessary that the definition of jets used in the actual experimental studies is consistent with such separation of modes.
We define a distance measure $d_{ij}$ in the $\eta$-$\phi$ space between particles $i$ and $j$ as
\beqa
d_{ij} 
\equiv 
\mathrm{Min} \bigl[ ({\pT}_i^\PD)^{2n} , ({\pT}_j^\PD)^{2n} \bigr] 
\frac{\sqrt{(\Delta \eta_{ij})^2 + (\Delta \phi_{ij})^2}}{R} 
\eeqa
with parameters $n$ and $R$.
We also define a distance measure between particle $i$ and the beam
\beqa
d_{i\mathrm{B}} \equiv ({\pT}_i^\PD)^{2n}
\eeqa
with the same $n$.
The choices $n=1$ and $n=0$ respectively give the $k_\mathrm{T}$ algorithm~\cite{kT}  and the Cambridge/Aachen algorithm~\cite{CA1, CA2}, while $n=-1$ corresponds to the anti-$k_\mathrm{T}$ algorithm~\cite{anti-kT} used by the relevant ATLAS and CMS studies of $WW$ prediction mentioned in \sref{intro},
with the jet-radius parameter $R$ taken to be $0.4$ by ATLAS and $0.5$ by CMS\@. Starting from the list of all $d_{ij}$'s and $d_{i\mathrm{B}}$'s,
we search for the smallest distance and if it is $d_{ij}$, we replace the particles $i$ and $j$ with a single, new particle (with a 4-momentum $p_i + p_j$),
while if it is $d_{i\mathrm{B}}$, we declare the particle $i$ a jet and remove it from the list. We recalculate the distances in the new list and repeat the procedure, until no particle is left in the list.

Since our factorization formula\erefn{the_xsec} is based on the separation of collinear and anticollinear modes in SCET, 
we must make sure that the jet algorithm does not cluster particles of 
different modes into a single jet. 
This is indeed the case as long as $|\log\la| \gg R$,  
because the rapidity difference between a collinear particle and an anticollinear particle is parametrically $\sim \ds{\log(1/\la) - \log\la} \sim |\log\la|$.
Another potential issue is that, 
since the jet algorithm introduces a new parameter $R$ to the theory, 
the jet-algorithm dependence of the clustering of two or more real gluon emissions 
can give rise to $\log R$.
Since we are not resumming $\log R$, we must take $R \sim \co(1)$ such that 
$|\log R| \lesssim 1$. 
Resummation of $\log R$ remains an open problem~\cite{logR:1, logR:2}, but for a color-singlet final state with a jet veto (such as our $WW$ case), it has been shown~\cite{logR:2} that the numerical impact of the $\ds{\log^2\! R}$ terms is small for $R \sim 0.5$.
We therefore assume that this conclusion holds to all orders in $\log R$, 
and take $|\log R| \lesssim 1$ and $|\log\la| \gg R$ parametrically.

\section{Analytical Calculations}
\label{s.analytical}

\subsection{The Wilson Coefficient}
\label{ss.Wilson-coeff}

\subsubsection{Matching SCET onto SM at $\mu \sim M$}
\label{sss.matching}
Since the hard coefficient $C_f^{(h)}$ in the factorized cross-section\erefn{averaged_xsec} is directly related to the Wilson coefficient $\Ga^{\mu\nu}$ of the SCET operator\erefn{current} 
through\erefn{Gamma_munu},\erefn{Ctilde}, and\erefn{C}, 
the first step is to determine $\Ga^{\mu\nu}$ at the \emph{hard} scale $\mu = \mu_\text{h} \sim \co(M)$ by integrating out guaranteed-off-shell SM physics with virtuality of $\co(M)$, such as the $t$-channel quark propagator between the $W^+$ and $W^-$ vertices.

Upon matching SCET and SM matrix elements to determine the Wilson coefficient $\Ga^{\mu\nu}$ at $\mu = \mu_\text{h}$, 
we can evaluate the matrix element of the SCET operator $\cj^{\mu\nu}$
for \emph{any} convenient states of our choice 
as long as their invariant mass is $\co(M)$.
Let us choose an obvious parton-level process $q\bar{q} \to W^+ W^-$ with\emph{out} a hadronic state $X$ (i.e., without any real gluon emission),
with $q$ and $\bar{q}$ having exactly collinear and anticollinear momenta $p_1$ and $p_2$, respectively, with $(p_1 + p_2)^2 = M^2$.
So, we first re-evaluate the matrix element\erefn{JX} with $\mu = \muh$ between the states $\bra{0}$ and $\ket{q^i(p_1)\, \bar{q}_j(p_2)}$ instead of $\bra{X}$ and $\ket{\proton(P_1) \, \proton(P_2)}$, where  $i$ and $j$ are color indices.
Let us denote this matrix element by $J_0^{\mu\nu i}{}_j(x, p_1,p_2, p_3,p_4)$.
Then, from\erefn{factorized}, it is given by
\beqa
&
J_0^{\mu\nu i}{}_j (x, p_1, p_2, p_3, p_4)
\\
&=
\bra{0} \chi_\acol^{k}(x^-\!, \vec{x}_\perp) \, u_\col \ket{\bar{q}_j(p_2)} \; 
\bra{0} u_\acol \, {\chi_\col}_{k}(x^+\!, \vec{x}_\perp) \ket{q^i(p_1)} \;
\tilde{C}^{\mu\nu}(p_{1+}, p_{2-}, p_{3+4\parallel}, p_{3-4}^\PD, \muh)
\\
&=
Z_q\, 
\bra{0} \chi_\acol^{k}(x^-\!, \vec{x}_\perp) \, u_\col \ket{\bar{q}_j(p_2)} \bigr|_\text{amp} \; 
\bra{0} u_\acol \, {\chi_\col}_{k}(x^+\!, \vec{x}_\perp) \ket{q^i(p_1)} \bigr|_\text{amp} \;
\tilde{C}^{\mu\nu}(p_{1+}, p_{2-}, p_{3+4\parallel}, p_{3-4}^\PD, \muh)
\,,\label{e.SCET_amp}
\eeqa
where $\tilde{C}^{\mu\nu}(p_{1+}, p_{2-}, p_{3+4\parallel}, p_{3-4}^\PD, \muh)$ is given by\erefn{Ctilde} 
with $p_{1+}$ and $p_{2-}$ being literally the $+$ and $-$ components of $p_1$ and $p_2$, respectively,
as given by\erefn{p1p2} with $P_{1,2} = p_{1,2}$ and $P_{X_\col} = P_{X_\acol} = 0$.
The symbol $|_\text{amp}$ indicates the amputated matrix element, 
and $Z_q$ is the product of wavefunction renormalization constants of $q$ and $\bar{q}$.
An example of the amputated diagrams in 
$\bra{0} u_\acol \, {\chi_\col}_{k}(x^+\!, \vec{x}_\perp) \ket{q^i(p_1)} \bigr|_\text{amp}$ 
is a 1-loop diagram with a gluon propagator with one end attached to a vertex from the collinear covariant derivative\erefn{col_cov_dev} and the other end to a vertex from the collinear Wilson line\erefn{col-Wilson}.   
On the other hand, the SM amplitude for $q\bar{q} \to WW$ is given by
\beqa
\cm_\SM^{(h)}{}^i_{~j} (p_1, \ldots, \muh) 
\equiv
Z_q \, \de^i_j \, J_\SM^{\mu\nu}(p_1, \ldots, \muh) \, 
\ep^*_\mu(p_3, s_3) \, \ep^*_\nu(p_4, s_4)
\label{e.M_SM}
\eeqa
with an amputated matrix element $J_\SM^{\mu\nu}$ and the product of wavefunction renormalization constants $Z_q$. 
This $Z_q$ here is the same $Z_q$ as that in\erefn{SCET_amp},
because we can always declare that the 4-momentum of a single on-shell quark is exactly collinear. 

Now, since the SM is a renormalizable theory 
and there exists no renormalizable $q$-$\bar{q}$-$W^+$-$W^-$ vertex,  
the SM matrix element $Z_q J_\SM^{\mu\nu}$ is actually UV finite.
On the other hand, 
both amputated SCET matrix elements in\erefn{SCET_amp} are UV divergent. 
Let $Z_\UV$ be the renormalization constant that absorbs those UV divergences in the SCET amplitude so that $\tilde{C}^{\mu\nu}$ has no divergences.
The matching condition is then given by
\beqa
&
Z_q \, \de^i_j \,  J_\SM^{\mu\nu}(p_1, \ldots, \muh)
\\
&=
Z_\UV\, Z_q\, 
\bra{0} \chi_\acol^{k}(0,0) \, u_\col \ket{\bar{q}_j(p_2)} \bigr|_\text{amp} \; 
\bra{0} u_\acol \, {\chi_\col}_{k}(0,0) \ket{q^i(p_1)} \bigr|_\text{amp} \;
\frac{1}{M} \, 
\tilde{C}^{\mu\nu}(p_1, \ldots, \muh)
\,,\label{e.matching:general}
\eeqa
where the factor of $1/M$ is from\erefn{Lint}.
Both SCET and SM amplitudes also have IR divergences, 
but they necessarily cancel out in the matching relation above, 
as the SCET amplitude has the same analytic structure in the IR as the SM amplitude by construction.

Let us evaluate the matching condition\erefn{matching:general} in dimensional regularization (DR) with the modified minimal subtraction scheme ($\overline{\text{MS}}$).
First, since loop integrals in the wavefunction renormalization for an on-shell massless fermion do not depend on any scales, they all vanish in DR\@.
We thus have $Z_q = 1$ exactly.   
Similarly,  
all loop integrals in 
$\bra{0} {\chi_\col}(0,0) \ket{q(p_1)}\bigr|_\text{amp}$ are scaleless for on-shell $p_1$ and thus vanish.
Therefore, in DR, we have
\beqa
\bra{0} u_\acol \, {\chi_\col}_{k} (0,0) \ket{q^i(p_1)} \bigr|_\text{amp}
= 
\bra{0} u_\acol \, {\psi_\col}_{k} (0,0) \ket{q^i(p_1)} 
= 
\de_k^i \,
u_\acol \, u_{p_1}
\,,
\eeqa
where $\psi_\col$ is the free collinear quark field.
The spinor wavefunction $u_p$ satisfies the Dirac equation ${\ds p \dt \sgb} u_p = 0$ 
and is normalized in the standard way as $u_p u_p^\dag = \ds{p \dt \sg}$.
With this convention,
the product of $u_\acol \, u_{p_1}$ and a similar factor from the anticollinear counterpart is given by
\beq
(u_{p_2} \, u_\col) \, (u_\acol \, u_{p_1})
= M
\,,
\eeq
due to the properties\erefn{ucol_uacol} and 
the fact that $p_1$ and $p_2$ are exactly collinear and anticollinear, respectively. 
Therefore, in DR, the matching condition\erefn{matching:general} reduces to a very simple form:
\beqa
J_\SM^{\mu\nu} (p_1, \ldots, \muh)
=  
Z_\UV \, \tilde{C}^{\mu\nu}(p_1, \ldots, \muh)
\,.\label{e.matching:DR}
\eeqa
Since the SM amplitude is UV finite as noted above, 
the $1 / \ep$ poles in $J_\SM^{\mu\nu}$ are all associated with IR divergences of the SM amplitude. 
In contrast, 
the $1 / \ep$ poles in the renormalization constant $Z_\UV$ are by definition 
all attributed to UV divergences of the SCET amplitude.
$\tilde{C}^{\mu\nu}$ has no UV or IR divergences as we noted above.

The DR matching relation\erefn{matching:DR} can be directly used to determine the SCET renormalization constant $Z_\UV$ from the $1 / \ep$ poles of the SM matrix element $J_\SM^{\mu\nu}$.
Let us parametrize $\cm_\SM^{(h)}$ defined in\erefn{M_SM} as
\beqa
\cm_\SM^{(h)}{}^i_{~j} 
= 
\cm_0^{(h)} \, \de^i_j
+
\frac{C_\mathrm{F} \, \al_\s(\muh)}{4\pi} 
\left(
\cm_\text{1, div}^{(h)} 
+ 
\cm_\text{1, reg}^{(h)} 
\right) \! 
\de^i_j
+
\co(\al_\s^2)
,\label{e.amp:NLO}
\eeqa
where $C_\mathrm{F} \equiv (N_\C^2 - 1) / (2N_\C) = 4/3$, 
and $\cm_0^{(h)}$ is the tree-level contribution
while $\cm_\text{1, div}^{(h)}$ and $\cm_\text{1, reg}^{(h)}$
are the divergent and regular parts of 1-loop contributions, respectively.
The divergent piece has the form
\beq
\cm_\text{1, div}^{(h)} 
= 
-
\left( \frac{4 \pi \muh^2 }{-M^2 - \I 0^+} \right)^{\!\!\ep} \,
\Gamma (1 + \ep) \!
\left( \frac{2}{\ep^2} + \frac{3}{\ep} \right) \!
\cm_0^{(h)} 
\,,\label{e.amp:div}
\eeq
Then, by identifying $1 / \ep$ poles of $Z_\UV$  with those of $\cm_\text{1, div}^{(h)}$,
we find
\beqa
Z_\UV
= 
1 
-\frac{C_\mathrm{F} \, \al_\s(\mu)}{4 \pi} \, (4 \pi \e^{-\ga})^\ep \!
\left( 
\frac{2}{\ep^2} + \frac{3 - 2 L_M(\mu)}{\ep} 
\right)
+
\co(\al_\s^2)
\,,\label{e.Z:NLO}
\eeqa
where the $\overline{\text{MS}}$ scheme has been used, and 
\beqa
L_M(\mu) \equiv \log\frac{-M^2 - \I 0^+}{\mu^2}
\,.
\eeqa
Note that we have actually obtained the form of $Z_\UV$ valid at all $\mu$, not just at the matching scale $\muh$, 
even though we determined it from the matching condition.
This is because $Z_\UV$ is a renormalization constant for UV divergences 
in the effective theory, 
which have nothing to do with where the EFT is superseded by the full theory.
In particular, it is valid even when the logarithm $L_M(\mu)$ is large, 
even though the matching calculation itself should be done at a scale $\mu$ where $L_M(\mu)$ is small.  
Another notable property of $Z_\UV$ is that it is independent of the quark's helicity $h$.
This can be understood by reinterpreting the $1 / \ep$ poles of $Z_\UV$ as those associated with the IR divergence of the SM amplitude.
For $q\bar{q} \to WW$, the IR divergent piece only involves QCD interactions on the initial $q$ and $\bar{q}$, so it clearly cannot depend on $h$. 

To calculate the Wilson coefficient $\tilde{C}^{\mu\nu}$, 
we multiply $Z_\UV^{-1}$ on both sides of\erefn{amp:NLO},
and then apply\erefn{M_SM} and\erefn{matching:DR} on the left-hand side, 
while using \erefn{Z:NLO} on the right-hand side and taking the $\ep \to 0$ limit. 
This gives us
\beqa
&
\tilde{C}^{\mu\nu} (p_1, \ldots, \muh) \,
\ep^*_\mu(p_3, s_3) \, \ep^*_\nu(p_4, s_4)  
\\
&= 
\left[ 
1 
-
\frac{C_\text{F} \al_\s(\muh)}{8\pi} \!
\left( 2 L_M^2(\muh) - 6 L_M(\muh) + \frac{\pi^2}{3} \right) \! 
\right] \!
\cm_0^{(h)}
+ 
\frac{C_\mathrm{F} \, \al_\s(\muh)}{4\pi} \, \cm_\text{1, reg}^{(h)}
+
\co(\al_\s^2) 
\,.
\eeqa
Squaring both sides of this relation and summing over the $W^\pm$ polarizations $s_3$ and $s_4$, 
we find that the hard coefficient $C_f^{(h)}$ (defined in\erefn{C}) is given at the scale $\muh$ by
\beqa 
C_f^{(h)}(p_1, \ldots, \muh)
= 
\sum_{s_3, s_4}
\biggl[
&
\left\{ 
1 
- 
\frac{C_\text{F} \, \al_\s(\muh)}{4\pi} \!
\left( 
2 L_M^2(\muh) - 6 L_M(\muh) + \frac{\pi^2}{3} 
\right) \! 
\right\} \!
\bigl| \cm_0^{(f,h)} \bigr|^2
\\
&
+
\frac{C_\mathrm{F} \, \al_\s(\muh)}{2\pi} \, 
\mathrm{Re} \Bigl( \cm_0^{(f,h)*} \cm_\text{1, reg}^{(f,h)} \Bigr)
+
\co(\al_\s^2)
\biggr]
\,,\label{e.Ch}
\eeqa
where we have put the flavor label $f$ back.
For the cross section\erefn{averaged_xsec}, we just need to know $\sum_h C_f^{(h)}$ (as opposed to $C_f^{(h)}$ itself),
which can be given in terms of
\beqa
\sum_{h}
\sum_{s_3, s_4}
\bigl| \cm_0^{(f,h)} \bigr|^2 
&= 
 c_f^{tt} F_f^{(0)\!}(\hat{s}, \hat{t})
-c_f^{ts} J_f^{(0)\!}(\hat{s}, \hat{t})
+c_f^{ss} K_f^{(0)\!}(\hat{s}, \hat{t})
\,,\\
\sum_{h}
\sum_{s_3, s_4}
\mathrm{Re} \Bigl( \cm_0^{(f,h)*} \cm_\text{1, reg}^{(f,h)} \Bigr)
&=
\frac12
\Bigl(
 c_f^{tt} F_f^{(1)\!}(\hat{s}, \hat{t})
-c_f^{ts} J_f^{(1)\!}(\hat{s}, \hat{t})
+c_f^{ss} K_f^{(1)\!}(\hat{s}, \hat{t})
\Bigr)
\,,
\eeqa
where $\hat{s} \equiv (p_1 + p_2)^2 = M^2$ and $\hat{t} \equiv (p_1 - p_3)^2$ are 
parton-level Mandelstam variables.
The expressions for the coefficients $c_f^{tt, ts, ss}$ and the functions $F_f^{(0,1)}$, $J_f^{(0,1)}$, and $K_f^{(0,1)}$ can be found in Ref.~\cite{WW:nlo2}. 

Finally, let us comment on a subtlety associated with the computation of $C^{(h)}_f$ for the $b \bar{b} \rightarrow W^+ W^-$ channel. 
In all of the above calculations for the matching, 
the massless quark limit is assumed not only for the initial $q$ and $\bar{q}$ but also for the $t$-channel quark propagator.  
This is not strictly correct for the $b\bar{b}$ initial state, for which a massive top quark would be exchanged in the $t$-channel.
However, we expect that errors from neglecting the top quark mass in the propagator should be small because the $b$-quark PDF is small.
We have checked using {\tt MadGraph\_aMC@NLO} \cite{aMC@NLO} that 
this is indeed the case at the LO\@.
Even at the 14-TeV LHC, the errors do not exceed $1 \%$ of the total $\proton \proton \to WW$ cross section with a $K$-factor of $1.5$.   
On dimensional grounds, 
a finite quark mass $m_q$ is expected to affect jet-veto cross sections through the powers and logarithms of the dimensionless ratio $m_q / \ptv$.
This issue was investigated thoroughly in~\cite{mass-effects}, which has shown that the quark mass effects can be treated as power corrections.

\subsubsection{RG-evolving the SCET down to $\mu \sim \ptv$}
\label{sss.hard-RG}
The expression of $C^{(h)}$ given in\erefn{Ch} 
is still not ready to be used in the factorized cross section\erefn{averaged_xsec}, 
because it is still evaluated at the hard scale $\mu = \muh \sim M$. 
With the jet veto, 
the actual scale of virtuality of can-be-on-shell modes in our process is at most $\co(\ptv)$, 
so we must evaluate the cross section at the factorization scale $\mu = \muf \sim \ptv$.
In other words, modes with virtuality between $\mu = \muh$ and $\mu = \muf$ are  guaranteed-off-shell modes in our process and hence must be integrated out from the effective theory.
Conceptually, this is done by matching the SCET with scale $\mu$ onto the SCET with scale $\ds{\mu - \dd \mu}$, 
which gives rise to RG equations that tell us how coefficients in the effective lagrangian should change with $\mu$ such that physical amplitudes do not depend on $\mu$. 
Being a matching calculation, 
the derivation of RG equations are free of IR divergences.
In particular, the RG evolution of $\tilde{C}^{\mu\nu}$ arises solely from the associated UV divergences, i.e., $Z_\UV$. 

So, from the pole structure of $Z_\UV$ in\erefn{Z:general},
the RG equation for the Wilson coefficient $\tilde{C}^{\mu\nu}$ can be directly read off as 
\beqa
\mu \frac{\dd}{\dd \mu}
\tilde{C}^{\mu\nu}(p_1, \ldots, \mu)
= 
\bigl( \Gcusp \, L_M(\mu) + 2 \ga_\mathrm{F} \bigr) \, 
\tilde{C}^{\mu\nu}(p_1, \ldots, \mu)
\,.\label{e.Ctilde_RG}
\eeqa
where the \emph{cusp} anomalous dimension $\Gcusp$ 
and the anomalous dimension $\ga_\mathrm{F}$ for the quark 
are defined via
\beqa
Z_\UV 
= 
1
- \frac{\al_\s(\mu)}{4 \pi} \, (4 \pi \e^{-\ga})^\ep 
\left[ 
\frac{\Gcusp}{2} \! 
\left( 
\frac{1}{\ep^2} 
- \frac{L_M(\mu)}{\ep}
\right) 
- \frac{\ga_\mathrm{F}}{\ep} 
\right].
\label{e.Z:general}
\eeqa
While we could read off the 1-loop expressions for $\Gcusp$ and $\ga_\mathrm{F}$ from\erefn{Z:NLO}, 
we need to know the combination $\bigl( \Gcusp \, L_M(\mu) + 2 \ga_\mathrm{F} \bigr)$ to $\co(\al_\s^2)$, because we would like to have the solutions of the RG equation to an $\co(\al_\s)$ accuracy. 
Note that $\Gcusp$ in\erefn{Ctilde_RG} is multiplied by $L_M(\mu) \sim \log(M / \ptv) \gg 1$, which should be parametrically counted as $\co(\al_\s^{-1})$,
because by definition we are regarding $\al_\s \log(M / \ptv)$ as an $\co(1)$ quantity that must be resummed in our problem. 
Therefore, 
in order for the solutions of\erefn{Ctilde_RG} to be parametrically at an $\co(\al_\s)$ accuracy, 
we must know $\Gcusp$ and $\ga_\mathrm{F}$ at the 3-loop and 2-loop levels, respectively.
The expressions for $\Gcusp$ and $\ga_\mathrm{F}$ at those loop orders in our notation can be found in Ref.~\cite{Becher:2010tm}.

Starting from the matching scale, $\mu = \muh \sim \co(M)$,
the Wilson coefficient is run down to a final scale $\mu \sim \co(\ptv)$.
The RG equation\erefn{Ctilde_RG} has an exact analytical solution
\beqa
\tilde{C}^{\mu\nu}(p_1, \ldots, \mu)
= 
\cu(\mu, \muh) \,
\tilde{C}^{\mu\nu}(p_1, \ldots, \muh) \,
\label{e.C:evolve}
\eeqa
where the evolution function $\cu$ is given by 
\beqa
\cu(\mu, \muh)
\equiv
\exp\! 
\left[ 
2 S(\mu, \muh) 
- a_\Ga^\PD(\mu, \muh) \, L_M(\muh)
- 2 a_\gamma (\mu, \muh) 
\right] 
\eeqa
where 
\beqa
S(\mu, \nu)
\equiv
-\int\limits_{\al_\s(\nu)}^{\al_\s(\mu)} \! 
 \frac{\dd\al}{\be(\al)} \,  
 \Gcusp (\al) \!\!
\int\limits_{\al_\s(\nu)}^{\al} \!\! \frac{\dd\al'}{\be(\al')} 
\,,\quad
a_\Gamma^\PD(\mu, \nu) 
\equiv 
-\int\limits_{\al_\s(\nu)}^{\al_\s(\mu)} \!
\frac{\dd \alpha}{\be(\al)} \,
\Gcusp (\alpha)
\eeqa
with $\be(\al_\s)$ being the $\be$-function for the QCD fine structure constant $\al_\s$.
The expression for $a_\gamma$ is given by that for $a_\Gamma^\PD$ with $\Gcusp$ replaced by $\ga_\mathrm{F}$.
Then, from\erefn{C} and\erefn{C:evolve}, 
we obtain the RG evolution of the hard coefficient:
\beqa
C^{(h)}(p_1, \ldots, \mu) 
= 
\bigl| \cu(\mu, \muh) \bigr|^2 \,
C^{(h)}(p_1, \ldots, \muh) 
\,.\label{e.Ch_RG}
\eeqa
Note that the evolution function $\cu$ is independent of $h$, 
because the RG equation\erefn{Ctilde_RG} directly derives from an $h$ independent function $Z_\UV$.

\subsection{The Beam Functions}
\label{ss.beam-func}
As we noted earlier, 
the beam function\erefn{summed-beam-func} would exactly coincide 
in the $\ptv \to \infty$ limit with the PDF:
\beqa
\phi_{\psi / \proton}(\xi, \mu)
=
\frac12 \sum_{\text{p spins}}
\frac{1}{2\pi}
\int\! \dd t \, \e^{-\I t \xi (n \cdot P)} 
\sum_{X} \,
\bra{\proton(P)} \chi_\psi^{\dag i\!} (t n) \ket{X} \,
\frac{\slashed{n}}{2} \,
\bra{X} \chi_{\psi i}^{\PD\!} (0) \ket{\proton(P)} 
\,,\label{e.PDF}
\eeqa
where $\sum_X$ goes over \emph{all} $X$, without any jet-veto constraints.
Notice that the dependence on the scale $\nu$ is absent in the PDF, 
because without a jet veto there is nothing in\erefn{PDF} that would require a cutoff in the rapidity space, so there are no rapidity divergences. 

To relate the beam function to PDFs,
note that the PDF $\phi_{\psi / \proton}(\xi, \mu)$ can be thought of a (spin-averaged) matrix element of the operator
\beqa
\hat{\phi}_{\psi}(\xi, \mu)
=
\frac{1}{2\pi}
\int\! \dd t \, \e^{-\I t \xi (n \cdot P)} 
\sum_{X} \,
\chi_\psi^{\dag i\!} (t n) 
\, \ket{X} \, \frac{\slashed{n}}{2} \, \bra{X} \, 
\chi_{\psi i}^{\PD\!} (0) 
\,.\label{e.PDF_op}
\eeqa
Similarly, the beam function can be thought of a matrix element of an operator $\hat{\cb}_\psi$ that is given by the right-hand side of\erefn{PDF_op} with $\sum_X$  replaced by $\sum'_{X}$. 
Assuming that the set of quark, antiquark, and gluon PDF operators ($\hat{\phi}_{q}$, $\hat{\phi}_{\bar{q}}$, and $\hat{\phi}_g$) form a complete set, 
we can perform an operator product expansion (OPE) on $\hat{\cb}_q$ to express it in terms of a linear combination of $\hat{\phi}_{i}(\xi, \mu)$, 
where the operator $\hat{\phi}$ is labelled by a discrete label $i = q, \bar{q}, g$ as well as a continuous label $\xi$.
Taking the matrix element of this OPE between $\bra{\proton(P)}$ and $\ket{\proton(P)}$, 
we obtain an expression of the beam function in terms of PDFs:
\beq
\cb_{q / \proton} (\xi, \ptv, \mu, \nu) 
= 
\sum_{i=q,\, \bar{q},\, g} 
\int_\xi^1 \! \frac{\dd z}{z} \,
\ci_{q \leftarrow i} (z, \ptv, \mu, \nu) \, 
\phi_{i / \proton} (\xi/z, \mu)
\,,\label{e.beam_OPE}
\eeq
where the kernel $\ci_{q \leftarrow i} (z, \ptv, \mu, \nu)$ is the OPE coefficient of the operator $\hat{\phi}$ with labels $i$ and $\xi/z$. 
The $z$ integral is bounded from below by $\xi$, 
in accord with the fact that when the parton $i$ splits into the parton $q$ and another parton, each parton has positive energy.


\subsubsection{Cancellations of Rapidity Divergences and Collinear Anomalies}
\label{sss.col_anom}
As discussed in \sssref{rap_div}, 
the beam functions have rapidity divergences that arise from artificially separating collinear and anticollinear modes
for the purpose of well-defined power counting.  
The divergences in the beam functions arise from the $\dd p_+$ and $\dd p_-$ integrations implicit in $\sum_{X_\col}$ and $\sum_{X_\acol}$, respectively, in the factorized cross section\erefn{factorized_xsec}.
We employ analytical regularization to regulate rapidity divergences,
which amounts to modifying the integration measure $\dd p_+$ in $\sum_{X_\col}$ as
\beqa
\int \frac{\dd p_+}{p_+}
\quad\Longrightarrow\quad
\int \frac{\dd p_+}{p_+} 
\left( \frac{\nu}{p_+} \right)^{\!\!\al}
\,,\label{e.AR:col}
\eeqa
and similarly the measure $\dd p_-$ in $\sum_{X_\acol}$ as
\beqa
\int \frac{\dd p_-}{p_-}
\quad\Longrightarrow\quad
\int \frac{\dd p_-}{p_-} 
\left( \frac{\nu}{p_+} \right)^{\!\!\al}
\,.\label{e.AR:acol}
\eeqa
Note that the $+ \leftrightarrow -$ exchange symmetry is now broken.
We have written the measures in dimensionless combinations $\dd p_\pm / p_\pm$ 
to highlight a feature of analytic regularization that 
it only gives logarithmic divergences like DR\@.
We could now go back to the expression\erefn{Sigma} where $\sum_X$ first appeared,
and verify that all the steps from there to here are unmodified by insertions of the analytic regulators\erefn{AR:col} and\erefn{AR:acol}.

Now, let $\cz$ be the renormalization constant to
cancel the $1 / \al$ poles in the product of the beam OPE coefficients 
$\ci_{q \leftarrow i} \, \ci_{\bar{q} \leftarrow j}$ 
inside the product of beam functions $\cb_q \, \cb_{\bar{q}}$, 
so that the $\al \to 0$ limit can be taken in $\cz \, \ci_{q \leftarrow i} \, \ci_{\bar{q} \leftarrow j}$.%
\footnote{
As we alluded in footnote~\ref{ftnote:soft-modes}, 
if we had introduced soft gluon modes $\sim (\la, \la, \la)M$ in our theory, 
the renormalization constant $\cz$ would be replaced by the \emph{soft function}%
\beqa 
\cs(\ptv, \mu, \nu) \equiv \frac{1}{N_\C}
\sum_{X_\s}'
\bra{0} [W_\col(0) \, W_\acol^\dag(0)]_{j}^{~i} \ket{X_\s} \,
\bra{X_\s} [W_\acol(0) \, W_\col^\dag(0)]_i^{~j} \ket{0}
\,,\nonumber
\eeqa
where $\ket{X_\s}$ is a hadronic state composed of soft modes only~\cite{beam-func-1, beam-func-2}.
Then, a renormalization constant would not be necessary, 
as the soft function would cancel the $1 / \al$ poles from the beam functions.
However, since this is the only place that soft gluons would ever enter in the calculation, 
we could as well call $\cs$ a renormalization constant $\cz$ for the rapidity divergences in the product $\cb_q \cb_{\bar{q}}$.
}
However, each of $\cz$, $\ci_{f \leftarrow i}$, and $\ci_{\bar{f} \leftarrow j}$ now depends on $\nu$.
Nevertheless, the $\nu$ dependence must cancel out in the product $\cz \, \ci_{q \leftarrow i} \, \ci_{\bar{q} \leftarrow j}$ 
so that physical observables such as the cross section\erefn{averaged_xsec} 
must be independent of the artificial scale $\nu$.
To derive the implications of the $\nu$ independence, 
notice that the $\nu$ dependences 
in $\ci_{q \leftarrow i}$ and $\ci_{\bar{q} \leftarrow j}$
are only through the combinations 
$\log(\nu / M)$ and $\log(M \nu / \mu^2)$, respectively, 
where the former directly follows from\erefn{AR:col}. 
The latter can be understood from\erefn{AR:acol} as well as with the fact that $p_+ \sim [\la(\mu)]^2 p_-$ in the anticollinear sector, 
where $\la(\mu) \sim \mu / M$ parametrizes the scale of virtuality just like $\la$ but at an arbitrary intermediate scale $\mu$ rather than the final scale $\sim \ptv$.
The $\nu$ dependence of $\cz$ then can be deduced from the fact that 
at the 1-loop level we get the sum of $\log(\nu / M)$ and $\log(M \nu / \mu^2)$, 
hence $\log(\nu / \mu)$, 
and this $\nu$ dependence can be cancelled only if $\cz$ is also a function of $\log(\nu / \mu)$. 
Therefore, to analyze the consequences of $\nu$-independence to all orders,   
we consider the product of three functions of the form
\beqa
P \equiv 
\cz \!\left( \ptv, \ptvbar, \mu, \log(\nu / \mu) \right) \,
\ci_{q} \!\left( \ptv, \mu, \log(\nu M/ \mu^2) \right) \,
\ci_{\bar{q}} \!\left( \ptvbar, \mu, \log(\nu / M) \right)
,
\eeqa
where the $\xi$ arguments and $i$/$j$ labels in $\ci_{q \leftarrow i}$ and $\ci_{\bar{q} \leftarrow j}$ have been suppressed as they are irrelevant for the discussions here.
Most importantly, their $\nu$ arguments have been replaced by the specific forms mentioned above. 
We have also recalled that the scale $\ptv$ could be chosen differently in the collinear and anticollinear sectors, and have let the former be $\ptv$ and the latter $\ptvbar$.

The above functional form of $P$ is identical to the function $P$ that appears in the analysis of collinear anomaly in $gg \to h$ with a jet veto~\cite{Becher:2012qa} 
except for the trivial replacement of the adjoint representation with the fundamental representation of $\SU(3)_\mathrm{C}$.
Hence, following Ref.~\cite{Becher:2012qa}, 
the solution of a set of RG equations expressing the $\nu$ independence of $P$ can be written for $\ptvbar = \ptv$ as
\beqa
&
\ci_{q \leftarrow i} (z_1, \ptv, \mu, \nu) \; 
\ci_{q \leftarrow j} (z_2, \ptv, \mu, \nu) 
\\
&= 
\left( \frac{M}{\ptv} \right)^{\!\!-2 F_{q \bar{q}}(\ptv, \mu)}
I_{q \leftarrow i} (z_1, \ptv, \mu) \; 
I_{q \leftarrow j} (z_2, \ptv, \mu)
\,,\label{e.col_anom_solved}
\eeqa
where it is now explicit in the right-hand side that there is no longer dependence on $\nu$.
As we are aiming at the $\co(\al_\s)$ accuracy, 
it suffices to have the expression for 
$I_{q \leftarrow i} (z, \ptv, \mu)$ at the 1-loop level, 
which can be found in Ref.~\cite{Becher:2010tm}.
On the other hand, 
like $\Gcusp$ and $\ga_\mathrm{F}$ discussed in \sssref{hard-RG}, 
we must know $F_{q \bar{q}}(\ptv, \mu)$ at the 2-loop level to achieve a parametrically $\co(\al_\s)$ accuracy. 
The 2-loop result can be written as
\beqa
F_{q \bar{q}}(\ptv, \mu) 
= 
\frac{\al_\s}{4\pi} 
\Gamma_0^\mathrm{F} L_\perp 
+
\left( \frac{\al_\s}{4\pi} \right)^{\!2}
\left[ \Gamma_0^\mathrm{F} \beta_0 \frac{L_\perp^2}{2} + \Gamma_1^\mathrm{F} L_\perp +  d_2^{\rm{veto}} (R)  \right]
\,,\label{e.Fqq}
\eeqa
where $L_\perp \equiv \ds{\log[\mu^2 / (\ptv)^2]}$ and $\be_0$ is the 1-loop QCD $\beta$ function. 
The expressions for the coefficients $\Ga_{0,1}^\mathrm{F}$ can be found in Ref.~\cite{Becher:2010tm}, which itself is a translation of the results of~\cite{Moch:2004pa} into our notation.
The expression for $d_2^{\rm{veto}} (R)$ can be found in Refs.~\cite{Higgs:veto, Higgs:veto2, Becher:2013xia}.
Note the dependence of $F_{q\bar{q}}$ on the jet-radius parameter $R$ at the 2-loop level due to the fact that the clustering of two real emissions necessarily depends on the jet algorithm discussed in \ssref{jet-algorithm}.

\subsubsection{RG Evolution of the Beam Functions}
\label{sss.beam-RG}
Having removed the $\nu$ dependence from $\ci_{q \leftarrow i} \, \ci_{\bar{q} \leftarrow j}$ inside the product of beam functions $\cb_q \, \cb_{\bar{q}}$, 
we are now left with the standard RG evolution of $I_{q \leftarrow i} \, 
I_{q \leftarrow j}$ with respect to $\mu$, which should run from $\mu = \muh \sim M$ down to $\mu = \muf \sim \ptv$.
The relevant RG equations can be translated from Ref.~\cite{Becher:2012qa}, which read
\beqa
\mu \frac{\dd}{\dd\mu} F_{q \bar{q}}(\ptv, \mu) 
=\>& 
2 \Gcusp
\,,\\
\mu \frac{\dd}{\dd\mu} I_{q \leftarrow i} (z, \ptv, \mu) 
=\>&
\bigl( \Gcusp L_\perp - 2 \ga_\mathrm{F}  \bigr) \, 
I_{q \leftarrow i} (z, \ptv, \mu)
\\
&-
\sum_j
\int_z^1 \! \frac{\dd u}{u} \,
I_{q \leftarrow j} (u, \ptv, \mu) \, 
\cp_{j \leftarrow i} (z/u, u)
\,,\label{e.beam-RG}
\eeqa
where $\cp_{j \leftarrow i}$ is the DGLAP splitting function defined through
the standard RG equation for the PDFs:
\beqa
\mu \frac{\dd}{\dd\mu} \phi_{i / \proton} (z,\mu) 
=
\sum_j
\int_z^1 \! \frac{\dd u}{u} \,
\cp_{i \leftarrow j} (z/u, u) \,
\phi_{j / \proton} (u,\mu)
\,.
\eeqa
Again, the 2-loop expression of $\Gcusp$ must be used in\erefn{beam-RG} 
to achieve a parametrically $\co(\al_\s)$ accuracy.

\section{Results and Discussions}
\label{s.numerical}

\subsection{Analytical NLL and NNLL Resummed Jet-Veto Cross-sections}
\label{ss.NLL-NNLL}

In this section, we present the jet-veto resummation results for the process $\proton \proton \rightarrow W^+ W^-$ at the LHC\@. 
We choose anti-$k_\mathrm{T}$ algorithm for jet-clustering with jet parameter, $R=0.4$. 
All cross-sections are evaluated using {\tt MSTW2008nnlo} PDFs unless otherwise specified. 
The choice of $\al_\s(M_Z)$ is set by the PDF itself. All calculations are performed in the massless quark limit,
so the CKM matrix is irrelevant and ignored in our analyses, 
as we have already commented on in \sssref{parton_picture}. 
We will present our resummed cross sections for both $\co(\al_\s^0)$ (``NLL'') and $\co(\al_\s^1)$ (``NNLL''), where the large logarithm $\ds{\log\bigl[ M^2 / (\ptv)^2 \bigr]}$ is counted as $\co(1 / \al_\s)$ in the $\al_\s$ power counting.

Let us first comment on the evaluation of the Wilson coefficient $C^{(h)}_f$ in\erefn{Ch} at the hard scale $\muh$. 
As one can see in\erefn{Ch}, we encounter 
logarithms of the form $\ds{\log\bigl[ (-M^2- \I 0^+) / \muh^2 \bigr]}$. 
To minimize those logarithms, 
an obvious choice for the matching scale may be $\muh^2 \sim M^2$. However, due to the presence of a branch cut, additional 
factors of $\pi^2$ arise when the logarithms are squared. As suggested in \cite{pisquare_1, pisquare_2, pisquare_3}, a better choice for the 
matching scale is $\muh^2 \sim -(M^2 + \I 0^+) < 0$ 
so that the $\pi^2$ terms are also resummed via RG evolution. 
To relate the QCD coupling constants at positive and negative values of $\mu^2$,
we use the following relation~\cite{ResumPi2:1}:
\beq
\frac{\al_\s(\mu^2)}{\al_\s(-\mu^2)} = 1 - \I a(\mu^2) + \frac{\beta_1}{\beta_0} 
\frac{\al_\s(\mu^2)}{4 \pi} \log[1 - \I a(\mu^2)] + \co(\al_\s^2)
\eeq
where $a(\mu^2) \equiv \beta_0 \al_\s(\mu^2) /4$.
We then evolve $C^{(h)}_f$ using\erefn{Ch_RG} down to the factorization scale $\muf^2 \sim +(\ptv)^2 > 0$, 
and substitute it into the cross section formula\erefn{averaged_xsec}. 

\begin{figure}[t]
        \centering
        \begin{subfigure}[b]{0.32\textwidth}
                \includegraphics[width=\textwidth]{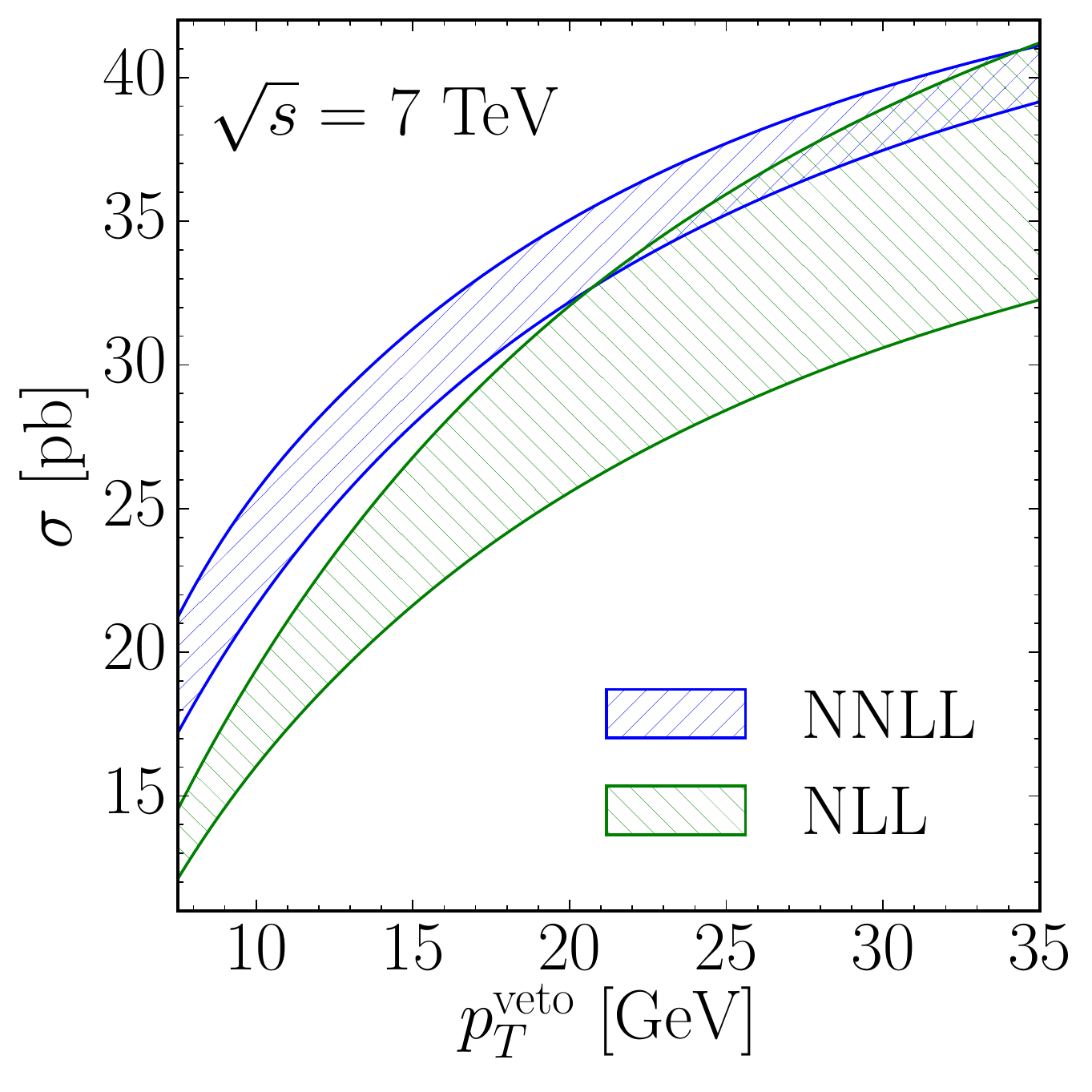}
                \caption{}
                \label{f.Xsection_7TeV}
        \end{subfigure}   
        \begin{subfigure}[b]{0.32\textwidth}
                \includegraphics[width=\textwidth]{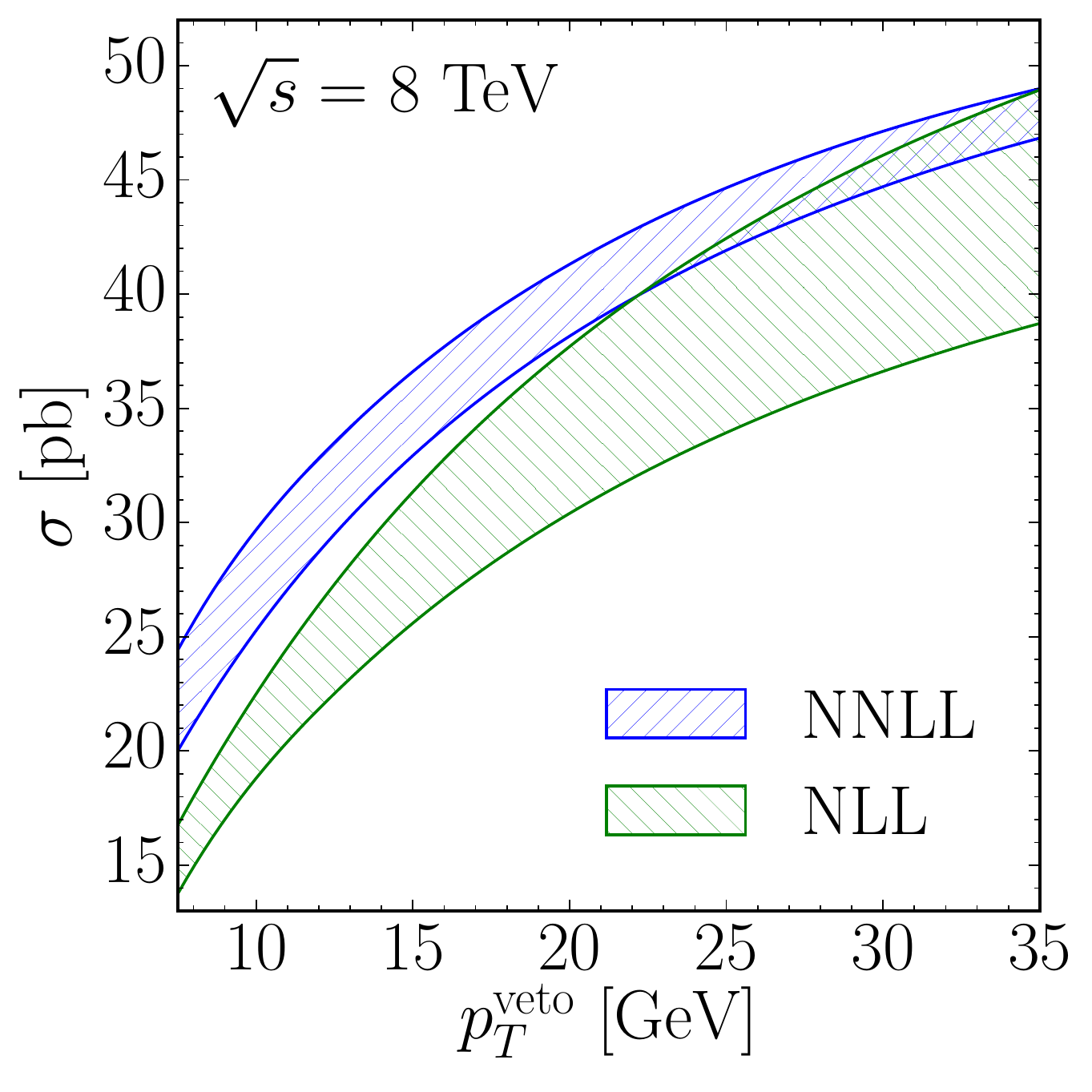}
                \caption{}
                \label{f.Xsection_8TeV}
        \end{subfigure}
        \begin{subfigure}[b]{0.32\textwidth}
                \includegraphics[width=\textwidth]{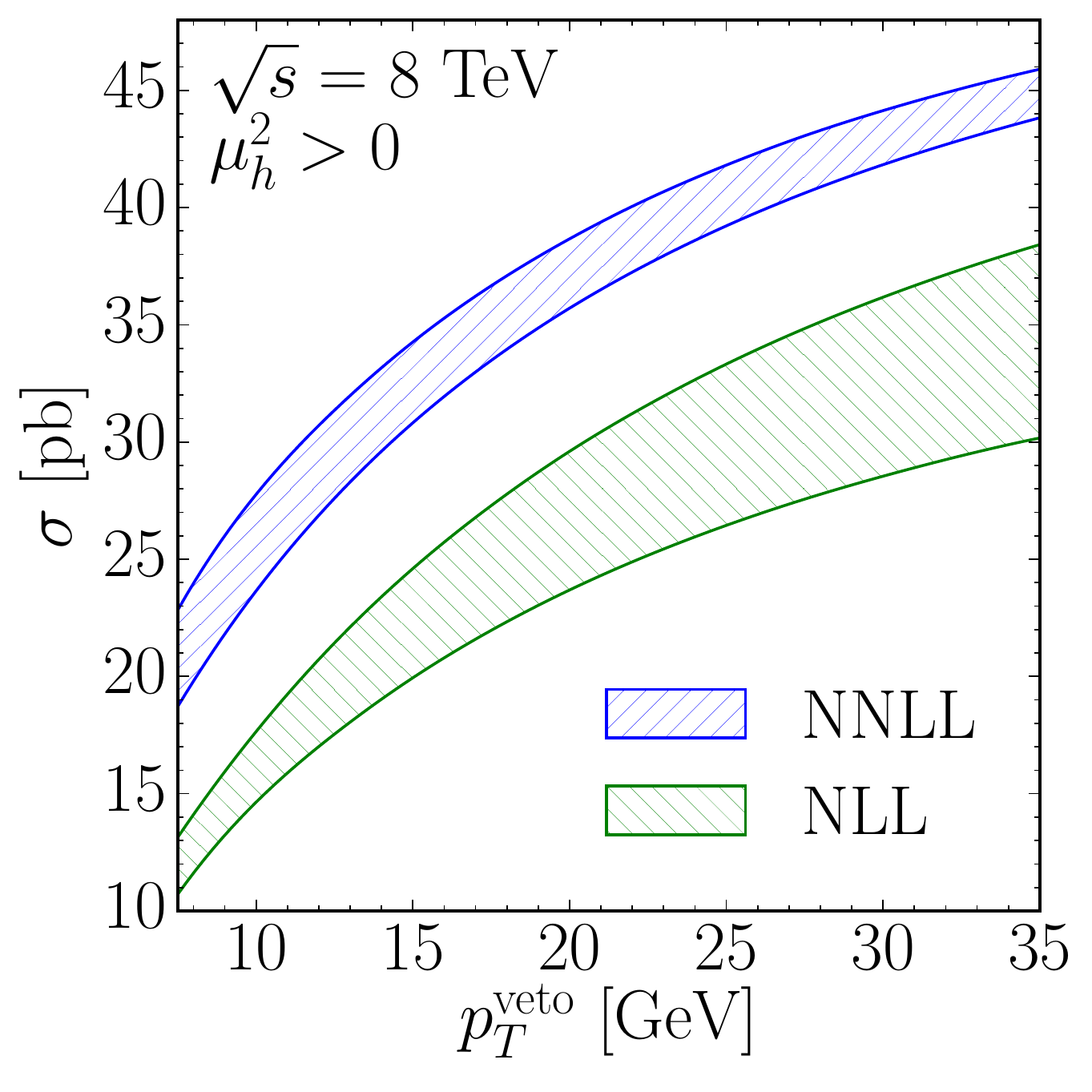}
                \caption{}
                \label{f.Xsection_noPi_8TeV}
        \end{subfigure}
        \caption{The NLL and NNLL jet-veto cross-sections for the process, 
        $\proton \proton \rightarrow W^+ W^-$ for (a) the 7-TeV  
        and (b) the 8-TeV LHC runs. 
        (c) is the same as (b) except that it is without $\pi^2$ resummation, 
        i.e., 
        it is evaluated with $\muh^2 > 0$ as opposed to $\muh^2 < 0$. 
        The shaded regions in all the plots indicate the scale uncertainties.}
        \label{f.Xsection}
\end{figure}
\begin{figure}[t]
        \centering
        \begin{subfigure}[b]{0.32\textwidth}
                \includegraphics[width=\textwidth]{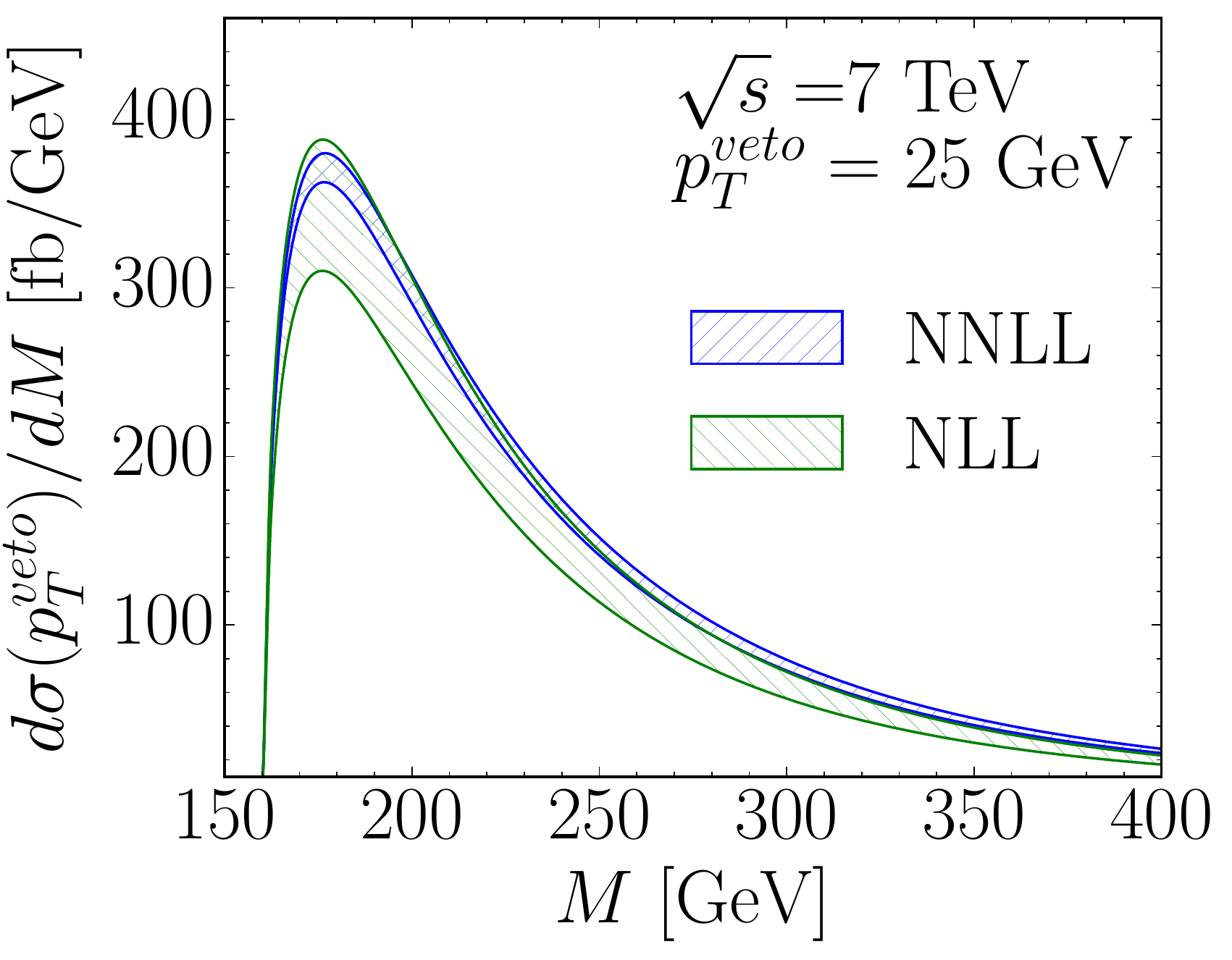}
                \caption{}
                \label{f.DiffXsec_7TeV_pT25}
        \end{subfigure}   
        \begin{subfigure}[b]{0.32\textwidth}
                \includegraphics[width=\textwidth]{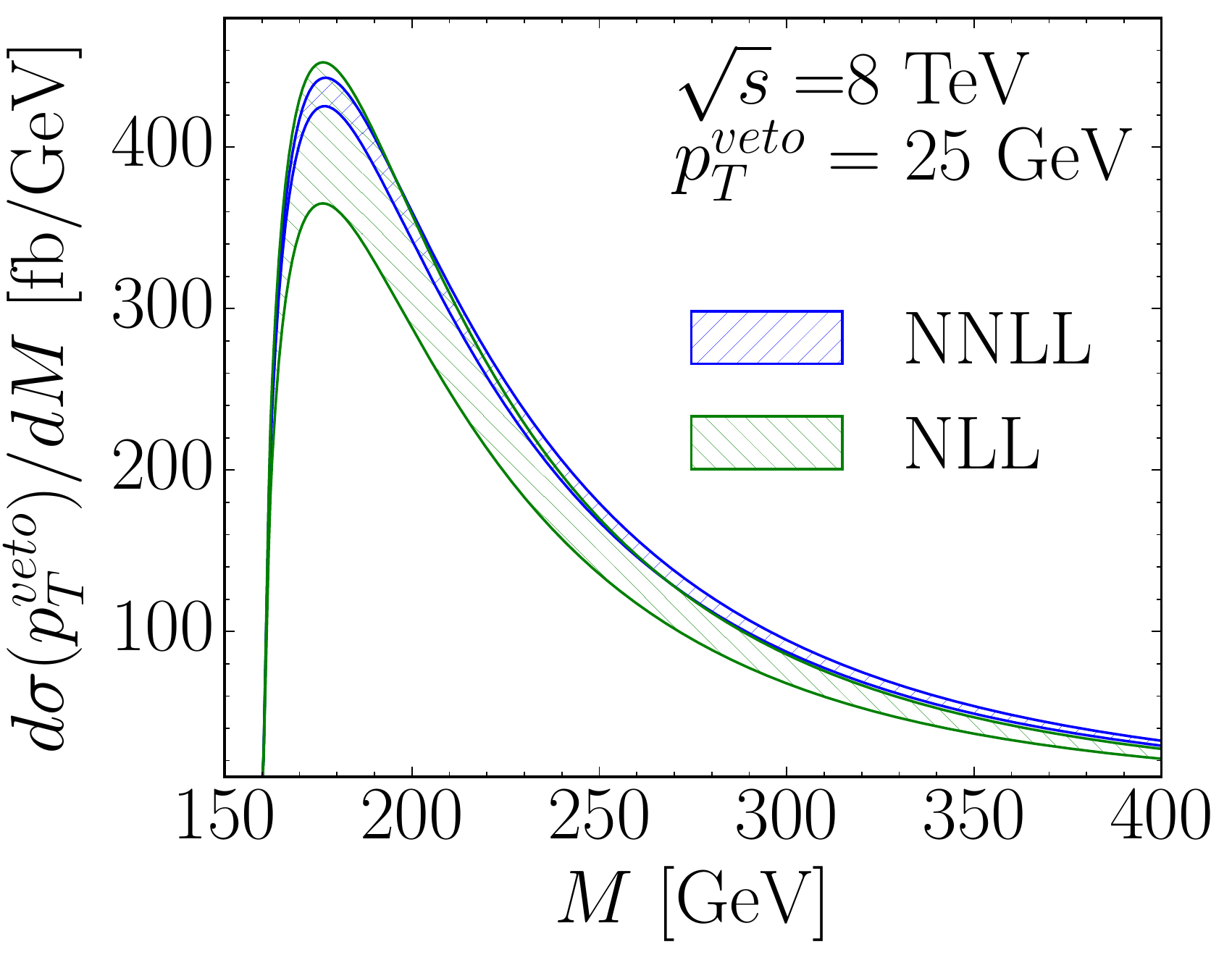}
                \caption{}
                \label{f.DiffXsec_8TeV_pT25}
        \end{subfigure}
         \begin{subfigure}[b]{0.32\textwidth}
                \includegraphics[width=\textwidth]{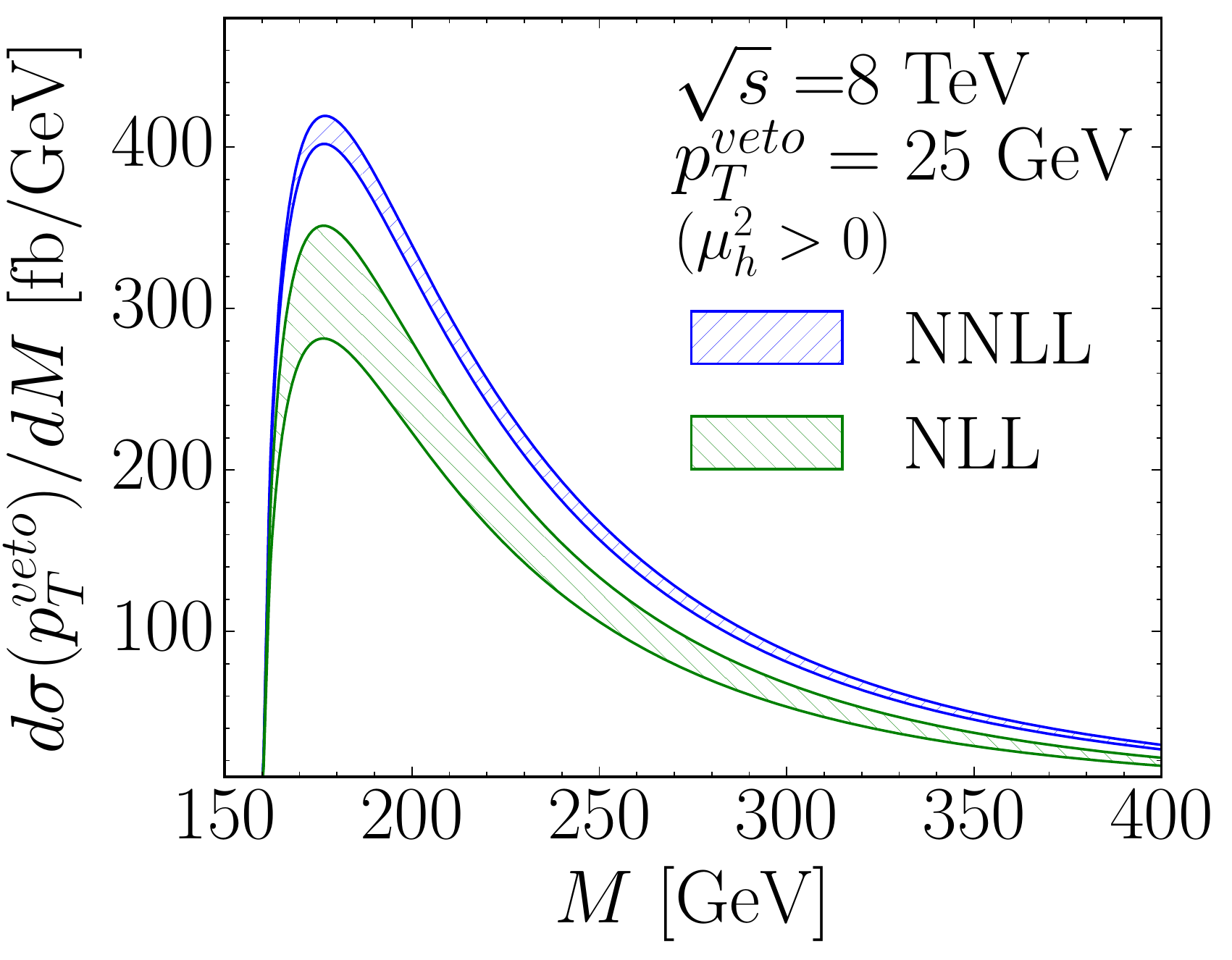}
                \caption{}
                \label{f.DiffXsec_8TeV_pT25_noPi}
        \end{subfigure}
        \caption{Same as \fref{Xsection} but for differential jet-veto cross-sections with respect to the $W^+ W^-$ invariant mass.}
        \label{f.diff_xsection}
\end{figure}

The resummed NLL and NNLL jet-veto
cross-section for the $\proton \proton \rightarrow W^+ W^-$ as a function of $\ptv$ are shown in~\fref{Xsection_7TeV} and~\fref{Xsection_8TeV} 
for $\sqrt{s}=7$ and $8$~TeV, respectively, in the 5-flavor number scheme. The error bands 
are obtained by separately varying the hard scale $\muh$ and the factorization scale $\muf$ by factors of $1/2$ and $2$, and adding
the uncertainties from those two variations in quadrature. 
The uncertainty from the hard-scale variation is generally found to be small 
($\sim 1\%$ of the total cross section) for NNLL cross-sections. 
In these plots, we have not included the contribution from $g g \rightarrow W^+ W^-$, 
which is not only in higher order in $\al_\s$ but also irrelevant for resummation as the leading contribution is already at 1-loop so it always (and trivially) passes the jet veto at the 1-loop level. 

In both~\fref{Xsection_7TeV} and~\fref{Xsection_8TeV}, 
the cross sections are evaluated for $\muh^2 \sim -M^2$ as discussed above.
For comparison, in~\fref{Xsection_noPi_8TeV}, 
the jet-veto cross-section for $\sqrt{s}=8$~TeV is 
evaluated using $\mu_h^2 \sim +M^2$ so that $\pi^2$ terms are not resummed. Indeed, we 
observe that the consistency of the NLL and NNLL results is poorer compared to 
\fref{Xsection_8TeV}. 
The impact of $\pi^2$ resummation is most significant for $M \lesssim 250$~GeV, while for larger $M$, the enhancement from the $\pi^2$ terms is partially cancelled by powers of $\log M/\ptv$, because in the square of $\log[(-M^2 - \I 0^+) / (\ptv)^2] = \log[M^2/(\ptv)^2] - \I \pi$, the logarithm and $\pi^2$ term come in the opposite signs.

The differential cross-section with respect to the invariant mass of the $W$-boson pair at NLL and NNLL are shown in~\fref{DiffXsec_7TeV_pT25} and~\fref{DiffXsec_8TeV_pT25} for 
$\sqrt{s} = 7$-TeV and $8$-TeV LHC runs using 5-flavor number scheme, 
but ignoring $gg$-initiated contributions as discussed above. 
The error bands from the scale uncertainties are obtained by exactly the same procedure as above.
We have fixed $\ptv = 25$~GeV in these plots.
In~\fref{DiffXsec_8TeV_pT25_noPi}, the differential cross-section is evaluated with $\muh^2 > 0$ as in~\fref{Xsection_noPi_8TeV}.
Again, we see a poor convergence due to the absence of $\pi^2$ resummation.

\subsection{Power Corrections}
\label{ss.power_corrections}

Our SCET is a power expansion in $\lambda \equiv \ptv / M$ 
and we have only considered leading-order terms, i.e., $\co(\la^0)$, 
so that $\la$ only appears as $\log\la$.  
Namely,
the hard coefficient matched as in\erefn{Ch} at $\mu = \muh$ has no $\la$ dependence as we are working at $\co(\la^0)$,
and the $\log\la$ dependence is introduced solely via the RG evolution\erefn{Ch_RG} down to $\mu = \muf$.
In this section, we study the effect of \emph{power corrections}, 
i.e., the impact of $\co(\lambda)$ terms, 
by incorporating $\co(\lambda)$ terms from what is normally referred to as an fixed-order NLO, but without redoing the whole SCET calculation by including $\la$-suppressed operators.

Referring to such NNLL resummed cross-sections with the inclusion of $\co(\al_\s \lambda)$ 
power corrections as NNLL+NLO, 
we have
\beq
\frac{\dd \sg^{\rm NNLL+NLO} }{\dd M} =
\frac{\dd \sg^{\rm NNLL} }{\dd M} 
+ \left[ \frac{\dd \sg^{\rm NLO} }{\dd M}  -
\frac{\dd \sg^{\rm NNLL} }{\dd M} 
\biggr|_\text{expanded to $\co(\al_\s)$}  \right],
\label{e.power_corrections}
\eeq
where ``Expanded to $\co(\al_\s)$'' 
in\erefn{power_corrections} means \emph{literally} expanding it to $\co(\al_\s)$ \emph{without} 
counting the $\log\la$ as $1 / \al_\s$. 
The power corrections are then given by the term inside the bracket, which is devoid of logarithms of the form $\ds{\log^n\!\la}$, and depends explicitly only on $\al_\s$ and $\la$ but it has an implicit dependence on the scale $\muf$ through the scale dependences of $\al_\s$ and PDF\@.
Since the scale of PDF is given by the factorization scale 
$\muf \sim \ptv$ in the presence of jet veto, 
we evaluate the power corrections and estimate the scale uncertainties by varying $\muf$ between factors of $1/2$ and $2$
around the central value $\muf = \ptv$.  
We use {\tt MadGraph5\_aMC@NLO} \cite{aMC@NLO}  to calculate $\sigma^\text{NLO}$ by setting the 
renormalization and factorization scales equal to $\muf$.
We do not include the $b\bar{b}$ channel for computing power corrections, as
the {\tt  MadGraph5\_aMC@NLO} program does not allow inclusion of $b$-quarks in the initial state.
However, given that the $b$ quark contribution is already very small due to the small $b$ PDF,
the exclusion of $b$-quarks in the initial states should be inconsequential for the study of power corrections. 

We would also like to briefly comment on a subtlety regarding the contribution from the $qg$ channel (and $\bar{q} g$). 
There are two types of diagrams in the $qg$ channel, 
depending on whether or not the diagram can be reinterpreted as the initial $g$ splitting into a nearly on-shell $q$ and $\bar{q}$, and this $\bar{q}$ subsequently annihilating with the initial $q$ to produce $WW$.  
The diagrams that permit this interpretation are included in our SCET calculation as the beam functions include the contributions from gluon splitting as can be seen in\erefn{beam_OPE}.
On the other hand, those that do not admit the gluon splitting interpretation are not included in our SCET calculation.
To include them, we would have to add new operators to $\cl_\text{int}$ that consists of \emph{three} (anti)collinear sectors corresponding to, e.g., 
the initial collinear $q$, the initial anticollinear $g$, and the final collinear $q$.
In principle, we must distinguish power corrections to those new operators from the corrections to our SCET operator\erefn{current}.  
However, in $qg \to WWq$, it is very difficult for the final state $q$ to pass the jet veto, as its $\pT$ is generically $\co(M)$,
so the contributions from the $qg$ channel to the \emph{jet-veto} cross section is highly suppressed.
Since we found the power corrections to be negligible, we just let the $qg$ contribution contaminate our calculation of power corrections to the $q\bar{q}$ SCET operator. 

We conclude this section by noting that we find the power corrections to be generally small, decreasing the unmatched NNLL cross-section by no more than a mere $\sim 0.5\%$ with a negligible impact on the scale uncertainty.

\begin{figure}[t]
        \centering
        \begin{subfigure}[b]{0.32\textwidth}
                \includegraphics[width=\textwidth]{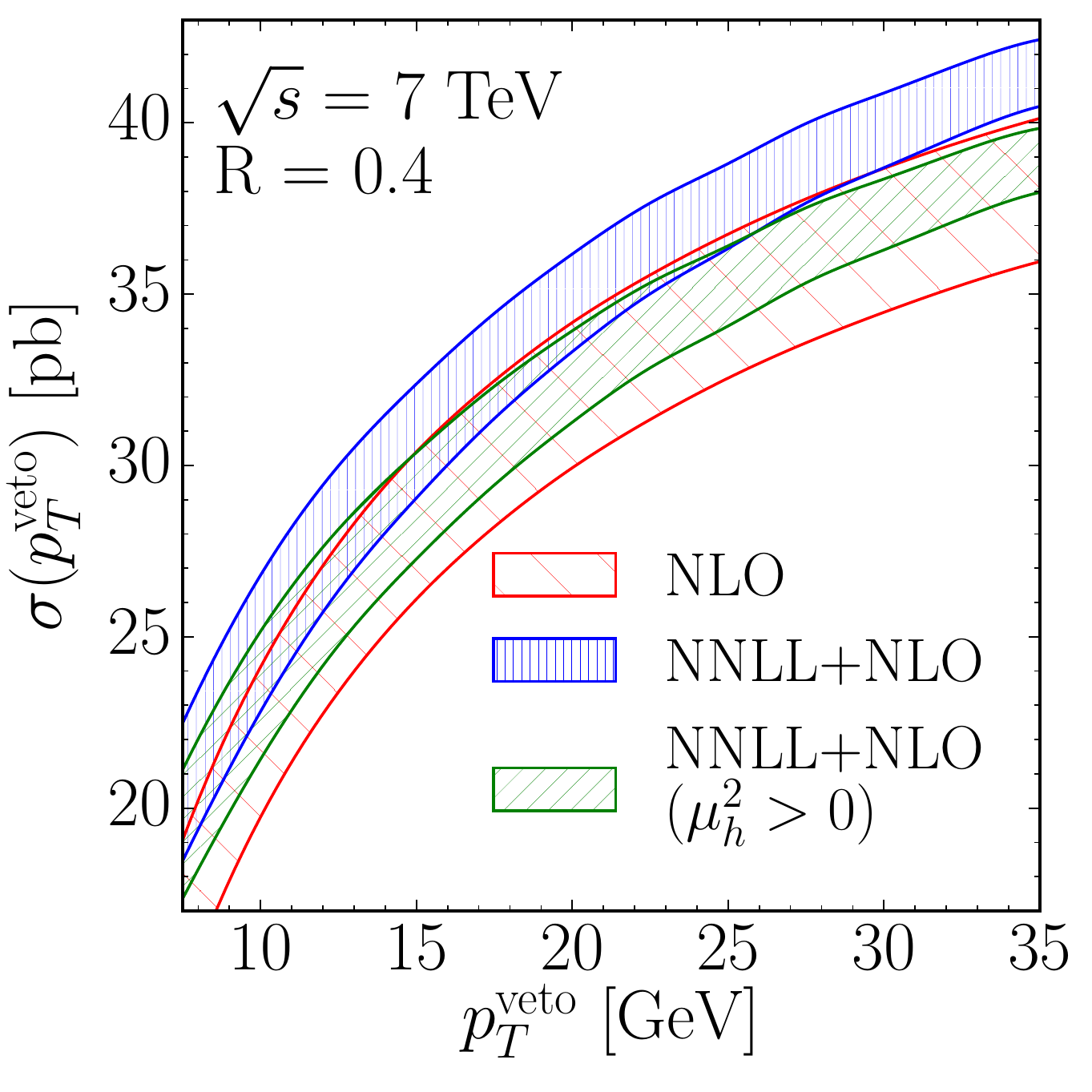}
                \caption{}
                \label{f.Xsection_NLO_R0.4_7TeV}
        \end{subfigure}   
        \begin{subfigure}[b]{0.32\textwidth}
                \includegraphics[width=\textwidth]{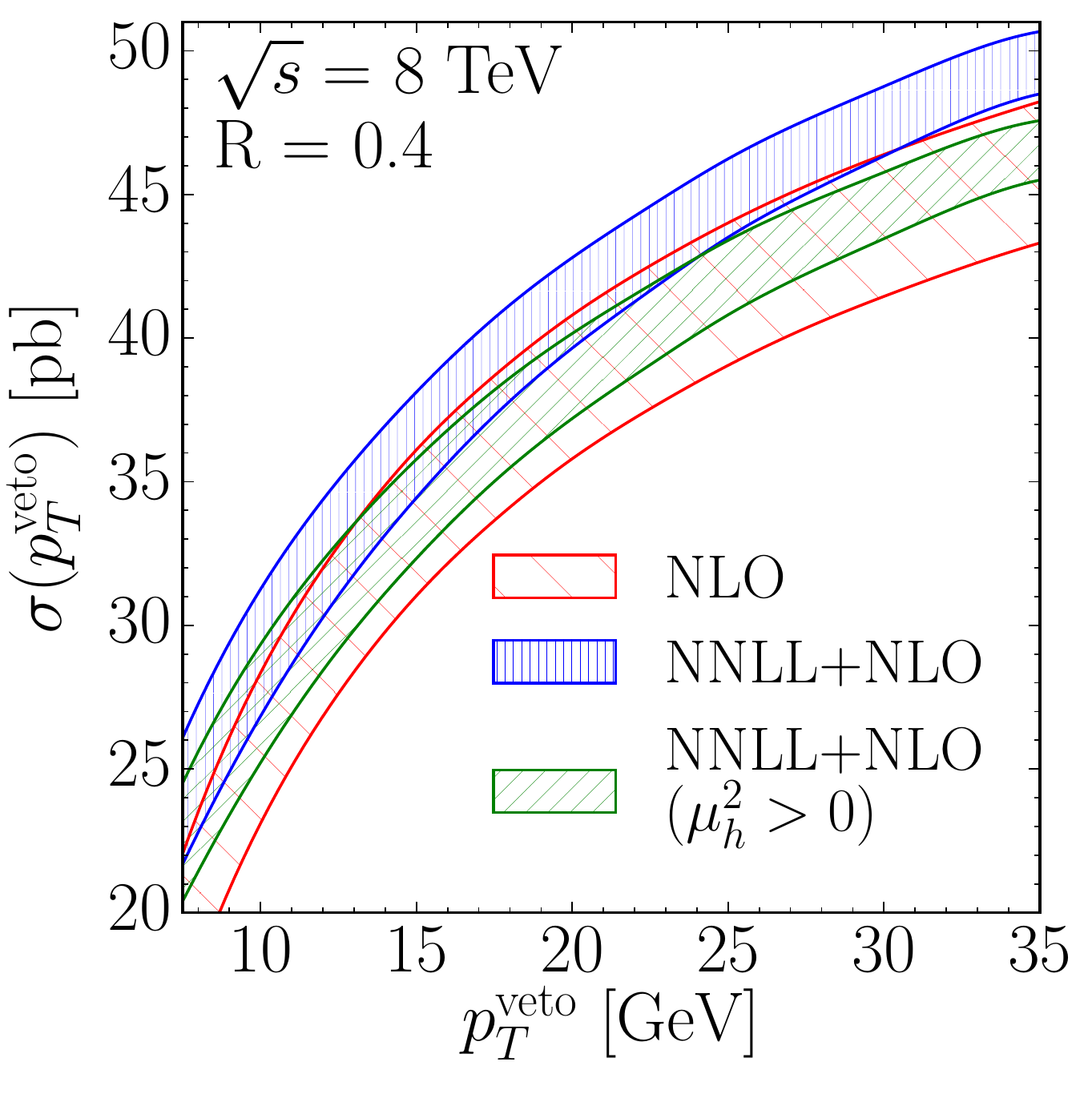}
                \caption{}
                \label{f.Xsection_NLO_R0.4_8TeV}
        \end{subfigure}
         \begin{subfigure}[b]{0.32\textwidth}
                \includegraphics[width=\textwidth]{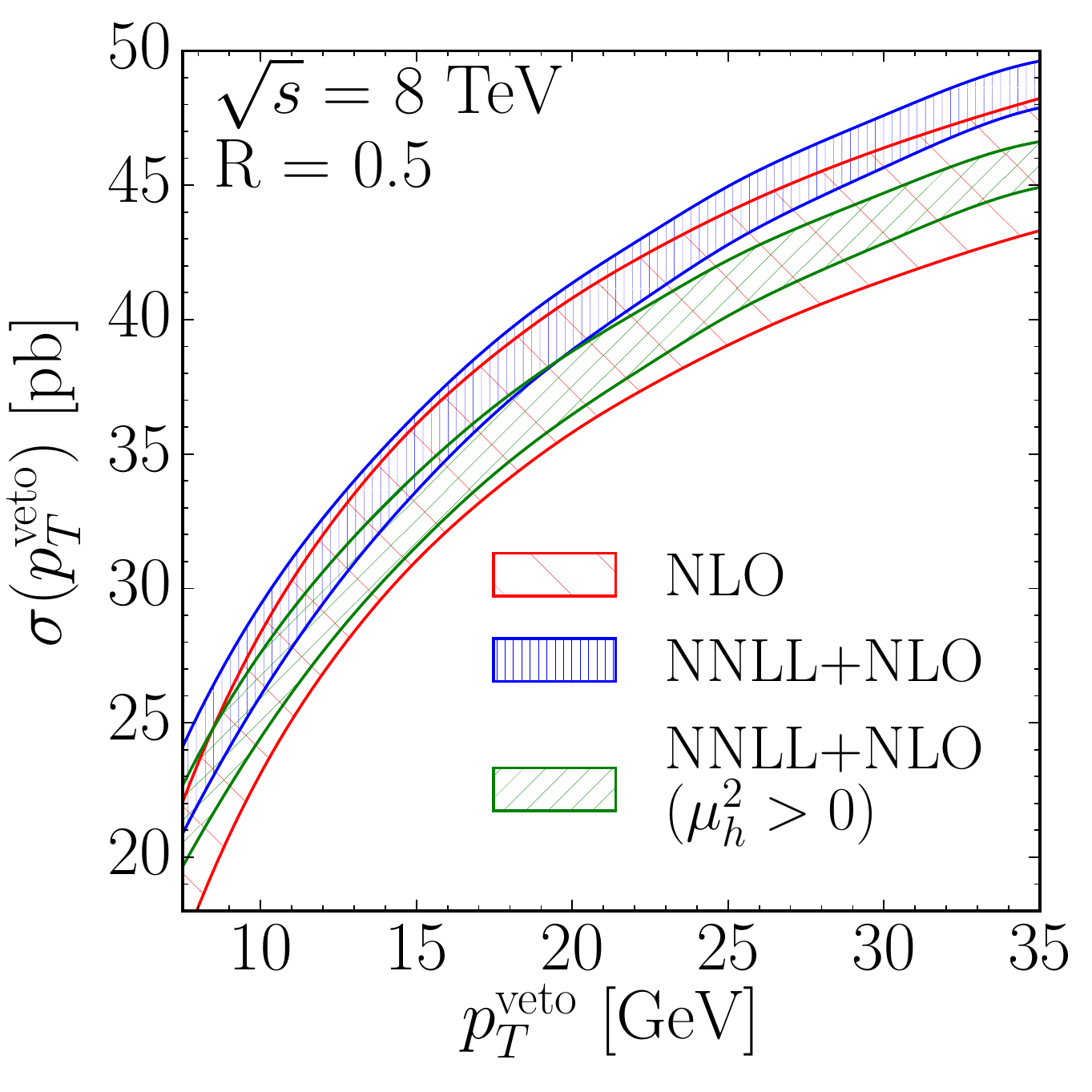}
                \caption{}
                \label{f.Xsection_NLO_R0.5_8TeV}
        \end{subfigure}
        \caption{Comparison of our NNLL+NLO resummed jet-veto cross-sections \emph{with and 
        without $\pi^2$ resummation} (in blue and green, respectively), with the fixed order NLO results
         obtained from {\tt MCFM} (in red) 
        for the (a) 7-TeV LHC and (b) 8-TeV LHC runs, with jet-radius parameter, $R=0.4$. In 
        figure (c), the same comparison is shown for the 8-TeV LHC run with $R=0.5$.
        }
        \label{f.NLO_vs_NNLL}
\end{figure}
%

\subsection{Comparison with Fixed-order NLO and Monte Carlo + Parton Shower Generators}
First, in \fref{NLO_vs_NNLL}, we compare our matched NNLL+NLO prediction (in blue) with the fixed-order NLO jet-veto cross-sections (in red)
obtained from {\tt MCFM}. The scale uncertainties in the NLO calculation are calculated using the procedure 
outlined in \sref{intro}. To gauge the effects of $\pi^2$ resummation, we have also presented our NNLL+NLO results for $\muh^2 > 0$ (in green).
With or without $\pi^2$ resummation, the reduction of scale uncertainties due to the resummation of the logarithms is evident. 
As we already noted in \ssref{NLL-NNLL}, 
the $\pi^2$ effects tend to cancel with the logarithms as the latter get larger 
(i.e., towards lower $\ptv$).  
The inclusive contribution of $g g \rightarrow W^+ W^-$ channel, which is formally NNLO, is obtained 
by running the \ttc{MCFM} program and included in both the fixed order and resummed cross-sections. 
The scale uncertainties associated with the $gg$ channel, while significant in isolation,
are expected to be sub-percent level relative to the total cross-section and therefore neglected.  

Since the ATLAS and CMS collaborations have used the Monte Carlo (MC) and Parton Shower (PS) generators to estimate the jet-veto efficiencies
in their $W^+ W^-$ cross-section measurement analyses, 
we would like to make comparisons with our analytical NNLL+NLO results by employing three sets of MC and PS generators for the process $q \bar{q} \rightarrow W^+ W^-$: 

\begin{itemize}
\item{{\tt MG+PY}} : The $WW+0/1/2$ jet parton-level matched samples were generated using the LO mode of  
{\tt MadGraph5\_aMC@NLO} \cite{aMC@NLO} followed by showering using the \ttc{Pythia6} PS generator\cite{Pythia6}.
 The matching is performed based on the default {\it $k_T$-jet MLM} scheme used in {\tt MadGraph5\_aMC@NLO}.  
\item{{\tt MC@NLO+HW}} : The parton level events were generated using the NLO MC generator, {\tt MC@NLO} \cite{MC@NLO} and 
showered by the \ttc{Herwig6} PS generator \cite{Herwig6}. Matching is automatically performed by the {\tt MC@NLO} program. 
\item{{\tt POWHEG+PY}} : The parton level events were generated using the {\tt POWHEG} NLO MC generator \cite{Powheg:0,
Powheg:1, Powheg:2, Powheg:3} interfaced with \ttc{Pythia6} for parton showering. Matching is automatically performed by the 
{\tt POWHEG} program. 
\end{itemize}

The MC samples were generated using {\tt CTEQ6L} ({\tt CT10nlo}) PDFs for the LO (NLO) MC generators. In each case, the total 
number of events are normalized to the NLO cross-section obtained using \ttc{MCFM}. 
For a fair comparison with our resummed theory predictions and to disentangle the effects of different choices of PDFs and $\al_\s$ 
to the overall normalization, we consistently use {\tt MSTW2008nnlo} PDFs 
for the MCFM calculation of the inclusive cross-section with $\al_\s$ set by the PDF itself. In the next step, 
we performed jet-clustering on the samples using the \ttc{FastJet} program \cite{FastJet:1,FastJet:2} 
employing the anti-$k_\mathrm{T}$ algorithm with 
jet-radius parameter, $R=0.4$. 
Given that the contribution from $g g \rightarrow W^+ W^-$ is formally NNLO, 
the jet-veto efficiency can be obtained to a good approximation as follows: %
\beq
\jv{\epsilon} = \frac{ \jv{\sigma}_{q \bar{q}} + \sigma_{gg}}{ \sigma_{q \bar{q}} + \sigma_{gg}} 
		      = \frac{ \jv{\epsilon}_{q \bar{q}} \sigma_{q \bar{q}} + \sigma_{gg}}{ \sigma_{q \bar{q}} + \sigma_{gg}}
\label{e.JVeff}		  
\eeq 
For our resummed calculations, 
we obtain the jet-veto efficiencies directly from the first expression in\erefn{JVeff},
since the jet-veto cross-section $\jv{\sigma}_{q \bar{q}}$ for the $q \bar{q}$ channel has already been calculated. 
While the scale uncertainties in the jet-veto cross-sections may partially cancel the scale uncertainties in the inclusive cross-section,
we make conservative estimates of the uncertainties by using the full scale uncertainty of the jet-veto cross-section while using the 
central value for the inclusive cross-section.

\begin{figure}[t]
        \centering
        \begin{subfigure}[b]{0.32\textwidth}
                \includegraphics[width=\textwidth]{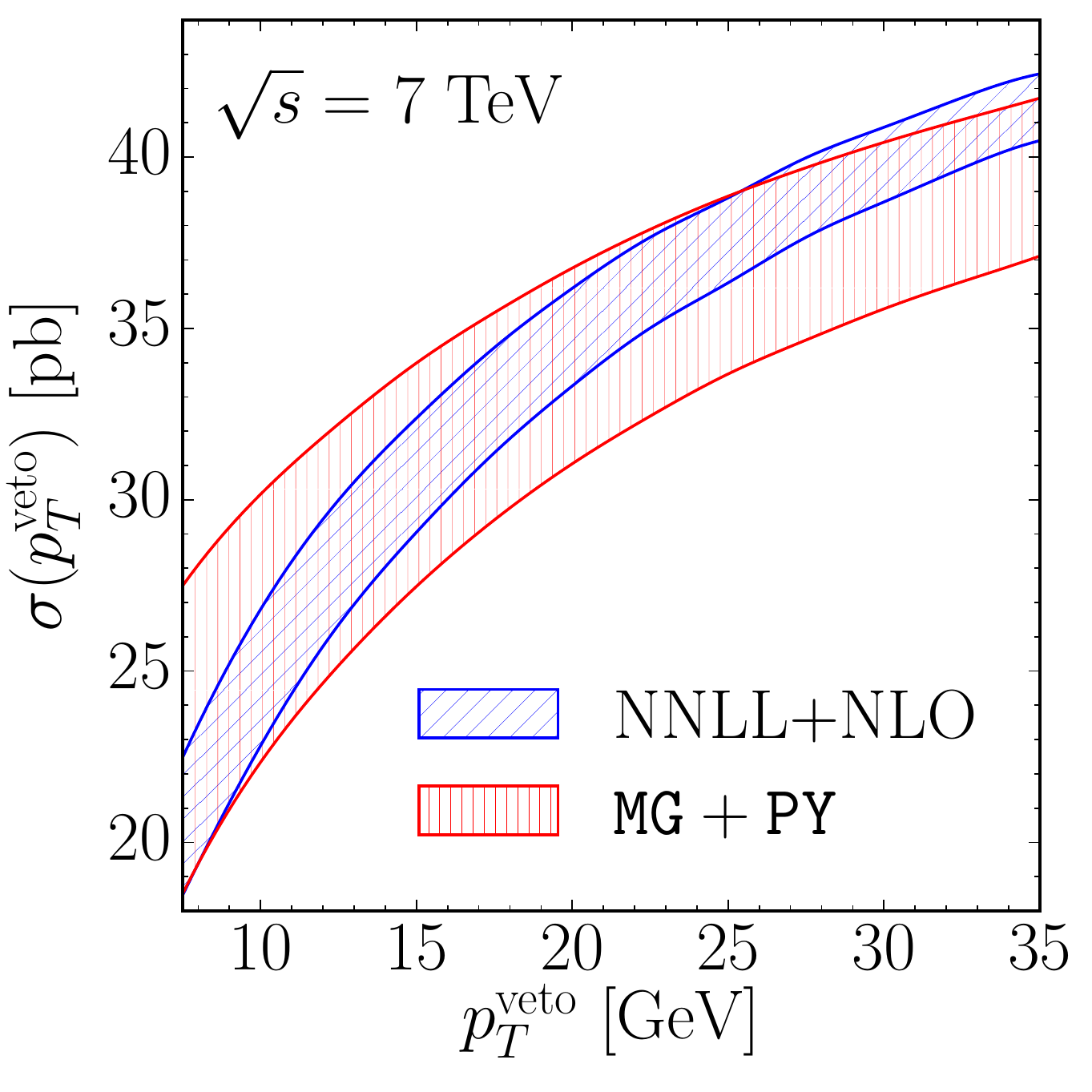}
                \caption{}
                \label{f.Xsection_Pythia_7TeV}
        \end{subfigure}   
        \begin{subfigure}[b]{0.32\textwidth}
                \includegraphics[width=\textwidth]{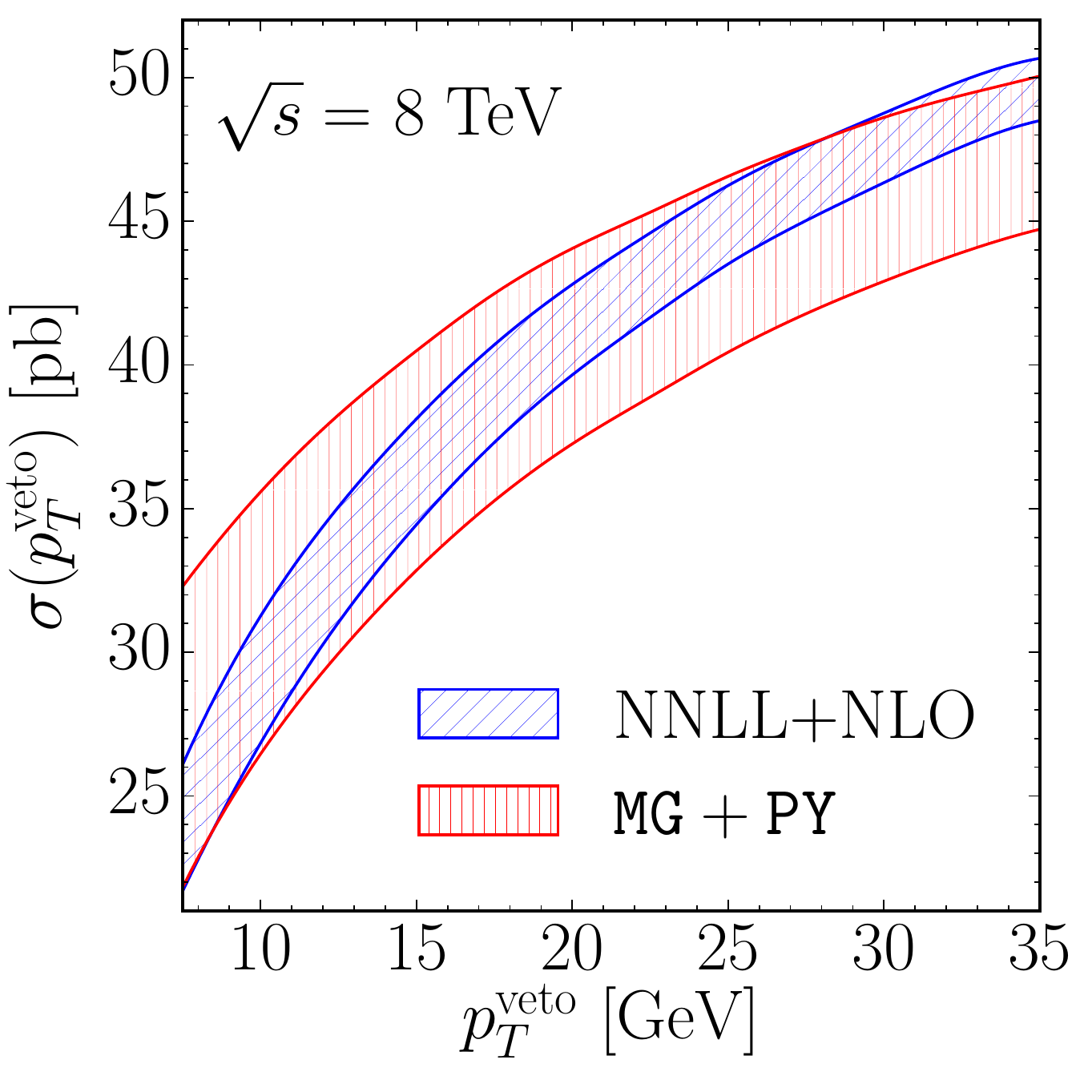}
                \caption{}
                \label{f.Xsection_Pythia_8TeV}
        \end{subfigure}
         \begin{subfigure}[b]{0.32\textwidth}
                \includegraphics[width=\textwidth]{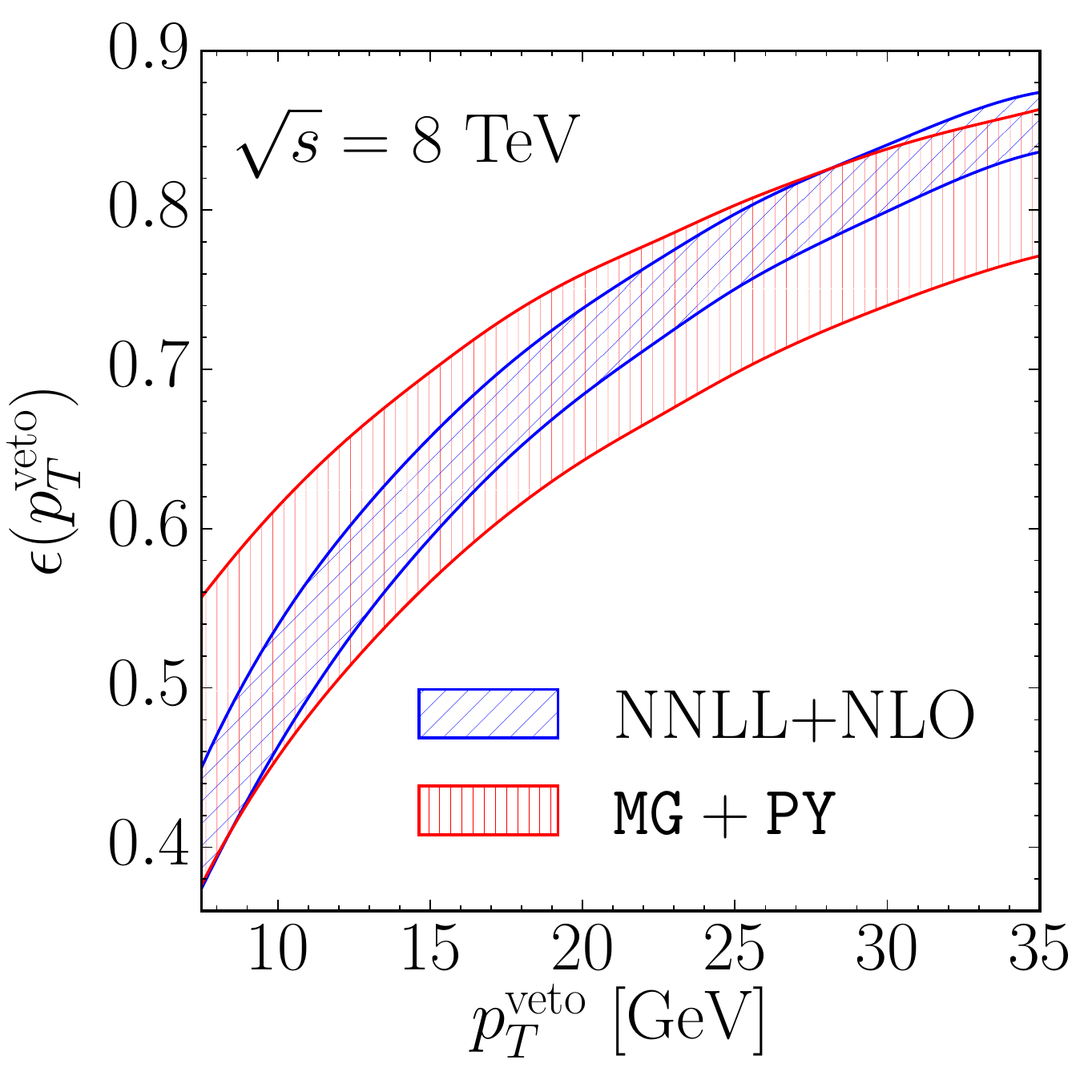}
                \caption{}
                \label{f.Eff_Pythia_8TeV}
        \end{subfigure}
        \caption{Comparison of our NNLL+NLO resummed jet-veto cross-sections with the MC samples generated using \ttc{MadGraph5}
                followed by showering with \ttc{Pythia6} parton-shower generator, with the total cross-section normalized to the inclusive NLO
                 cross-section obtained from \ttc{MCFM} for the (a) $7$-TeV and (b) $8$-TeV LHC runs. In figure (c), the jet-veto 
                 efficiencies from our analytical results are compared with that from MC samples at the $8$-TeV LHC run.}
        \label{f.Xsection_Pythia}
\end{figure}
\begin{figure}[t]
        \centering
        \begin{subfigure}[b]{0.32\textwidth}
                \includegraphics[width=\textwidth]{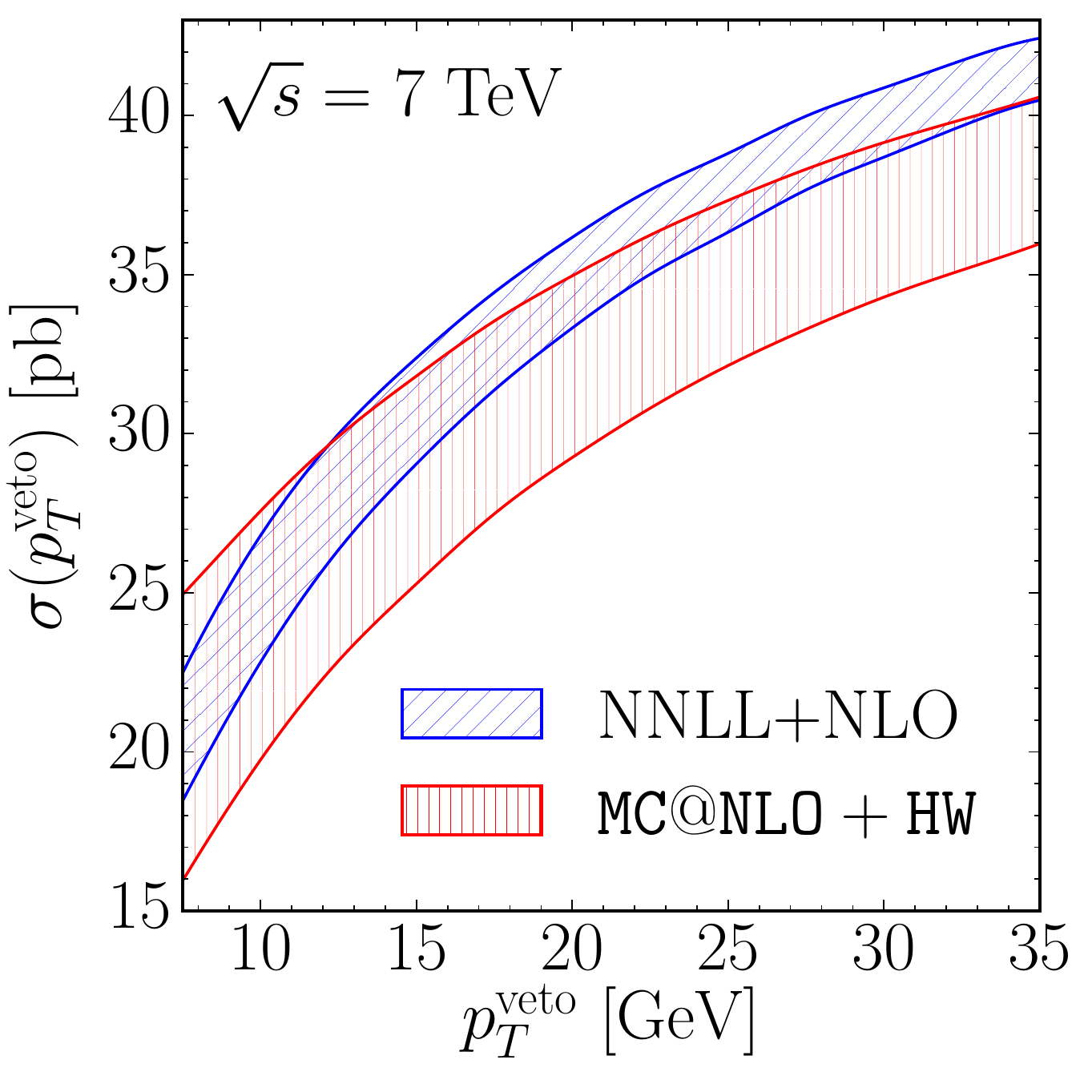}
                \caption{}
                \label{f.Xsection_MCatNLO_7TeV}
        \end{subfigure}   
        \begin{subfigure}[b]{0.32\textwidth}
                \includegraphics[width=\textwidth]{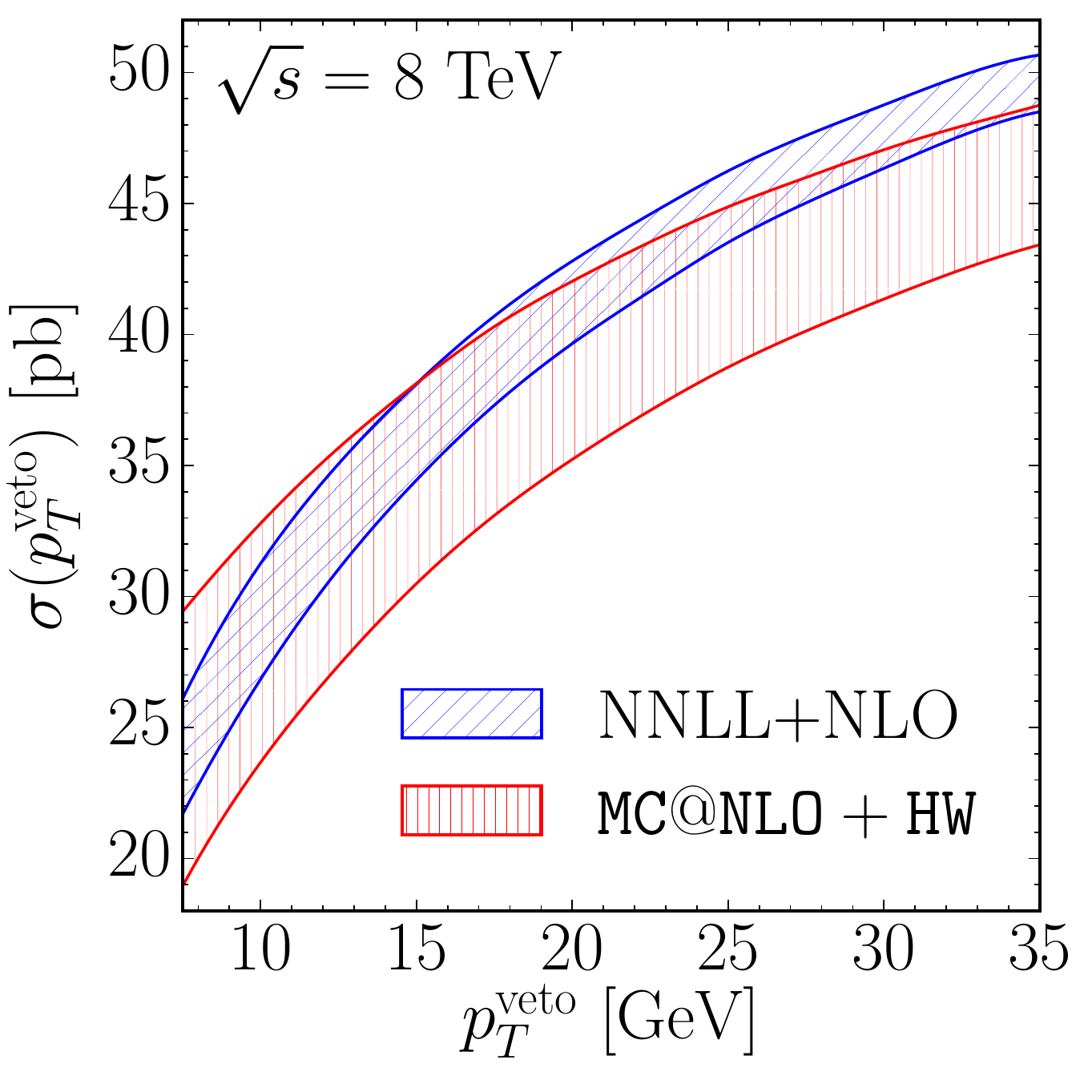}
                \caption{}
                \label{f.Xsection_MCatNLO_8TeV}
        \end{subfigure}
         \begin{subfigure}[b]{0.32\textwidth}
                \includegraphics[width=\textwidth]{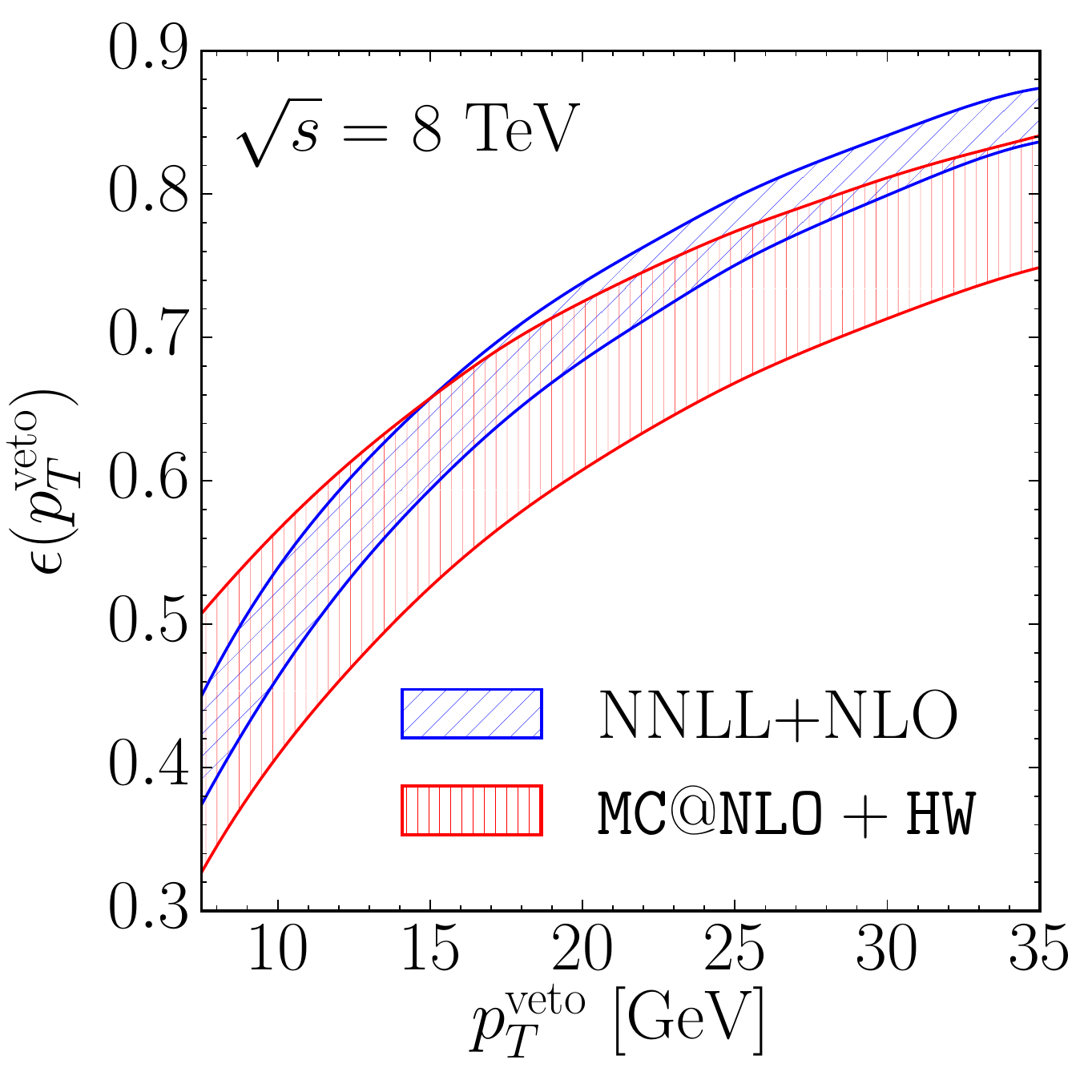}
                \caption{}
                \label{f.Eff_MCatNLO_8TeV}
        \end{subfigure}
        \caption{Same as \fref{Xsection_Pythia} but with MC samples generated using \ttc{MC@NLO} interfaced with \ttc{HERWIG6} for 
        parton-showering.}
        \label{f.Xsection_MCatNLO}
\end{figure}
\begin{figure}[t]
        \centering
        \begin{subfigure}[b]{0.32\textwidth}
                \includegraphics[width=\textwidth]{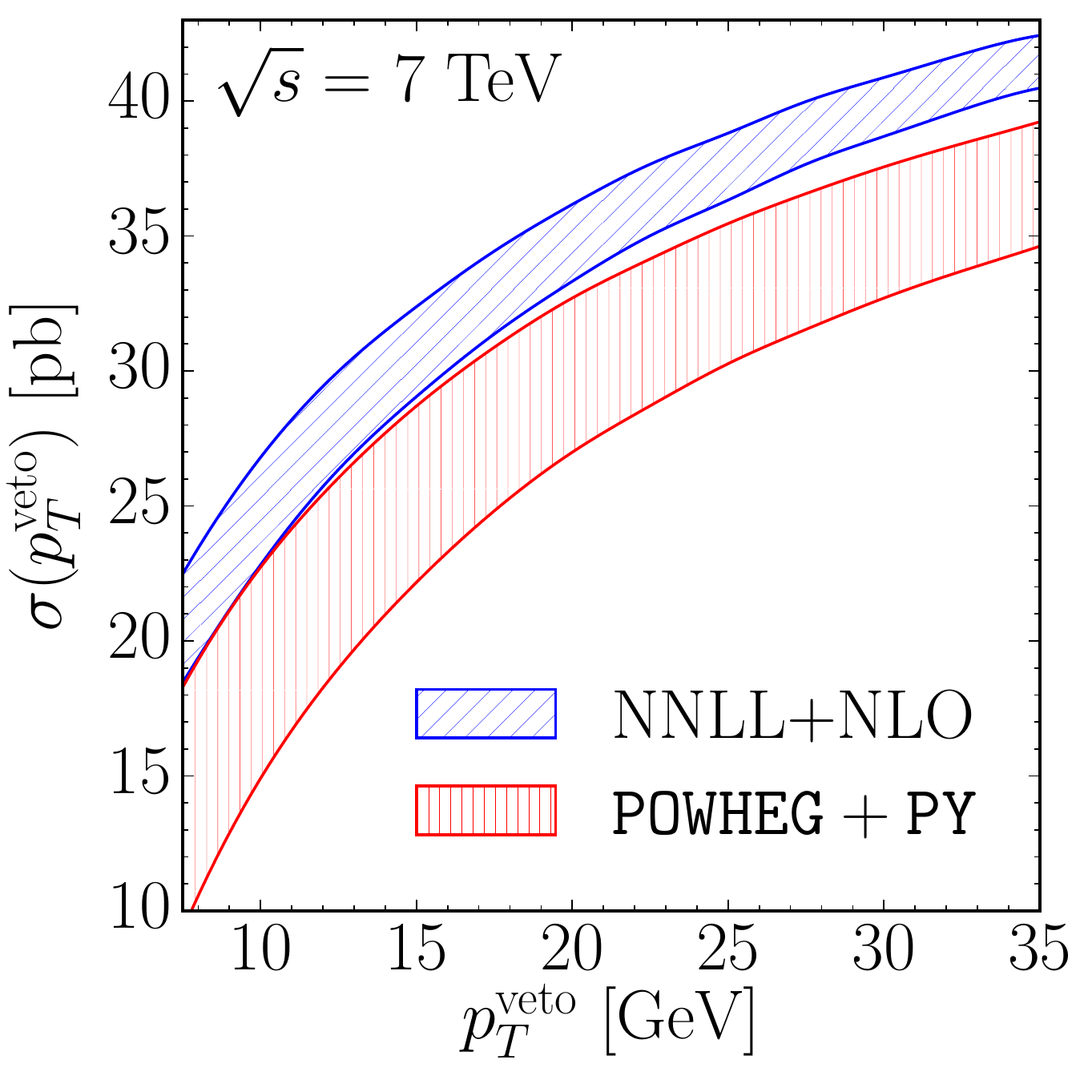}
                \caption{}
                \label{f.Xsection_POWHEG_7TeV}
        \end{subfigure}   
        \begin{subfigure}[b]{0.32\textwidth}
                \includegraphics[width=\textwidth]{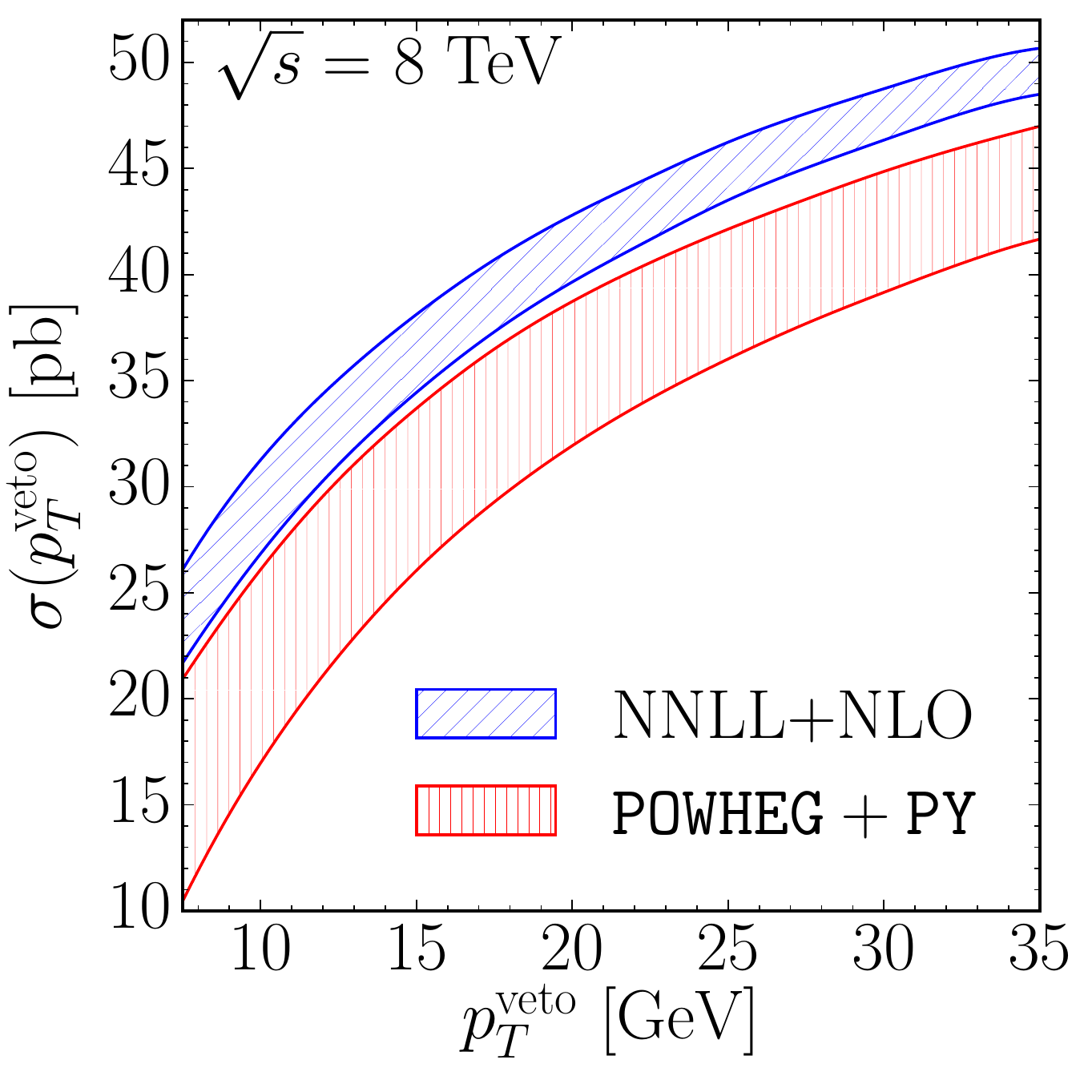}
                \caption{}
                \label{f.Xsection_POWHEG_8TeV}
        \end{subfigure}
         \begin{subfigure}[b]{0.32\textwidth}
                \includegraphics[width=\textwidth]{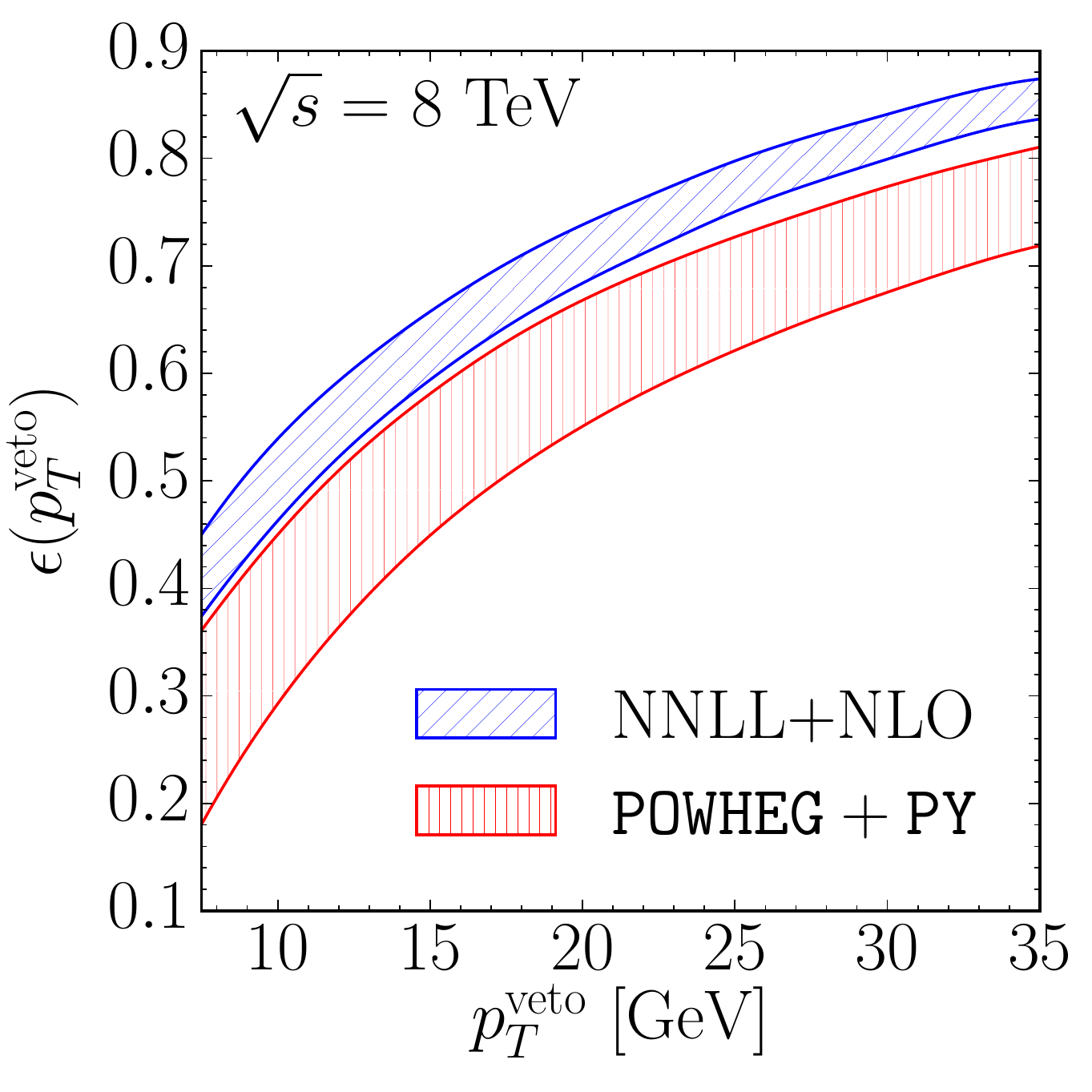}
                \caption{}
                \label{f.Eff_POWHEG_8TeV}
        \end{subfigure}
        \caption{Same as \fref{Xsection_Pythia} but with MC samples generated using \ttc{POWHEG} interfaced with \ttc{Pythia6} for 
        parton-showering.}
        \label{f.Xsection_POWHEG}
\end{figure}
\begin{figure}[t]
        \centering
        \begin{subfigure}[b]{0.32\textwidth}
                \includegraphics[width=\textwidth]{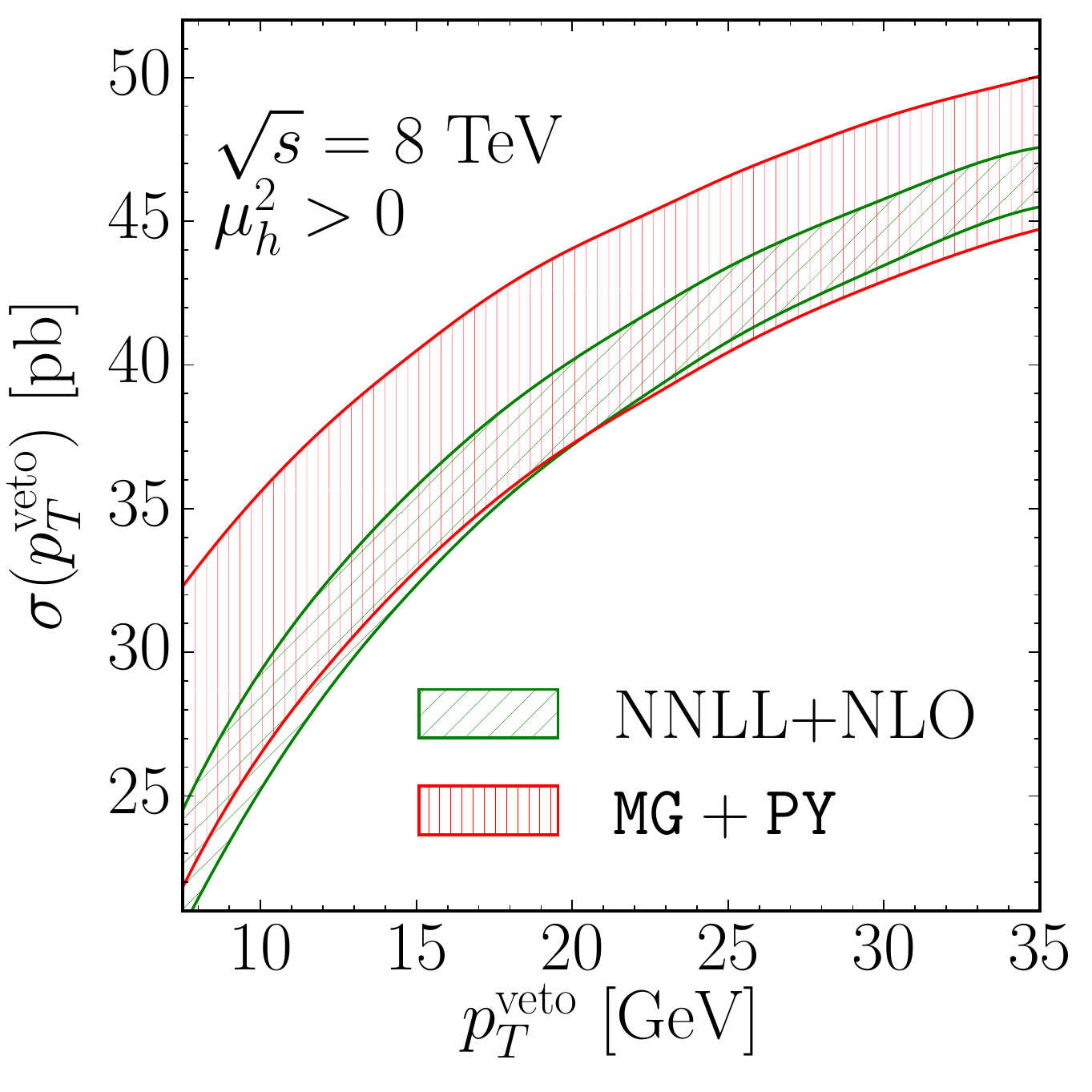}
                \caption{}
                \label{f.Xsection_Pythia_8TeV_NoPiSq}
        \end{subfigure}   
        \begin{subfigure}[b]{0.32\textwidth}
                \includegraphics[width=\textwidth]{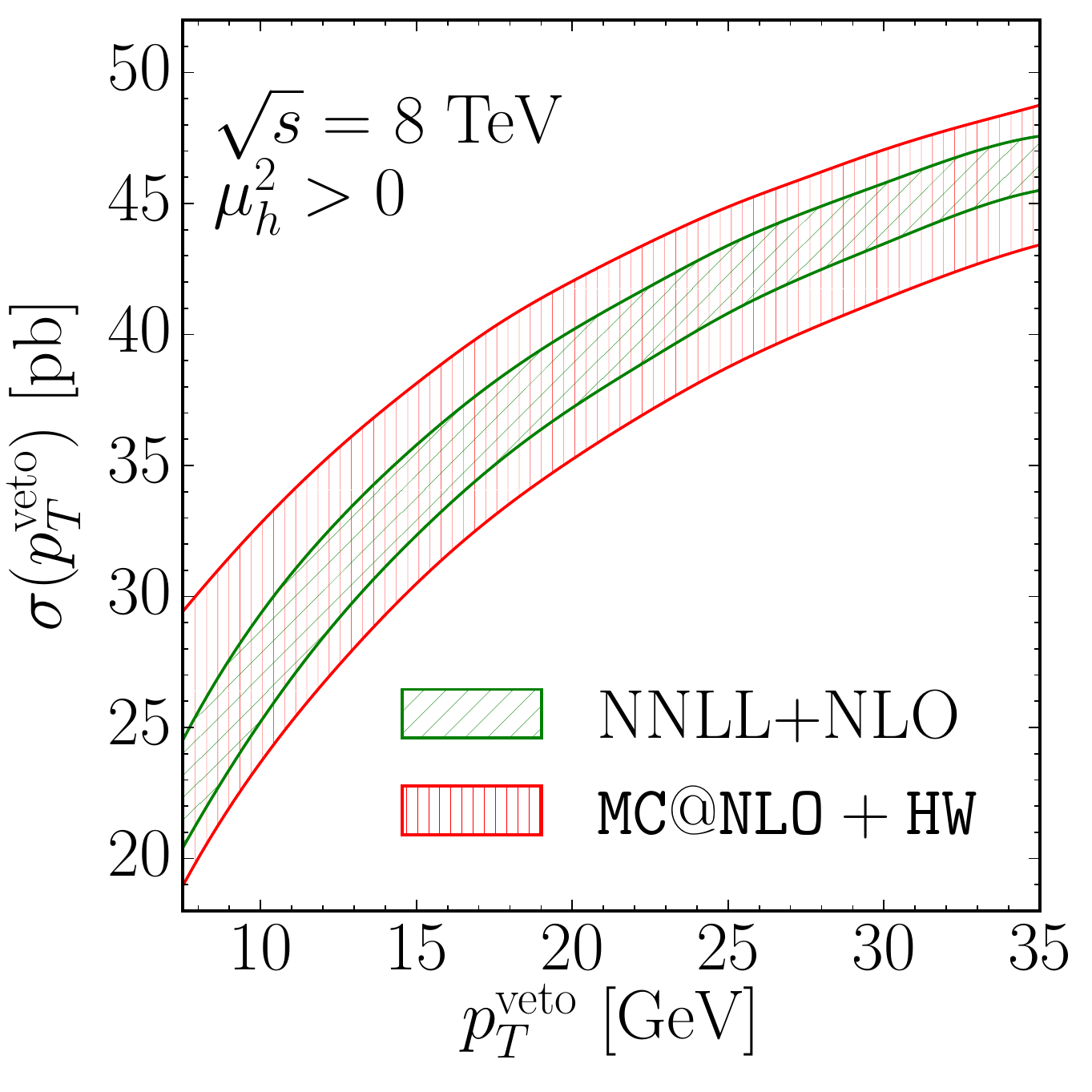}
                \caption{}
                \label{f.Xsection_MCatNLO_8TeV_NoPiSq}
        \end{subfigure}
         \begin{subfigure}[b]{0.32\textwidth}
                \includegraphics[width=\textwidth]{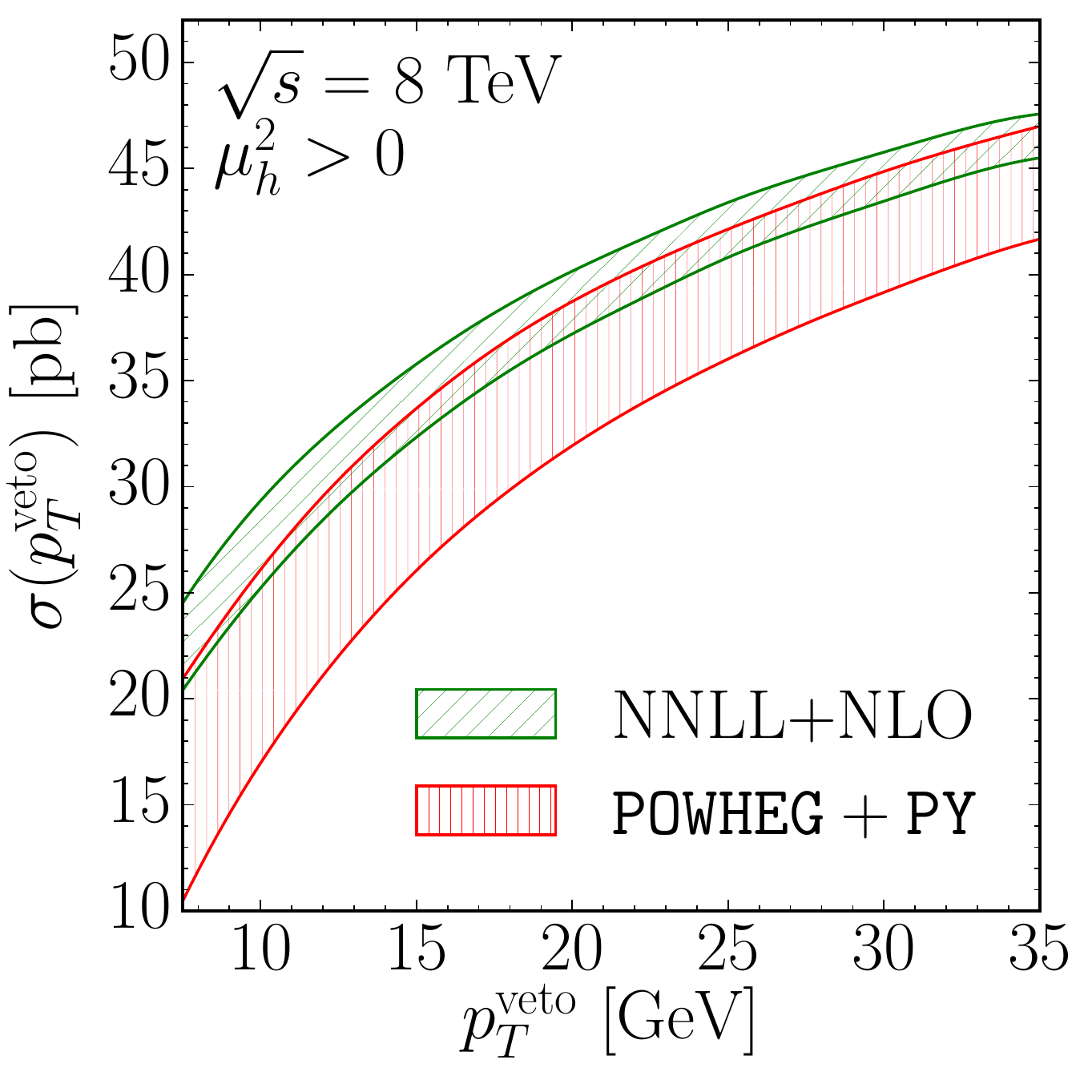}
                \caption{}
                \label{f.Xsection_POWHEG_8TeV_NoPiSq}
        \end{subfigure}
        \caption{Comparison of our NNLL+NLO resummed jet-veto cross-sections \emph{without $\pi^2$ resummation} at the 8-TeV LHC with those from (a) \ttc{MG+PY}, (b) \ttc{MC@NLO+HW}, and (c) \ttc{POWHEG+PY}.}
        \label{f.Xsection_NoPisq_vs_MC+PS}
\end{figure}

For the MC+PS generators, on the other hand, we use the second expression in\erefn{JVeff}, where the inclusive cross-sections in all the 
channels are obtained from \ttc{MCFM} while the jet-veto efficiency 
$\jv{\epsilon}_{q \bar{q}}$ for the $q \bar{q}$ channel are estimated using the respective MC+PS samples. 
As such, there is no well defined procedure to estimate the scale uncertainties
in this case. 
However, as discussed in \sref{intro}, to a good approximation, the uncertainties in 
$\sigma (n_\text{jet} \geq 0)$ and $\sigma (n_\text{jet} \geq 1)$ are expected to be uncorrelated, as their perturbative expansion start at 
different orders in $\al_\s$.  Therefore, as a reasonable estimate, we vary the renormalization and factorization scales 
in the inclusive $W^+ W^-$ production and $W^+ W^- + \text{jet}$ production and add the scale uncertainties from the two 
processes in quadrature to obtain the total scale uncertainty.  

In \fref{Xsection_Pythia}, \fref{Xsection_MCatNLO} and \fref{Xsection_POWHEG}, we have compared our analytical resummed cross-section
and jet-veto efficiency with those obtained from the different MC+PS samples discussed above. We have checked that different choices of PDF 
sets or variation between different eigen-directions within a given PDF set do not affect the jet-veto efficiencies significantly, except when 
comparing LO PDFs to NLO PDFs, for which differences of $\sim 3 \%$ can arise. 
We have also found that the underlying 
events in showering generators have negligible impact on the efficiencies,
and have checked that hadronization effects on the efficiency are $\simlt  1\%$  for $\ptv \simgt 15 \gev$. 
Finally, in \fref{Xsection_NoPisq_vs_MC+PS}, we compare the same MC+PS predictions (in red) with the analytical NNLL+NLO prediction \emph{without $\pi^2$ resummation} (in green). These figures should be compared with \fref{Xsection_Pythia_8TeV}, \fref{Xsection_MCatNLO_8TeV}, and \fref{Xsection_POWHEG_8TeV}, for which $\pi^2$ resummation is included.

To conclude, we observe that {\tt MG+PY} generator tend to produce softer jets compared to {\tt POWHEG+PY} generator which have a 
much harder jet $p_T$ spectrum. The MC+PS predictions in general underestimate the jet-veto cross-sections compared to our best resummed calculations with both the logarithms and $\pi^2$ terms resummed, where the difference is most significant $\sim 11 \%$ for the {\tt POWHEG} samples in the region 
$\ptv \simgt 20 \gev$.  Moreover, our NNLL+NLO results significantly reduce the scale uncertainties by almost a factor of 2.

\subsection{Comparison with Experimental Results}
\label{ss.results}
Even though both ATLAS and CMS experiments present their measurements as the inclusive $\proton \proton \to W^+ W^-$ cross section, 
we have seen that the jet-veto efficiencies they use to extrapolate from the measured jet-veto cross sections to the quoted inclusive cross sections 
suffer from the large logarithms that are not properly resummed by the MC+PS generators.
Since both what they actually measured and what we calculated from SCET are the jet-veto cross section, 
not the inclusive cross section,   
we first must \emph{undo} the jet-veto efficiencies from the inclusive cross sections quoted by the ATLAS and CMS collaborations:   
\beq
\sigma_{WW}^\text{veto} = \sigma_{WW} \times \ep_{WW}^\text{veto}
\,.
\eeq
To estimate the jet-veto efficiency in $W^+ W^-$ production, both ATLAS and CMS experiments rely on MC+PS simulations (will simply be referred to as ``MC'' from now on), 
with a data-to-MC correction factor measured from Drell-Yan process in the $Z$ peak region, so that the jet-veto efficiency is obtained as
\beq
\ep_{WW}^\text{veto} = \frac{\ep_{Z}}{\ep_{Z}^\text{MC}} \times \ep_{WW}^\text{MC}
\,.\label{e.Eff_exp}
\eeq
The reasoning behind multiplying such a scaling factor is that the experimental systematic  uncertainties cancel out when the ratio of 
two MC efficiencies are considered. Throughout this discussion, we will assume this scaling factor to be $1$, as indicated by both 
ATLAS \cite{WW:ATLAS_8TeV} and CMS \cite{WW:CMS_8TeV} experiments. The total inclusive 
cross-section is then estimated as
\beq
\sigma_{WW} = \frac{\cn^\text{obs}-\cn^\text{bkg}}{\epsilon_{WW}^\text{veto} \times \ca_{WW}}
\,,\label{e.InclXs}
\eeq
where $\cn^{obs}$ and $\cn^{bkg}$ are the observed and estimated background number of events, respectively, and $\ca_{WW}$ 
include all the acceptances and efficiencies other than the jet-veto efficiency. 

The ATLAS collaboration uses {\tt MC@NLO} interfaced with \ttc{Herwig6} for their MC samples at $\sqrt{s}=7$~TeV \cite{WW:ATLAS_7TeV}, 
while \ttc{Powheg} interfaced with \ttc{Pythia6} are used to generate MC samples at $\sqrt{s}=8$~TeV \cite{WW:ATLAS_8TeV}.
In both cases, the {\tt CT10nlo} PDFs are used and the jets are clustered 
 using anti-$k_\mathrm{T}$ algorithm with jet radius parameter $R=0.4$. 
 The CMS collaboration, on the other hand, states that ``$q \bar{q} \rightarrow W^+ W^-$'' samples were generated using  LO generator 
 \ttc{MadGraph5} interfaced with \ttc{Pythia6} for showering for both 7-TeV and 8-TeV runs \cite{WW:CMS_7TeV,WW:CMS_8TeV}. 
 The choice of PDFs in their analysis is {\tt CTEQ6L} LO PDFs, and the anti-$k_\mathrm{T}$ jet
 clustering algorithm with $R=0.5$ is used. 
 Assuming CMS uses matched samples in their analysis, we generated $p p \rightarrow W^+ W^- + 0/1/2$ parton samples matched in
  {\tt MadGraph5} and showered in \ttc{Pythia6}, 
  using the default $k_\mathrm{T}$-jet MLM matching scheme. 
 As a final step towards undoing the jet-veto efficiency from the experimental results, we remove the jet-veto uncertainties, which 
 were provided by both ATLAS and CMS as part of their systematic uncertainties. The remaining systematic uncertainties in both experiments are mostly dominated by the experimental uncertainties.  

\begin{center} 
\begin{table}
\begin{minipage}{\textwidth}
\begin{minipage}[b]{0.65\linewidth}
\centering
\begin{tabular}{c|c|c|}
\cline{2-3}
		&  	\multicolumn{2} {c|} {$\sqrt{s}=7\tev$}  	\\ \cline{2-3} 
		& $R = 0.4$		 & $R=0.5$ 		  	\\ 
		& $\ptv=25\gev$	 & $\ptv=30\gev$ 	 	\\ \hline
\multicolumn{1} {|m{2.2cm}|}{ATLAS $\sigma_{WW}^\text{veto}$ [pb]} 
		& $37.9^{+3.8\% + 5.0\%+3.8\%}_{-3.8\% -5.0\% -3.8\%}$		& $-$ 			  \\ \hline 
\multicolumn{1} {|m{2.2cm}|}{CMS $\qquad \sigma_{WW}^\text{veto}$ [pb]} 
		&  $-$ 			 & $41.5^{+3.8\% +7.2\% +2.3\%}_{-3.8\% -7.2\% -2.3\%}$		\\ \hline 
\multicolumn{1} {|m{2.4cm}|}{Theory $\quad \sigma_{WW}^\text{veto}$ [pb]} 
		&$37.4^{+3.8\%}_{-3.0\% }$		 & $39.0^{+2.4\%}_{-2.3\% }$				  	\\ \hline
\multicolumn{1} {|m{2.5cm}|}{Theory $\sigma_{h \rightarrow WW}^\text{veto}$ [pb]} 
		& $2.1^{+13.5\%}_{-11.4\%}$		 &$2.3^{+11.5\%}_{-10.6\%}	$				 \\ \hline  
\end{tabular}
\par\vspace{0pt}
\end{minipage}
\hfill
\begin{minipage}[b]{0.34\linewidth}
\centering
\includegraphics[width=\textwidth]{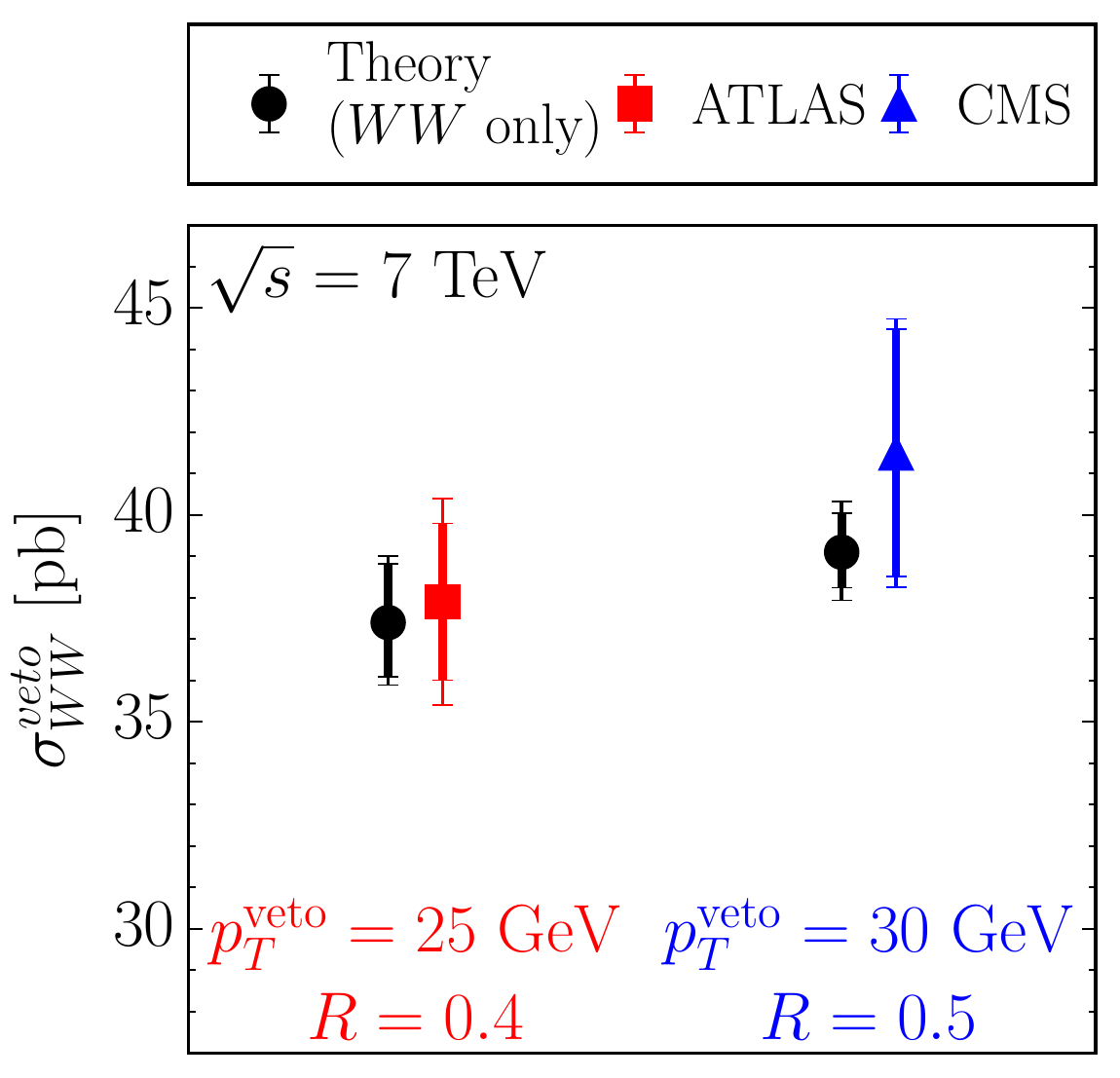}
\par\vspace{0pt}
\end{minipage}
\end{minipage}

\vspace{5mm}
\begin{minipage}{\textwidth}
\begin{minipage}[b]{0.65\linewidth}
\centering
\begin{tabular}{c|c|c|}
\cline{2-3}
		& 	\multicolumn{2} {|c|} {$\sqrt{s}=8\tev$} 	\\ \cline{2-3} 
		& $R = 0.4$ 		 & $R=0.5$			\\ 
		& $\ptv=25\gev$	 & $\ptv=30\gev$		\\ \hline
\multicolumn{1} {|m{2.2cm}|}{ATLAS $\sigma_{WW}^\text{veto}$ [pb]}		  
		&  $48.1^{+1.7\% +6.2\% +3.1\%}_{-1.7\% -5.2\% -2.9\%}$  		& $-$ 			\\ \hline 
\multicolumn{1} {|m{2.2cm}|}{CMS $\qquad \sigma_{WW}^\text{veto}$ [pb]} 			  
		&  	$-$ 			 & $54.2^{+4.0\% +6.5\% +4.4\%}_{-4.0\% -6.5\% -4.4\%}$ 		\\ \hline 
\multicolumn{1} {|m{2.4cm}|}{Theory $\quad \sigma_{WW}^\text{veto}$ [pb]} 			  
		&$44.7^{+3.5\%}_{-2.8\% }$   		 & $46.6^{+2.2\%}_{-2.1\% }$					\\ \hline
\multicolumn{1} {|m{2.5cm}|}{Theory $\sigma_{h \rightarrow WW}^\text{veto}$ [pb]} 			 
		&$2.6^{+13.3\%}_{-11.7\%}$   		 &$2.9^{+11.5\%}_{-11.5\%}$					 \\ \hline  
\end{tabular}
\par\vspace{0pt}
\end{minipage}
\hfill
\begin{minipage}[b]{0.34\linewidth}
\centering
\includegraphics[width=\textwidth]{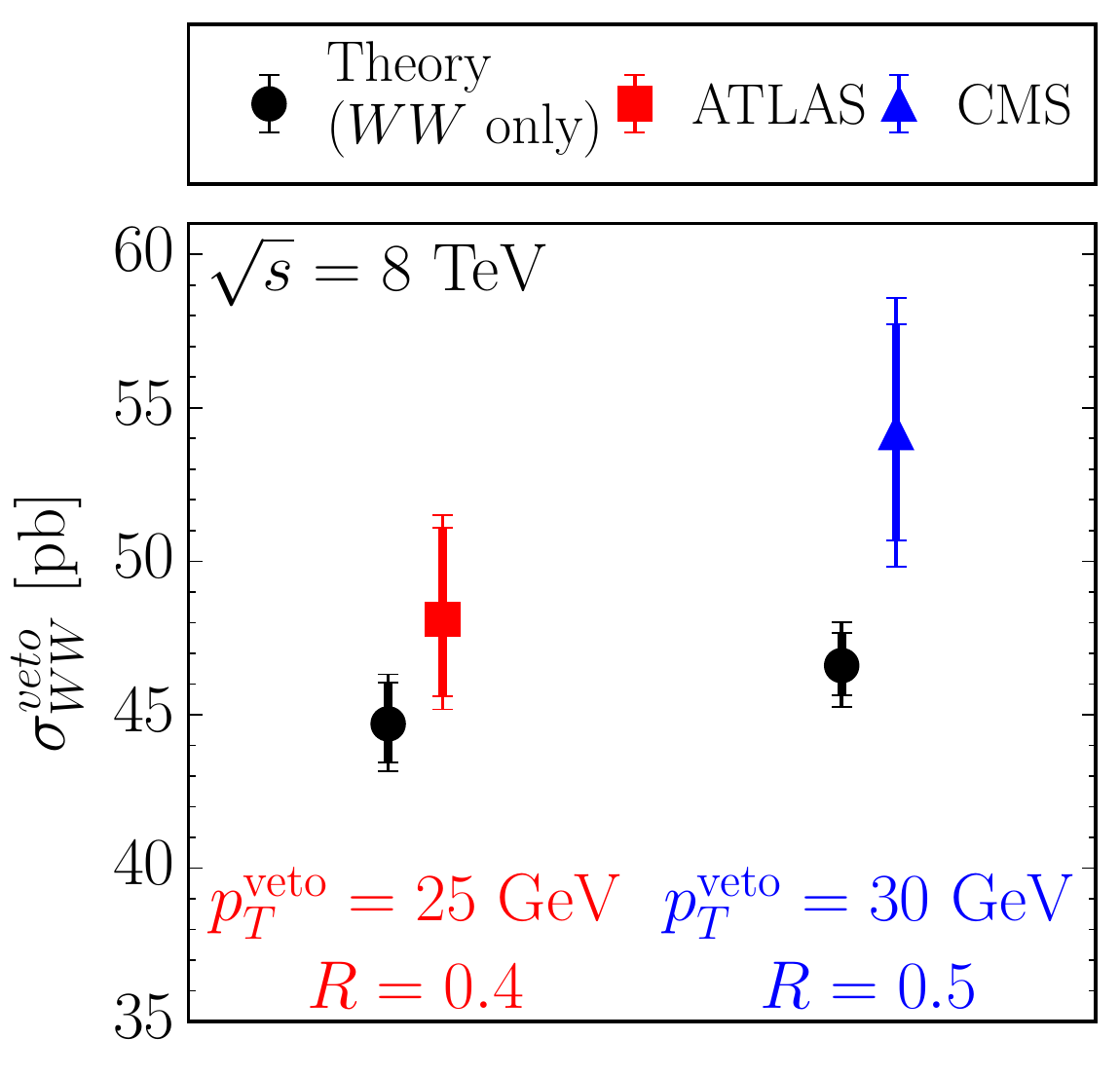}
\par\vspace{0pt}
\end{minipage}
\end{minipage}
\caption{Comparison of our theory predictions for jet-veto cross-section with those measured by the
ATLAS and CMS experiments at $\sqrt{s}=7$- and $8$-TeV LHC runs. The Higgs jet-veto cross-sections 
are taken from \cite{Higgs:veto}.
As in the rest of the paper, the scale uncertainties in the theory predictions here correspond to the standard convention of varying $\muh$ and $\muf$ by a factor of 2 above and below $M$ and $\ptv$, respectively.
It should be noted that they may be somewhat smaller than the theory uncertainties  
estimated from comparing the NLL to NNLL calculations in \fref{Xsection_7TeV} and \fref{Xsection_8TeV}.}
\label{t.Results}
\end{table}
\end{center}

In \tref{Results}, we have shown a comparison of our resummed theory predictions for jet-veto cross-section with those measured by 
the ATLAS and CMS experiments at $\sqrt{s}=7$ and $8 \tev$ LHC runs. 
The experimental veto cross-sections were 
obtained by removing the jet-veto efficiency factors from their reported inclusive cross-sections while simultaneously 
removing the jet-veto uncertainties from their respective systematic uncertainties, as explained above. The errors for the 
theory results represent the scale uncertainties while the PDF uncertainties are estimated to be $\sim 2 \%$. The errors 
in the experimental cross-sections are the scale, systematic (excluding the jet-veto uncertainty) and luminosity uncertainties 
in that order. On the right-hand-side column of \tref{Results}, we give a visual presentation of the error bars, with the thick lines 
for theory (experiment) representing the scale (systematic) uncertainties while the thin lines represent the total error in both 
the cases. The jet-veto cross-sections from the Higgs production \cite{Higgs:veto, Becher:2012qa, Becher:2013xia} are also 
shown in the table, but we refrain from adding
them to our theory prediction since one of the $W$ in Higgs decays is off-shell leading to softer leptons in general, which 
may not have the same acceptance as those from the on-shell $W$ decays. 
Excluding the Higgs contribution, our resummed theory 
predictions for jet-veto cross-sections are in excellent agreement with experiments at $\sqrt{s} = 7$~TeV for 
both ATLAS and CMS\@. At $\sqrt{s} = 8$~TeV, a slight discrepancy of $\sim 1 \sigma$ is present for the CMS 
experiment, while the ATLAS result is compatible with our prediction with a discrepancy $< 1 \sigma$. We conclude this 
section by making a few remarks:
\begin{itemize}
\item{
While our resummed calculations for the $WW$ jet-veto cross-section as well as estimates of the scale uncertainties are extremely robust, 
our reinterpretation of the experimental results to obtain the jet-veto cross-sections may not be. In particular, both ATLAS and 
CMS collaborations use specific tunes for PS generators,
while we have used the default tunes that come with the packages. 
Reproducing those tunes, however, is beyond the scope of this work.
It would be beneficial to both theoretical and experimental communities if the jet-veto cross-sections were directly presented by 
the collaborations.
} 
\item{
Although the leptons from the Higgs decay $h \rightarrow W W^* \rightarrow 2 \ell 2\nu$ are expected to be softer compared
to those from on-shell $W$-pair production, there would be some contamination from this channel. It is conceivable
that Higgs decays could lead to a further increase by $\sim 1$ -- $2$ pb in the theory prediction.
}
\item{
The $gg \rightarrow WW$ process without involving the higgs, which are considered at the LO without resummation in our work, is 
estimated to be $\sim 3\%$ at $\sqrt{s} \sim 8$~TeV\@. 
The NLO contributions to this channel can further influence the theory 
prediction and needs to be studied.
}   
\item{
Finally and possibly most importantly, we would like to point out that some of the background processes to $W$ pair production may also have been 
incorrectly estimated from the MC+PS simulations in the 0-jet bin. 
This applies to many di-boson backgrounds that are purely 
estimated from fixed-order MC, but also some of the data-driven methods such as $t \bar{t}$ and $t W$, which too rely on MC partially.
This may be particularly important for the slight discrepancy between our calculation and the 8-TeV CMS measurement.
}
\end{itemize}
A more detailed study of $WW$ differential cross-sections with the inclusion of $W$ leptonic decays%
\footnote{
For a study of jet-veto resummation for 
$gg \rightarrow h \rightarrow 2 \ell 2 \nu$ interfering with the 
$gg \rightarrow WW \rightarrow 2 \ell 2 \nu$ process, see \cite{HiggsIntf:veto}.
}
in the zero-jet bin will be presented in a future
publication and a public code will be made available shortly.  
A similar study for 13- and 14-TeV LHC runs will be also presented in our future work.

{\bf Note added in version 4:} 
Through private communication with T.~Becher and L.~Rothen and a detailed numerical comparison, we found a minor bug in the NNLL beam functions in our code. 
Fixing the bug has resulted in slightly better convergence between the NLL and NNLL resummations in Figures~\ref{f.Xsection} and~\ref{f.diff_xsection}.
The effects of the bug are nearly undone by the corresponding changes of power corrections, 
so our final results (Figures~\ref{f.NLO_vs_NNLL}--\ref{f.Xsection_NoPisq_vs_MC+PS} as well as  Table~\ref{t.Results}) are all practically unchanged, differing at most by 0.5\% from those in the published version of our paper. 
We should also note that since the new power corrections are too small to plot, we have removed the related figures from section~\ref{ss.power_corrections} and streamlined the discussion on the power corrections.

\subsection*{Acknowledgment}
We would like to thank David Curtin, Sally Dawson, Marat Freytsis, Ian Lewis, Seth Quackenbush, Gavin Salam, Nobuo Sato, 
Ding Yu Shao, Martin Schmaltz, Frank Tackmann and Jonathan Walsh for discussions and/or comments on the manuscript. 
We would also like to thank \ttc{MadGraph} team for answering our queries regarding the {\tt MadGraph\_aMC@NLO} package. 
This work was supported by the US Department of Energy under grant DE-FG02-13ER41942,
and also in part by the National Science Foundation under Grant No.~PHY-1066293 and the hospitality of the Aspen Center of Physics.


\begin{thebibliography}{10}

\bibitem{WW:ATLAS_7TeV}
{\bf ATLAS Collaboration} Collaboration, G.~Aad {\em et.~al.}, {\it
  {Measurement of $W^+W^-$ production in pp collisions at $\sqrt{s}$=7 TeV with
  the ATLAS detector and limits on anomalous $WWZ$ and $WW\gamma$ couplings}},
  {\em Phys.Rev.} {\bf D87} (2013), no.~11 112001,
  [\href{http://xxx.lanl.gov/abs/1210.2979}{{\tt arXiv:1210.2979}}].

\bibitem{WW:CMS_7TeV}
{\bf CMS Collaboration} Collaboration, S.~Chatrchyan {\em et.~al.}, {\it
  {Measurement of the $W^+W^-$ Cross section in $pp$ Collisions at $\sqrt{s} =
  7$ TeV and Limits on Anomalous $WW\gamma$ and $WWZ$ couplings}},  {\em
  Eur.Phys.J.} {\bf C73} (2013) 2610,
  [\href{http://xxx.lanl.gov/abs/1306.1126}{{\tt arXiv:1306.1126}}].

\bibitem{WW:ATLAS_8TeV}
{\it {Measurement of the $W^+W^-$ production cross section in proton-proton
  collisions at $\sqrt{s} =8$ TeV with the ATLAS detector}},  Tech. Rep.
  ATLAS-CONF-2014-033, CERN, Geneva, Jul, 2014.

\bibitem{WW:CMS_8TeV}
{\bf CMS Collaboration} Collaboration, S.~Chatrchyan {\em et.~al.}, {\it
  {Measurement of W+W- and ZZ production cross sections in pp collisions at
  sqrt(s) = 8 TeV}},  {\em Phys.Lett.} {\bf B721} (2013) 190--211,
  [\href{http://xxx.lanl.gov/abs/1301.4698}{{\tt arXiv:1301.4698}}].

\bibitem{HWW_LHC_ATLAS}
{\bf ATLAS Collaboration} Collaboration, G.~Aad {\em et.~al.}, {\it
  {Measurements of Higgs boson production and couplings in diboson final states
  with the ATLAS detector at the LHC}},  {\em Phys.Lett.} {\bf B726} (2013)
  88--119, [\href{http://xxx.lanl.gov/abs/1307.1427}{{\tt arXiv:1307.1427}}].

\bibitem{HWW_LHC_CMS}
{\bf CMS Collaboration} Collaboration, S.~Chatrchyan {\em et.~al.}, {\it
  {Search for the standard model Higgs boson produced in association with a $W$
  or a $Z$ boson and decaying to bottom quarks}},  {\em Phys.Rev.} {\bf D89}
  (2014) 012003, [\href{http://xxx.lanl.gov/abs/1310.3687}{{\tt
  arXiv:1310.3687}}].

\bibitem{MCFM1}
J.~M. Campbell and R.~K. Ellis, {\it {An update on vector boson pair production
  at hadron colliders}},  {\em Phys.Rev.} {\bf D60} (1999) 113006,
  [\href{http://xxx.lanl.gov/abs/hep-ph/9905386}{{\tt hep-ph/9905386}}].

\bibitem{MCFM2}
J.~M. Campbell, R.~K. Ellis, and C.~Williams, {\it {Vector boson pair
  production at the LHC}},  {\em JHEP} {\bf 1107} (2011) 018,
  [\href{http://xxx.lanl.gov/abs/1105.0020}{{\tt arXiv:1105.0020}}].

\bibitem{MSTW}
A.~Martin, W.~Stirling, R.~Thorne, and G.~Watt, {\it {Parton distributions for
  the LHC}},  {\em Eur.Phys.J.} {\bf C63} (2009) 189--285,
  [\href{http://xxx.lanl.gov/abs/0901.0002}{{\tt arXiv:0901.0002}}].

\bibitem{CT10}
H.-L. Lai, M.~Guzzi, J.~Huston, Z.~Li, P.~M. Nadolsky, {\em et.~al.}, {\it {New
  parton distributions for collider physics}},  {\em Phys.Rev.} {\bf D82}
  (2010) 074024, [\href{http://xxx.lanl.gov/abs/1007.2241}{{\tt
  arXiv:1007.2241}}].

\bibitem{WW:NewPhysics1}
B.~Feigl, H.~Rzehak, and D.~Zeppenfeld, {\it {New physics backgrounds to the $H
  \to WW$ search at the LHC?}},  {\em Phys.Lett.} {\bf B717} (2012) 390--395,
  [\href{http://xxx.lanl.gov/abs/1205.3468}{{\tt arXiv:1205.3468}}].

\bibitem{WW:NewPhysics2}
D.~Curtin, P.~Jaiswal, and P.~Meade, {\it {Charginos hiding in plain sight}},
  {\em Phys.Rev.} {\bf D87} (2013), no.~3 031701,
  [\href{http://xxx.lanl.gov/abs/1206.6888}{{\tt arXiv:1206.6888}}].

\bibitem{WW:NewPhysics3}
P.~Jaiswal, K.~Kopp, and T.~Okui, {\it {Higgs production amidst the LHC
  detector}},  {\em Phys.Rev.} {\bf D87} (2013), no.~11 115017,
  [\href{http://xxx.lanl.gov/abs/1303.1181}{{\tt arXiv:1303.1181}}].

\bibitem{WW:NewPhysics4}
K.~Rolbiecki and K.~Sakurai, {\it {Light stops emerging in WW cross section
  measurements?}},  {\em JHEP} {\bf 1309} (2013) 004,
  [\href{http://xxx.lanl.gov/abs/1303.5696}{{\tt arXiv:1303.5696}}].

\bibitem{WW:NewPhysics5}
D.~Curtin, P.~Jaiswal, P.~Meade, and P.-J. Tien, {\it {Casting light on BSM
  physics with SM standard candles}},  {\em JHEP} {\bf 1308} (2013) 068,
  [\href{http://xxx.lanl.gov/abs/1304.7011}{{\tt arXiv:1304.7011}}].

\bibitem{WW:NewPhysics6}
D.~Curtin, P.~Meade, and P.-J. Tien, {\it {Natural SUSY in Plain Sight}},
  \href{http://xxx.lanl.gov/abs/1406.0848}{{\tt arXiv:1406.0848}}.

\bibitem{WW:nlo1}
J.~Ohnemus, {\it {An order $\alpha_s$ calculation of hadronic $W^{-} W^{+}$
  production}},  {\em Phys.Rev.} {\bf D44} (1991) 1403--1414.

\bibitem{WW:nlo2}
S.~Frixione, {\it {A next-to-leading order calculation of the cross-section for
  the production of $W^+ W^-$ pairs in hadronic collisions}},  {\em Nucl.Phys.}
  {\bf B410} (1993) 280--324.

\bibitem{WW:nloLepton1}
L.~J. Dixon, Z.~Kunszt, and A.~Signer, {\it {Helicity amplitudes for
  O($\alpha_s$) production of $W^{+} W^{-}$, $W^\pm Z$, $Z Z$, $W^\pm \gamma$,
  or $Z \gamma$ pairs at hadron colliders}},  {\em Nucl.Phys.} {\bf B531}
  (1998) 3--23, [\href{http://xxx.lanl.gov/abs/hep-ph/9803250}{{\tt
  hep-ph/9803250}}].

\bibitem{WW:nloLepton2}
J.~Ohnemus, {\it {Hadronic $Z Z$, $W^{-} W^{+}$, and $W^\pm Z$ production with
  QCD corrections and leptonic decays}},  {\em Phys.Rev.} {\bf D50} (1994)
  1931--1945, [\href{http://xxx.lanl.gov/abs/hep-ph/9403331}{{\tt
  hep-ph/9403331}}].

\bibitem{ggWW1}
D.~A. Dicus, C.~Kao, and W.~Repko, {\it {Gluon Production of Gauge Bosons}},
  {\em Phys.Rev.} {\bf D36} (1987) 1570.

\bibitem{ggWW2}
E.~N. Glover and J.~van~der Bij, {\it {Vectpr boson pair production via gluon
  fusion}},  {\em Phys.Lett.} {\bf B219} (1989) 488.

\bibitem{ggWW3}
T.~Binoth, M.~Ciccolini, N.~Kauer, and M.~Kramer, {\it {Gluon-induced WW
  background to Higgs boson searches at the LHC}},  {\em JHEP} {\bf 0503}
  (2005) 065, [\href{http://xxx.lanl.gov/abs/hep-ph/0503094}{{\tt
  hep-ph/0503094}}].

\bibitem{WW:ew1}
A.~Bierweiler, T.~Kasprzik, J.~H. KÃŒhn, and S.~Uccirati, {\it {Electroweak
  corrections to $W$-boson pair production at the LHC}},  {\em JHEP} {\bf 1211}
  (2012) 093, [\href{http://xxx.lanl.gov/abs/1208.3147}{{\tt
  arXiv:1208.3147}}].

\bibitem{WW:ew2}
J.~Baglio, L.~D. Ninh, and M.~M. Weber, {\it {Massive gauge boson pair
  production at the LHC: a next-to-leading order story}},  {\em Phys.Rev.} {\bf
  D88} (2013) 113005, [\href{http://xxx.lanl.gov/abs/1307.4331}{{\tt
  arXiv:1307.4331}}].

\bibitem{WW:threshold}
S.~Dawson, I.~M. Lewis, and M.~Zeng, {\it {Threshold resummed and approximate
  next-to-next-to-leading order results for $W^+W^-$ pair production at the
  LHC}},  {\em Phys.Rev.} {\bf D88} (2013), no.~5 054028,
  [\href{http://xxx.lanl.gov/abs/1307.3249}{{\tt arXiv:1307.3249}}].

\bibitem{WW:MC1}
T.~Melia, P.~Nason, R.~Rontsch, and G.~Zanderighi, {\it {$W^+W^-$, $WZ$ and
  $ZZ$ production in the POWHEG BOX}},  {\em JHEP} {\bf 1111} (2011) 078,
  [\href{http://xxx.lanl.gov/abs/1107.5051}{{\tt arXiv:1107.5051}}].

\bibitem{WW:MC2}
K.~Hamilton, {\it {A positive-weight next-to-leading order simulation of weak
  boson pair production}},  {\em JHEP} {\bf 1101} (2011) 009,
  [\href{http://xxx.lanl.gov/abs/1009.5391}{{\tt arXiv:1009.5391}}].

\bibitem{WW:MC3}
F.~Cascioli, S.~H\"oche, F.~Krauss, P.~Maierhfer, S.~Pozzorini, {\em et.~al.},
  {\it {Precise Higgs-background predictions: merging NLO QCD and squared
  quark-loop corrections to four-lepton + 0,1 jet production}},  {\em JHEP}
  {\bf 1401} (2014) 046, [\href{http://xxx.lanl.gov/abs/1309.0500}{{\tt
  arXiv:1309.0500}}].

\bibitem{WW:pTresum1}
M.~Grazzini, {\it {Soft-gluon effects in WW production at hadron colliders}},
  {\em JHEP} {\bf 0601} (2006) 095,
  [\href{http://xxx.lanl.gov/abs/hep-ph/0510337}{{\tt hep-ph/0510337}}].

\bibitem{WW:pTresum2}
Y.~Wang, C.~S. Li, Z.~L. Liu, D.~Y. Shao, and H.~T. Li, {\it
  {Transverse-Momentum Resummation for Gauge Boson Pair Production at the
  Hadron Collider}},  {\em Phys.Rev.} {\bf D88} (2013) 114017,
  [\href{http://xxx.lanl.gov/abs/1307.7520}{{\tt arXiv:1307.7520}}].

\bibitem{0Jet}
I.~W. Stewart and F.~J. Tackmann, {\it {Theory Uncertainties for Higgs and
  Other Searches Using Jet Bins}},  {\em Phys.Rev.} {\bf D85} (2012) 034011,
  [\href{http://xxx.lanl.gov/abs/1107.2117}{{\tt arXiv:1107.2117}}].

\bibitem{WW:nloVeto}
F.~Campanario, M.~Rauch, and S.~Sapeta, {\it {$W^+W^-$ production at high
  transverse momenta beyond NLO}},  {\em Nucl.Phys.} {\bf B879} (2014) 65--79,
  [\href{http://xxx.lanl.gov/abs/1309.7293}{{\tt arXiv:1309.7293}}].

\bibitem{pisquare_1}
G.~Parisi, {\it {Summing Large Perturbative Corrections in QCD}},  {\em
  Phys.Lett.} {\bf B90} (1980) 295.

\bibitem{pisquare_2}
G.~F. Sterman, {\it {Summation of Large Corrections to Short Distance Hadronic
  Cross-Sections}},  {\em Nucl.Phys.} {\bf B281} (1987) 310.

\bibitem{pisquare_3}
L.~Magnea and G.~F. Sterman, {\it {Analytic continuation of the Sudakov
  form-factor in QCD}},  {\em Phys.Rev.} {\bf D42} (1990) 4222--4227.

\bibitem{beam-func-1}
I.~W. Stewart, F.~J. Tackmann, and W.~J. Waalewijn, {\it {Factorization at the
  LHC: From PDFs to Initial State Jets}},  {\em Phys.Rev.} {\bf D81} (2010)
  094035, [\href{http://xxx.lanl.gov/abs/0910.0467}{{\tt arXiv:0910.0467}}].

\bibitem{beam-func-2}
I.~W. Stewart, F.~J. Tackmann, and W.~J. Waalewijn, {\it {The Quark Beam
  Function at NNLL}},  {\em JHEP} {\bf 1009} (2010) 005,
  [\href{http://xxx.lanl.gov/abs/1002.2213}{{\tt arXiv:1002.2213}}].

\bibitem{Higgs:veto1}
C.~F. Berger, C.~Marcantonini, I.~W. Stewart, F.~J. Tackmann, and W.~J.
  Waalewijn, {\it {Higgs Production with a Central Jet Veto at NNLL+NNLO}},
  {\em JHEP} {\bf 1104} (2011) 092,
  [\href{http://xxx.lanl.gov/abs/1012.4480}{{\tt arXiv:1012.4480}}].

\bibitem{Higgs:veto2}
A.~Banfi, G.~P. Salam, and G.~Zanderighi, {\it {NLL+NNLO predictions for
  jet-veto efficiencies in Higgs-boson and Drell-Yan production}},  {\em JHEP}
  {\bf 1206} (2012) 159, [\href{http://xxx.lanl.gov/abs/1203.5773}{{\tt
  arXiv:1203.5773}}].

\bibitem{Becher:2012qa}
T.~Becher and M.~Neubert, {\it {Factorization and NNLL Resummation for Higgs
  Production with a Jet Veto}},  {\em JHEP} {\bf 1207} (2012) 108,
  [\href{http://xxx.lanl.gov/abs/1205.3806}{{\tt arXiv:1205.3806}}].

\bibitem{logR:1}
F.~J. Tackmann, J.~R. Walsh, and S.~Zuberi, {\it {Resummation Properties of Jet
  Vetoes at the LHC}},  {\em Phys.Rev.} {\bf D86} (2012) 053011,
  [\href{http://xxx.lanl.gov/abs/1206.4312}{{\tt arXiv:1206.4312}}].

\bibitem{Higgs:veto}
A.~Banfi, P.~F. Monni, G.~P. Salam, and G.~Zanderighi, {\it {Higgs and Z-boson
  production with a jet veto}},  {\em Phys.Rev.Lett.} {\bf 109} (2012) 202001,
  [\href{http://xxx.lanl.gov/abs/1206.4998}{{\tt arXiv:1206.4998}}].

\bibitem{Becher:2013xia}
T.~Becher, M.~Neubert, and L.~Rothen, {\it {Factorization and
  $N^{3}LL_{p}$+NNLO predictions for the Higgs cross section with a jet veto}},
   {\em JHEP} {\bf 1310} (2013) 125,
  [\href{http://xxx.lanl.gov/abs/1307.0025}{{\tt arXiv:1307.0025}}].

\bibitem{most-complete}
I.~W. Stewart, F.~J. Tackmann, J.~R. Walsh, and S.~Zuberi, {\it {Jet $p_T$
  Resummation in Higgs Production at $NNLL'+NNLO$}},  {\em Phys.Rev.} {\bf D89}
  (2014) 054001, [\href{http://xxx.lanl.gov/abs/1307.1808}{{\tt
  arXiv:1307.1808}}].

\bibitem{logR:2}
S.~Alioli and J.~R. Walsh, {\it {Jet Veto Clustering Logarithms Beyond Leading
  Order}},  {\em JHEP} {\bf 1403} (2014) 119,
  [\href{http://xxx.lanl.gov/abs/1311.5234}{{\tt arXiv:1311.5234}}].

\bibitem{mass-effects}
A.~Banfi, P.~F. Monni, and G.~Zanderighi, {\it {Quark masses in Higgs
  production with a jet veto}},  {\em JHEP} {\bf 1401} (2014) 097,
  [\href{http://xxx.lanl.gov/abs/1308.4634}{{\tt arXiv:1308.4634}}].

\bibitem{VH-1}
D.~Y. Shao, C.~S. Li, and H.~T. Li, {\it {Resummation Prediction on Higgs and
  Vector Boson Associated Production with a Jet Veto at the LHC}},  {\em JHEP}
  {\bf 1402} (2014) 117, [\href{http://xxx.lanl.gov/abs/1309.5015}{{\tt
  arXiv:1309.5015}}].

\bibitem{VH-2}
Y.~Li and X.~Liu, {\it {High precision predictions for exclusive $VH$
  production at the LHC}},  {\em JHEP} {\bf 1406} (2014) 028,
  [\href{http://xxx.lanl.gov/abs/1401.2149}{{\tt arXiv:1401.2149}}].

\bibitem{tagged-jets-1}
X.~Liu and F.~Petriello, {\it {Resummation of jet-veto logarithms in hadronic
  processes containing jets}},  {\em Phys.Rev.} {\bf D87} (2013) 014018,
  [\href{http://xxx.lanl.gov/abs/1210.1906}{{\tt arXiv:1210.1906}}].

\bibitem{tagged-jets-2}
X.~Liu and F.~Petriello, {\it {Reducing theoretical uncertainties for exclusive
  Higgs-boson plus one-jet production at the LHC}},  {\em Phys.Rev.} {\bf D87}
  (2013), no.~9 094027, [\href{http://xxx.lanl.gov/abs/1303.4405}{{\tt
  arXiv:1303.4405}}].

\bibitem{tagged-jets-3}
R.~Boughezal, X.~Liu, F.~Petriello, F.~J. Tackmann, and J.~R. Walsh, {\it
  {Combining Resummed Higgs Predictions Across Jet Bins}},  {\em Phys.Rev.}
  {\bf D89} (2014) 074044, [\href{http://xxx.lanl.gov/abs/1312.4535}{{\tt
  arXiv:1312.4535}}].

\bibitem{future-jet-veto}
R.~Boughezal, C.~Focke, Y.~Li, and X.~Liu, {\it {Jet vetoes for Higgs
  production at future hadron colliders}},
  \href{http://xxx.lanl.gov/abs/1405.4562}{{\tt arXiv:1405.4562}}.

\bibitem{Bauer_SCET_1}
C.~W. Bauer, S.~Fleming, and M.~E. Luke, {\it {Summing Sudakov logarithms in $B
  \to X_s \gamma$ in effective field theory}},  {\em Phys.Rev.} {\bf D63}
  (2000) 014006, [\href{http://xxx.lanl.gov/abs/hep-ph/0005275}{{\tt
  hep-ph/0005275}}].

\bibitem{Bauer_SCET_2}
C.~W. Bauer, S.~Fleming, D.~Pirjol, and I.~W. Stewart, {\it {An Effective field
  theory for collinear and soft gluons: Heavy to light decays}},  {\em
  Phys.Rev.} {\bf D63} (2001) 114020,
  [\href{http://xxx.lanl.gov/abs/hep-ph/0011336}{{\tt hep-ph/0011336}}].

\bibitem{Bauer_SCET_3}
C.~W. Bauer and I.~W. Stewart, {\it {Invariant operators in collinear effective
  theory}},  {\em Phys.Lett.} {\bf B516} (2001) 134--142,
  [\href{http://xxx.lanl.gov/abs/hep-ph/0107001}{{\tt hep-ph/0107001}}].

\bibitem{Bauer_SCET_4}
C.~W. Bauer, D.~Pirjol, and I.~W. Stewart, {\it {Soft collinear factorization
  in effective field theory}},  {\em Phys.Rev.} {\bf D65} (2002) 054022,
  [\href{http://xxx.lanl.gov/abs/hep-ph/0109045}{{\tt hep-ph/0109045}}].

\bibitem{Beneke_SCET_1}
M.~Beneke, A.~Chapovsky, M.~Diehl, and T.~Feldmann, {\it {Soft collinear
  effective theory and heavy to light currents beyond leading power}},  {\em
  Nucl.Phys.} {\bf B643} (2002) 431--476,
  [\href{http://xxx.lanl.gov/abs/hep-ph/0206152}{{\tt hep-ph/0206152}}].

\bibitem{Beneke_SCET_2}
M.~Beneke and T.~Feldmann, {\it {Multipole expanded soft collinear effective
  theory with non-Abelian gauge symmetry}},  {\em Phys.Lett.} {\bf B553} (2003)
  267--276, [\href{http://xxx.lanl.gov/abs/hep-ph/0211358}{{\tt
  hep-ph/0211358}}].

\bibitem{Luke_SCET}
S.~M. Freedman and M.~Luke, {\it {SCET, QCD and Wilson lines}},  {\em
  Phys.Rev.} {\bf D85} (2012) 014003,
  [\href{http://xxx.lanl.gov/abs/1107.5823}{{\tt arXiv:1107.5823}}].

\bibitem{Matt_SCET_1}
I.~Feige and M.~D. Schwartz, {\it {An on-shell approach to factorization}},
  {\em Phys.Rev.} {\bf D88} (2013), no.~6 065021,
  [\href{http://xxx.lanl.gov/abs/1306.6341}{{\tt arXiv:1306.6341}}].

\bibitem{Matt_SCET_2}
I.~Feige and M.~D. Schwartz, {\it {Hard-soft-collinear factorization to all
  orders}},  \href{http://xxx.lanl.gov/abs/1403.6472}{{\tt arXiv:1403.6472}}.

\bibitem{Sterman:2004pd}
G.~F. Sterman, {\it {QCD and jets}},
  \href{http://xxx.lanl.gov/abs/hep-ph/0412013}{{\tt hep-ph/0412013}}.

\bibitem{analytic_reg}
T.~Becher and G.~Bell, {\it {Analytic Regularization in Soft-Collinear
  Effective Theory}},  {\em Phys.Lett.} {\bf B713} (2012) 41--46,
  [\href{http://xxx.lanl.gov/abs/1112.3907}{{\tt arXiv:1112.3907}}].

\bibitem{Chiu:2012ir}
J.-Y. Chiu, A.~Jain, D.~Neill, and I.~Z. Rothstein, {\it {A Formalism for the
  Systematic Treatment of Rapidity Logarithms in Quantum Field Theory}},  {\em
  JHEP} {\bf 1205} (2012) 084, [\href{http://xxx.lanl.gov/abs/1202.0814}{{\tt
  arXiv:1202.0814}}].

\bibitem{Becher:2010tm}
T.~Becher and M.~Neubert, {\it {Drell-Yan production at small $q_T$, transverse
  parton distributions and the collinear anomaly}},  {\em Eur.Phys.J.} {\bf
  C71} (2011) 1665, [\href{http://xxx.lanl.gov/abs/1007.4005}{{\tt
  arXiv:1007.4005}}].

\bibitem{Manohar:2006nz}
A.~V. Manohar and I.~W. Stewart, {\it {The Zero-Bin and Mode Factorization in
  Quantum Field Theory}},  {\em Phys.Rev.} {\bf D76} (2007) 074002,
  [\href{http://xxx.lanl.gov/abs/hep-ph/0605001}{{\tt hep-ph/0605001}}].

\bibitem{Hill:2002vw}
R.~J. Hill and M.~Neubert, {\it {Spectator interactions in soft collinear
  effective theory}},  {\em Nucl.Phys.} {\bf B657} (2003) 229--256,
  [\href{http://xxx.lanl.gov/abs/hep-ph/0211018}{{\tt hep-ph/0211018}}].

\bibitem{Manohar:2002fd}
A.~V. Manohar, T.~Mehen, D.~Pirjol, and I.~W. Stewart, {\it {Reparameterization
  invariance for collinear operators}},  {\em Phys.Lett.} {\bf B539} (2002)
  59--66, [\href{http://xxx.lanl.gov/abs/hep-ph/0204229}{{\tt
  hep-ph/0204229}}].

\bibitem{Sundrum:1997ut}
R.~Sundrum, {\it {Reparameterization invariance to all orders in heavy quark
  effective theory}},  {\em Phys.Rev.} {\bf D57} (1998) 331--336,
  [\href{http://xxx.lanl.gov/abs/hep-ph/9704256}{{\tt hep-ph/9704256}}].

\bibitem{Collins:1981ta}
J.~C. Collins and G.~F. Sterman, {\it {Soft Partons in {QCD}}},  {\em
  Nucl.Phys.} {\bf B185} (1981) 172.

\bibitem{Bodwin:1981fv}
G.~T. Bodwin, S.~J. Brodsky, and G.~P. Lepage, {\it {Initial State Interactions
  and the Drell-Yan Process}},  {\em Phys.Rev.Lett.} {\bf 47} (1981) 1799.

\bibitem{Bauer:2010cc}
C.~W. Bauer, B.~O. Lange, and G.~Ovanesyan, {\it {On Glauber modes in
  Soft-Collinear Effective Theory}},  {\em JHEP} {\bf 1107} (2011) 077,
  [\href{http://xxx.lanl.gov/abs/1010.1027}{{\tt arXiv:1010.1027}}].

\bibitem{glauber_cancel_1}
J.~C. Collins, D.~E. Soper, and G.~F. Sterman, {\it {Factorization for One Loop
  Corrections in the {Drell-Yan} Process}},  {\em Nucl.Phys.} {\bf B223} (1983)
  381.

\bibitem{glauber_cancel_2}
G.~T. Bodwin, {\it {Factorization of the Drell-Yan Cross-Section in
  Perturbation Theory}},  {\em Phys.Rev.} {\bf D31} (1985) 2616.

\bibitem{glauber_cancel_3}
J.~C. Collins, D.~E. Soper, and G.~F. Sterman, {\it {Factorization for Short
  Distance Hadron - Hadron Scattering}},  {\em Nucl.Phys.} {\bf B261} (1985)
  104.

\bibitem{glauber_cancel_4}
S.~M. Aybat and G.~F. Sterman, {\it {Soft-Gluon Cancellation, Phases and
  Factorization with Initial-State Partons}},  {\em Phys.Lett.} {\bf B671}
  (2009) 46--50, [\href{http://xxx.lanl.gov/abs/0811.0246}{{\tt
  arXiv:0811.0246}}].

\bibitem{kT}
S.~D. Ellis and D.~E. Soper, {\it {Successive combination jet algorithm for
  hadron collisions}},  {\em Phys.Rev.} {\bf D48} (1993) 3160--3166,
  [\href{http://xxx.lanl.gov/abs/hep-ph/9305266}{{\tt hep-ph/9305266}}].

\bibitem{CA1}
Y.~L. Dokshitzer, G.~Leder, S.~Moretti, and B.~Webber, {\it {Better jet
  clustering algorithms}},  {\em JHEP} {\bf 9708} (1997) 001,
  [\href{http://xxx.lanl.gov/abs/hep-ph/9707323}{{\tt hep-ph/9707323}}].

\bibitem{CA2}
M.~Wobisch and T.~Wengler, {\it {Hadronization corrections to jet
  cross-sections in deep inelastic scattering}},
  \href{http://xxx.lanl.gov/abs/hep-ph/9907280}{{\tt hep-ph/9907280}}.

\bibitem{anti-kT}
M.~Cacciari, G.~P. Salam, and G.~Soyez, {\it {The Anti-k(t) jet clustering
  algorithm}},  {\em JHEP} {\bf 0804} (2008) 063,
  [\href{http://xxx.lanl.gov/abs/0802.1189}{{\tt arXiv:0802.1189}}].

\bibitem{aMC@NLO}
J.~Alwall, R.~Frederix, S.~Frixione, V.~Hirschi, F.~Maltoni, {\em et.~al.},
  {\it {The automated computation of tree-level and next-to-leading order
  differential cross sections, and their matching to parton shower
  simulations}},  \href{http://xxx.lanl.gov/abs/1405.0301}{{\tt
  arXiv:1405.0301}}.

\bibitem{Moch:2004pa}
S.~Moch, J.~Vermaseren, and A.~Vogt, {\it {The Three loop splitting functions
  in QCD: The Nonsinglet case}},  {\em Nucl.Phys.} {\bf B688} (2004) 101--134,
  [\href{http://xxx.lanl.gov/abs/hep-ph/0403192}{{\tt hep-ph/0403192}}].

\bibitem{ResumPi2:1}
V.~Ahrens, T.~Becher, M.~Neubert, and L.~L. Yang, {\it {Origin of the Large
  Perturbative Corrections to Higgs Production at Hadron Colliders}},  {\em
  Phys.Rev.} {\bf D79} (2009) 033013,
  [\href{http://xxx.lanl.gov/abs/0808.3008}{{\tt arXiv:0808.3008}}].

\bibitem{Pythia6}
T.~Sjostrand, S.~Mrenna, and P.~Z. Skands, {\it {PYTHIA 6.4 Physics and
  Manual}},  {\em JHEP} {\bf 0605} (2006) 026,
  [\href{http://xxx.lanl.gov/abs/hep-ph/0603175}{{\tt hep-ph/0603175}}].

\bibitem{MC@NLO}
S.~Frixione and B.~R. Webber, {\it {Matching NLO QCD computations and parton
  shower simulations}},  {\em JHEP} {\bf 0206} (2002) 029,
  [\href{http://xxx.lanl.gov/abs/hep-ph/0204244}{{\tt hep-ph/0204244}}].

\bibitem{Herwig6}
G.~Corcella, I.~Knowles, G.~Marchesini, S.~Moretti, K.~Odagiri, {\em et.~al.},
  {\it {HERWIG 6: An Event generator for hadron emission reactions with
  interfering gluons (including supersymmetric processes)}},  {\em JHEP} {\bf
  0101} (2001) 010, [\href{http://xxx.lanl.gov/abs/hep-ph/0011363}{{\tt
  hep-ph/0011363}}].

\bibitem{Powheg:0}
T.~Melia, P.~Nason, R.~Rontsch, and G.~Zanderighi, {\it {W+W-, WZ and ZZ
  production in the POWHEG BOX}},  {\em JHEP} {\bf 1111} (2011) 078,
  [\href{http://xxx.lanl.gov/abs/1107.5051}{{\tt arXiv:1107.5051}}].

\bibitem{Powheg:1}
P.~Nason, {\it {A New method for combining NLO QCD with shower Monte Carlo
  algorithms}},  {\em JHEP} {\bf 0411} (2004) 040,
  [\href{http://xxx.lanl.gov/abs/hep-ph/0409146}{{\tt hep-ph/0409146}}].

\bibitem{Powheg:2}
S.~Frixione, P.~Nason, and C.~Oleari, {\it {Matching NLO QCD computations with
  Parton Shower simulations: the POWHEG method}},  {\em JHEP} {\bf 0711} (2007)
  070, [\href{http://xxx.lanl.gov/abs/0709.2092}{{\tt arXiv:0709.2092}}].

\bibitem{Powheg:3}
S.~Alioli, P.~Nason, C.~Oleari, and E.~Re, {\it {A general framework for
  implementing NLO calculations in shower Monte Carlo programs: the POWHEG
  BOX}},  {\em JHEP} {\bf 1006} (2010) 043,
  [\href{http://xxx.lanl.gov/abs/1002.2581}{{\tt arXiv:1002.2581}}].

\bibitem{FastJet:1}
M.~Cacciari and G.~P. Salam, {\it {Dispelling the $N^{3}$ myth for the $k_t$
  jet-finder}},  {\em Phys.Lett.} {\bf B641} (2006) 57--61,
  [\href{http://xxx.lanl.gov/abs/hep-ph/0512210}{{\tt hep-ph/0512210}}].

\bibitem{FastJet:2}
M.~Cacciari, G.~P. Salam, and G.~Soyez, {\it {FastJet User Manual}},  {\em
  Eur.Phys.J.} {\bf C72} (2012) 1896,
  [\href{http://xxx.lanl.gov/abs/1111.6097}{{\tt arXiv:1111.6097}}].

\bibitem{HiggsIntf:veto}
I.~Moult and I.~W. Stewart, {\it {Jet Vetoes Interfering with $H \rightarrow
  WW$}},  \href{http://xxx.lanl.gov/abs/1405.5534}{{\tt arXiv:1405.5534}}.

\end{thebibliography}

\providecommand{\href}[2]{#2}\begingroup\raggedright\endgroup

\end{document}